\documentclass[a4paper,twoside,prd,showpacs,preprintnumbers,twocolumn,superscriptaddress,nofootinbib]{revtex4-1}
\usepackage{amssymb,amsmath,bm,natbib}
\usepackage{color}
\usepackage{slashed}
\usepackage{graphics}
\usepackage{graphicx}
\usepackage[utf8]{inputenc}
\usepackage[caption=false]{subfig}
\usepackage{hyperref}
\usepackage{url}
\usepackage{dsfont}
\usepackage{float} 
\usepackage{cancel}
\usepackage{xcolor,colortbl}
\usepackage{rotating}
\usepackage[export]{adjustbox}
\usepackage{subfig}

\newcommand{\B}{{\mathcal B}}
\newcommand{\Lcal}{{\mathcal L}}
\newcommand{\Ft}{\mathcal{F}t}

\newcommand{\eq}[1]{Eq.~\eqref{#1}}

\newcommand{\e}{\ensuremath{\mathrm{e}}}

\definecolor{Gray}{gray}{0.7}

\begin{document}
\preprint{PSI-PR-20-07,   ZU-TH  19/20}

	\title{Explaining $b\to s\ell^+\ell^-$ and the Cabibbo Angle Anomaly with a Vector Triplet}

\author{Bernat Capdevila}
\email{bernat.capdevilasoler@unito.it}
\affiliation{Universita di Torino and INFN Sezione di Torino, Via P. Giuria 1, Torino I-10125, Italy}

\author{Andreas Crivellin}
\email{andreas.crivellin@cern.ch}
\affiliation{Physik-Institut, Universit\"at Z\"urich, Winterthurerstrasse 190, CH--8057 Z\"urich, Switzerland}
\affiliation{Paul Scherrer Institut, CH--5232 Villigen PSI, Switzerland}

\author{Claudio Andrea Manzari}
\email{claudioandrea.manzari@physik.uzh.ch}
\affiliation{Physik-Institut, Universit\"at Z\"urich, Winterthurerstrasse 190, CH--8057 Z\"urich, Switzerland}
\affiliation{Paul Scherrer Institut, CH--5232 Villigen PSI, Switzerland}

\author{Marc Montull}
\email{marc.montull@gmail.com}
\affiliation{Physik-Institut, Universit\"at Z\"urich, Winterthurerstrasse 190, CH--8057 Z\"urich, Switzerland}
\affiliation{Paul Scherrer Institut, CH--5232 Villigen PSI, Switzerland}

\begin{abstract}
The  most statistically significant hints for new physics in the flavor sector are discrepancies between theory and experiment in $B$ decays to lepton pairs ($b\to s\ell^+\ell^-$) and a deficit in $1^{\rm st}$ row CKM unitarity (the Cabibbo Angle anomaly). We propose that these anomalies can be reconciled by a simplified model with massive gauge bosons transforming in the adjoint representation of $SU(2)_L$. After calculating the impact of this model on $B$ decays, observables testing charged current lepton flavor universality (LFU), electro-weak precision observables and LHC searches we perform a global fit to all available data. We find that our model can provide a consistent common explanation of both anomalies and that the fit to the data is more than $7\,\sigma$ better than the fit of the SM. The model also predicts interesting correlations between LFU violation in the charged current and $b\to s\ell^+\ell^-$ data which can be tested experimentally in the near future.
\end{abstract}

\maketitle

\section{Introduction}
\label{Introduction}

In 2012, the LHC confirmed the predictions of the Standard Model (SM) of particle physics by discovering the (Brout-Englert) Higgs boson~\cite{Aad:2012tfa,Chatrchyan:2012xdj}. However, so far no particles beyond those of the SM have been observed in high energy searches. Therefore, great hopes of finding physics beyond the SM rest on the low energy precision frontier. Here, fortunately, flavor experiments have accumulated intriguing hints for new physics (NP) within the recent years. Among them, the most statistically significant are discrepancies between the SM predictions and experiments in semileptonic $B$ meson decays involving $b\to s\ell^+\ell^-$ transitions and a (apparent) violation  of 1$^{\rm st}$ row CKM unitarity, known as the ``Cabibbo Angle Anomaly'' (CAA).

In decays involving $b\!\to\! s\ell^+\ell^-$ transitions at the constituent level, LHCb measurements~\cite{Aaij:2017vbb,Aaij:2019wad} indicate a deficit in muons with respect to electrons, i.e. a violation of lepton flavor universality (LFU) with a combined significance of $\approx4\sigma$~\cite{Capdevila:2017bsm, Altmannshofer:2017yso, DAmico:2017mtc, Ciuchini:2017mik, Hiller:2017bzc, Geng:2017svp,Hurth:2017hxg,Alguero:2019ptt,Aebischer:2019mlg,Ciuchini:2019usw,Arbey:2019duh}\footnote{We do not consider LFU violation in charged current $B$ decays here, were hints for LFU violation in $b\to c\tau\nu$ transitions at the $3\,\sigma$ level were observed~\cite{Amhis:2019ckw}. Explaining these observables would require NP at the 10\% level while we are here considering effects below the percent level.}. Furthermore, this observation is consistent with many other measurements involving the same current, in particular angular observables~\cite{Matias:2012xw,Descotes-Genon:2013vna} where the data also shows a deficit in muonic channels~\cite{Aaij:2015oid,Aaij:2020nrf}. In fact, the data gives rise to a consistent pattern such that the most up-to-date global analyses find several NP scenarios to be preferred over the SM at the $5-6\sigma$ level~\cite{Alguero:2019ptt,Aebischer:2019mlg,Ciuchini:2019usw}.

The CAA is due to the fact that $1^{\rm st}$ row CKM unitarity is violated, i.e. one observes $|V_{ud}^2|+|V_{us}^2|+|V_{ub}^2|<1$ with a significance of $\approx 4\,\sigma$~\cite{Belfatto:2019swo,Grossman:2019bzp,Shiells:2020fqp}. Equivalently, this means that there is a disagreement between the CKM element $V_{us}$ extracted from kaon and tau decays and the one determined from beta decays (using CKM unitarity). Interestingly, this discrepancy can also be interpreted as a sign of LFUV~\cite{Coutinho:2019aiy,Crivellin:2020lzu,Endo:2020tkb}
where the sensitivity to NP in the determination via beta decays is enhanced by a factor of $V_{ud}^2/V_{us}^2$ compared to the NP sensitivity of $V_{us}$ from kaon or tau decays~\cite{Crivellin:2020lzu}~\footnote{Alternatively, it can be interpreted as a sign of (apparent) CKM unitarity violation~\cite{Belfatto:2019swo,Cheung:2020vqm}. However, a sizeable violation of CKM unitarity is in general difficult due to the strong bounds from flavor-changing neutral currents, such as kaon mixing (see  Ref.~\cite{Bobeth:2016llm}) Furthermore, a right-handed $W$ coupling~\cite{Bernard:2007cf,Crivellin:2009sd} can only partially account for it~\cite{Grossman:2019bzp}.}.

Since both the $b\!\to\! s\ell^+\ell^-$ data and the CAA are related to LFU violation in the muon/electron sector, it seems plausible that a connection between them exists and it is both interesting and important to explore which NP models can provide a common explanation. In order to account for the CAA, NP must in some way be related to the charged current, which can be achieved in the form of modified $W\mu\nu$ couplings and/or by effects in $\bar u d e\nu$ operators. Both of these possibilities can be realized with a $W^\prime$ boson coupling to left-handed SM fermions; the first one via $W-W^\prime$ mixing, the second one through a tree-level contribution. Furthermore, due to $SU(2)_L$ gauge invariance, a left-handed $W^\prime$ boson always comes together with a left-handed $Z^\prime$~\cite{PerezVictoria:2011uj} which is a prime candidate for an explanation of the $b\to s\ell^+\ell^-$ anomalies~\cite{Buras:2013qja,Gauld:2013qba,Gauld:2013qja,Altmannshofer:2014cfa,Crivellin:2015mga,Crivellin:2015lwa,Niehoff:2015bfa,Carmona:2015ena,Falkowski:2015zwa,Celis:2015eqs,Celis:2015ara,Crivellin:2015era,Crivellin:2016ejn,GarciaGarcia:2016nvr,Altmannshofer:2016oaq,Faisel:2017glo,King:2017anf,Chiang:2017hlj,DiChiara:2017cjq,Ko:2017lzd,Sannino:2017utc,Falkowski:2018dsl,Benavides:2018rgh,Maji:2018gvz,Singirala:2018mio,Guadagnoli:2018ojc,Allanach:2018lvl,Duan:2018akc,King:2018fcg,Kohda:2018xbc,Dwivedi:2019uqd,Foldenauer:2019vgn,Ko:2019tts,Allanach:2019iiy,Altmannshofer:2019xda,Calibbi:2019lvs,Aebischer:2019blw}. This obviously opens up the possibility of addressing both discrepancies simultaneously.

A minimal dynamical model of a left-handed $Z^\prime$ and $W^\prime$ is obtained by extending the SM with massive vector bosons transforming in the adjoint representation (or equivalently as a triplet) of $SU(2)_L$ and with zero hyper-charge~\cite{delAguila:2010mx,deBlas:2012qp,Pappadopulo:2014qza}. This Lagrangian can be generated by various NP models, for instance composite Higgs and extradimensional models~\cite{ArkaniHamed:1998rs,Rizzo:1999en,Csaki:2004ay,Rizzo:2009pu,Bella:2010vi,Bella:2010sc,Contino:2006nn,Contino:2011np,Bellazzini:2012tv} or models based on $SU(2)_1\times SU(2)_2$~\cite{Li:1981nk,Muller:1996dj,Morrissey:2005uza,Chiang:2009kb,Fuentes-Martin:2014fxa}~\footnote{In the last years, such models have also been studied in an effort to explain $b\to s\ell^+\ell^-$ data together with the $R(D^{(*)})$ anomalies~\cite{He:2012zp,Bhattacharya:2014wla,Alonso:2014csa,Greljo:2015mma,Boucenna:2016wpr,Boucenna:2016qad,Bhattacharya:2016mcc,Buttazzo:2017ixm,Kumar:2018kmr,Blanke:2018sro}.}. Because of the many possible UV completions, we find that in order to understand the effects of generic vector triplets, it is convenient to focus on a simplified model. To determine the viability of such a heavy vector $SU(2)_L$ triplet, we will perform a global fit including all relevant observables that are modified, i.e. $b\to s\ell^+\ell^-$, the CAA, LFU tests of the charged current (e.g.
 $\pi\to\mu\nu/\pi\to e\nu$), as well as electroweak (EW) precision data (e.g. $Z$ pole data), {LEP-II constrains on four-fermion contact interactions} and direct LHC searches for vector resonances.

\section{Simplified Model}
\label{Setup}

\begin{figure}[t!]
	\centering
	\includegraphics[width=0.95\linewidth]{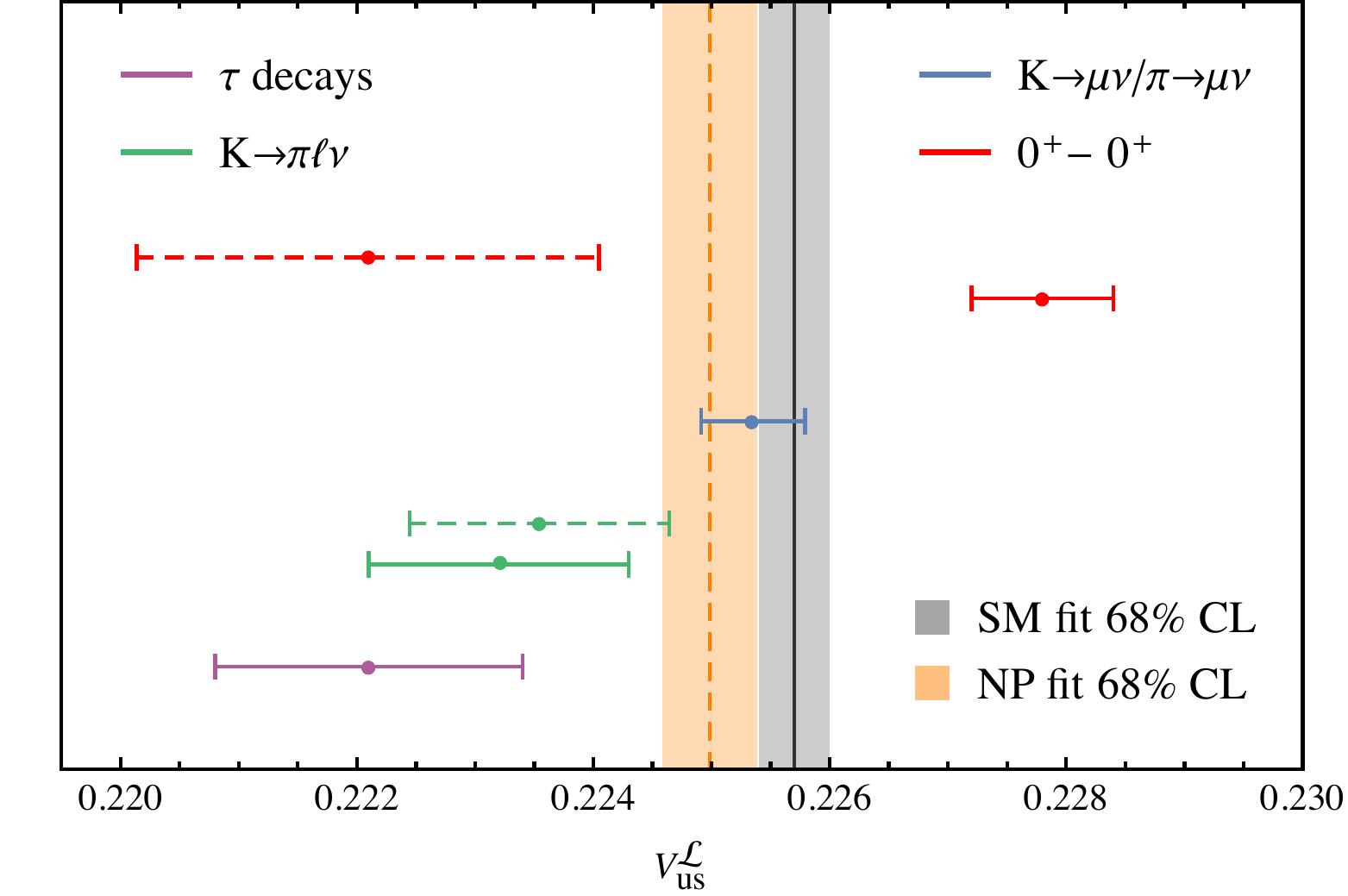}
	\caption{Comparison of the different determinations of $V_{us}$ from kaon decays~\cite{Tanabashi:2018oca}, tau decays~\cite{Amhis:2019ckw} and superallowed beta decays~\cite{Hardy:2018zsb} 
		resulting in the CAA. For the latter, due to its smaller error, we will use the SGPR determination, which was recently confirmed by Ref.~\cite{Seng:2020wjq}, in our numerical analysis. However, we confirmed that using CMS instead only has a minor impact on our results. The dashed lines indicate the posteriors for $V_{us}^{\mathcal L}$ in our NP fit extracted from the corresponding modes.}
	\label{VusPlot}
\end{figure}

In our pursue of a common explanation of $b\to s\ell^+\ell^-$ data and the CAA, we supplement the SM by an $SU(2)_L$ triplet of heavy vector bosons $X_\mu^a$ (with $a=1,2,3$)with  zero hypercharge~\cite{deBlas:2012qp,Pappadopulo:2014qza} and mass $M_X$\footnote{Neglecting small $SU(2)$ breaking effects, the masses of the $Z^\prime$ and the $W^\prime$ bosons are the same since they originate from the same SM representation.}. Following the conventions of Ref.~\cite{deBlas:2012qp} we write 
\begin{align}
\Lcal_X &=  -\frac{1}{2}\left[D_{\mu} X_{\nu}\right]^{a}\left[D^{\mu} X^{\nu}\right]_{a}+\frac{1}{2}\left[D_{\mu} X_{\nu}\right]^{a}\left[D^{\nu} X^{\mu}\right]_{a} \nonumber \\
&+ \frac{\mu_{X}^{2}}{2} X_{\mu}^{a} X_{a}^{\mu} - g_{ji}^\ell X_a^\mu {{\bar \ell }_j}{\gamma _\mu }\frac{{{\sigma ^a}}}{2} {\ell _i} - g_{ji}^qX_a^\mu {{\bar q}_j}{\gamma _\mu }\frac{{{\sigma ^a}}}{2} {q_i}   \nonumber\\
& - \Big( {ig_X^{D\phi }X_a^\mu {\phi ^\dag }\frac{{{\sigma ^a}}}{2}{D_\mu }\phi  + h.c.} \Big) +g_X^\phi X_\mu ^aX_a^\mu {\phi ^\dag }\phi \,,
\label{eq:LX} 
\end{align}
where $D_\mu = \partial_\mu + i g_2 \sigma^a W^{a (0)}_{\mu} /2 + i g_1 Y B^{(0)}_\mu $,  $\sigma^a$ are the Pauli matrices and $W^{a (0)}, B^{(0)}$ correspond, in the absence of $SU(2)_L$ breaking, to the SM gauge bosons. The first two terms in $\Lcal_X$ generate the interactions of the new gauge bosons with the SM ones while the third term gives them masses even before  EW symmetry breaking (EWSB). The terms proportional to $g_{ji}^{q(\ell)}$ parametrize the couplings of the new gauge bosons to left-handed quarks (leptons) and the term containing $g_X^{D\phi}$ gives rise to a mass mixing between $X^\mu_a$ and the SM gauge bosons after EWSB. The last term in $\Lcal_X$ creates interactions between $X_\mu^a$ and the SM Higgs, which gives an additional contribution to their mass after EWSB.

The mass spectrum of the gauge bosons contains a zero mass eigenstate identified with the photon $A_\mu$ which does not mix with $X^a_\mu$. Hence the SM relation between $g, g'$ and the measured fine structure constant $\alpha$ is not modified~\cite{Pappadopulo:2014qza}. Thus, one can consider the mass matrices after EWSB in the basis {$(Z^{(0)},X^3)$}. Taking $\langle \phi \rangle = ( 0 ,\, v/\sqrt{2})^T$, we have
\begin{equation}
\begin{array}{l}
M_0^2 = \left( {\begin{array}{*{20}{c}}
	{M_{{Z^{(0)}}}^2}&{\frac{x}{{c_W}}}\\
	{\frac{x^*}{{c_W}}}&{M_X^2}
	\end{array}} \right)\,,\;\;
M_ \pm ^2 = \left( {\begin{array}{*{20}{c}}
	{M_{{W^{(0)}}}^2}&x\\
	x^*&{M_X^2}
	\end{array}} \right)
\end{array},
\label{eq:massMatrix}\nonumber
\end{equation}
where the superscript $(0)$ refers to the SM fields in the absence of mixing,
$M_{{W^{(0)}}} = g_2 v/2$ and  $M_{{Z^{(0)}}}^{} = M_{{W^{(0)}}}^{}/{c_W}$. {The mass squared of the new gauge boson is $M_X^2$, $x = M_{W^{(0)}} (g_X^{D\phi} v/2)$ and $c_W \equiv g_2/\sqrt{g_2^2+g_1^2}$ is the cosine of the Weinberg angle.} Provided that $|x| \ll M_X$ one can work in the approximation $M_{W^\prime}\approx M_{Z^\prime} \approx M_X$ while the $M_W$ and $M_Z$ masses are shifted by
\begin{align}
\frac{M_W^2}{M_{{W^{(0)}}}^2} &\approx  \frac{M_Z^2}{M_{{Z^{(0)}}}^2}  \approx \Big( 1 - \frac{M_X^2}{{{ M_{Z^{(0)}}^2}}} \sin^2\alpha_{ZZ^{\prime}} \Big)\,.
\label{eq:MassRatios}
\end{align}
respecting the SM tree-level relation $M_{W^{(0)}}=c_W M_{Z^{(0)}}$. When mixing is present, the eigenvalues are linear combinations of  $( Z^{(0)},X^3)$ and $(W^{(0)},X^\pm)$ which for $|x| \ll M_X$ yield the following mixing angles
\begin{equation}
\sin \alpha_{ZZ^\prime}  \approx {x}/({{M_X^2c_W}}),\qquad \sin \alpha_{WW^\prime}  \approx {x}/({{M_X^2}})\,.
\end{equation}
The mass eigenstates $Z^{(\prime)}$ can then be expressed as
\begin{equation}
\left( {\begin{array}{*{20}{c}}
	{Z^\prime} \\
	Z
	\end{array}} \right) = \left( {\begin{array}{*{20}{c}}
	{{X^3}\cos \alpha_{ZZ^\prime} \; - Z^{(0)} \sin \alpha_{ZZ^\prime} }\\
	{{X^3}\sin \alpha_{ZZ^\prime}  + Z^{(0)} \cos \alpha_{ZZ^\prime} }
	\end{array}} \right)
\label{eq:mixing}
\end{equation}
and similarly for the charged gauge bosons $W$ and $W^\prime$. In the end we have two additional (compared to the SM) mass eigenstates, a charged $W^\prime$ and a neutral $Z^\prime$. The couplings of the new $W^\prime$ and $Z^\prime$ to fermions (in the down-quark basis) are given by
\begin{align}
\label{eq:ZpWpQuarkCouplings}
\Lcal_{q\ell}^{Z^\prime W^\prime} =  & \frac{g_{ji}^q}{2} ({\bar d}_j {\gamma ^\mu }{P_L}{d_i}) \, Z^\prime_\mu  - \frac{V_{jk} g_{kk'}^qV_{ik'}^*}{2} ({\bar u}_j {\gamma ^\mu }{P_L}{u_i}) \, {Z^\prime_\mu } \nonumber \\
&- \Big( \frac{{V_{jk}}g_{ki}^q}{\sqrt{2}} ({{\bar u}_j}{\gamma ^\mu }{P_L}{d_i}) \, {W_\mu^{\prime} } + h.c. \Big) \\
& +\frac{g_{ji}^\ell}{2} ({\bar \ell}_j {\gamma ^\mu }{P_L}{\ell_i}) \, Z^\prime_\mu  - \frac{g_{ji}^\ell}{2} ({\bar \nu}_j {\gamma ^\mu }{P_L}{\nu_i}) \, {Z^\prime_\mu } \nonumber \\
&- \Big( \frac{g_{ji}^\ell}{\sqrt{2}} ({{\bar \nu}_j}{\gamma ^\mu }{P_L}{\ell_i}) \, {W_\mu^{\prime} } + h.c. \Big) \,,\nonumber
\end{align}
where $V_{ij}$ is the CKM matrix. Note that one can neglect gauge boson mixing effects in the couplings of the new heavy bosons to SM fermions as this would lead to dim-8 operators. These operators can be neglected in our phenomenological analysis where we will assume that the $Z'$, $W'$ couplings to quarks respect an (approximate) $U(2)^3$ flavor symmetry~\cite{Barbieri:1995uv,Barbieri:1997tu,Barbieri:2011fc,Barbieri:2011ci,Crivellin:2011fb,Barbieri:2012uh,Barbieri:2012bh,Buras:2012sd} \footnote{Note that this differs from ``standard" minimal flavor violation~\cite{Chivukula:1987fw,Hall:1990ac,Buras:2000dm} (MFV) which is based on $U(3)^3$~\cite{DAmbrosio:2002vsn}, however, $U(3)^3$ is anyway strongly broken to $U(2)^3$ by the large third-generation Yukawa couplings.}. This means that to a good approximation
\begin{equation}
g^{ud}_{11,22}\approx g_{11,22}^q \equiv g^q,\; g^{ud}_{12} = V_{us} g^q,\; g^{d}_{23}= {\cal O}(V_{cb}).
\end{equation}
Note that in this setup the $Z^\prime$  coupling $g^d_{12}$ is of third order in the Wolfenstein parameter~\cite{Crivellin:2015lwa,Calibbi:2019lvs}, i.e. $O(10^{-3})$ so that with this ansatz for the $Z^\prime$ couplings dangerously large effects in $K-\bar K$ and/or $D-\bar D$ mixing are avoided. Since the coupling $g_{33}^q$ does not affect the observables that we consider (except for a suppressed effects in LHC searches) we disregard it from this time forward.

Concerning the couplings to fermions of the SM gauge boson mixing effects are important and for leptons we have
\begin{eqnarray}
\Lcal_{W, Z}&=&\dfrac{g_2}{2 c_{W}} \Big[\bar{\ell}_{j} \gamma^{\mu} ( \Delta_{ji} \, P_{L} -2 s_{W}^{2} \delta_{j i} )  \, \ell_{i} Z_{\mu}  \label{eq:mixing}  \\
&-&\sqrt{2} \, \Delta_{ji} \, (\bar{\nu}_{j} \gamma^{\mu} P_{L} \ell_{i}) W_{\mu} - \Delta_{ji} (\bar{\nu}_{j} \gamma^{\mu} P_{L} \nu_{i}) Z_{\mu}\Big] \,,  \nonumber
\end{eqnarray}
with 
\begin{eqnarray}
\Delta_{j i}= \delta_{j i}+c_W\frac{g^\ell_{ji}}{g_2}\sin \alpha_{ZZ^\prime}\,.
\end{eqnarray}
The analogous expressions for quarks are obtained in a straightforward way. Note that since the $Z^\prime$ couples to $\bar s b$, flavor violating couplings for the $Z$ are induced.

\begin{figure*}[t!]
	\centering
	\includegraphics[width=\linewidth]{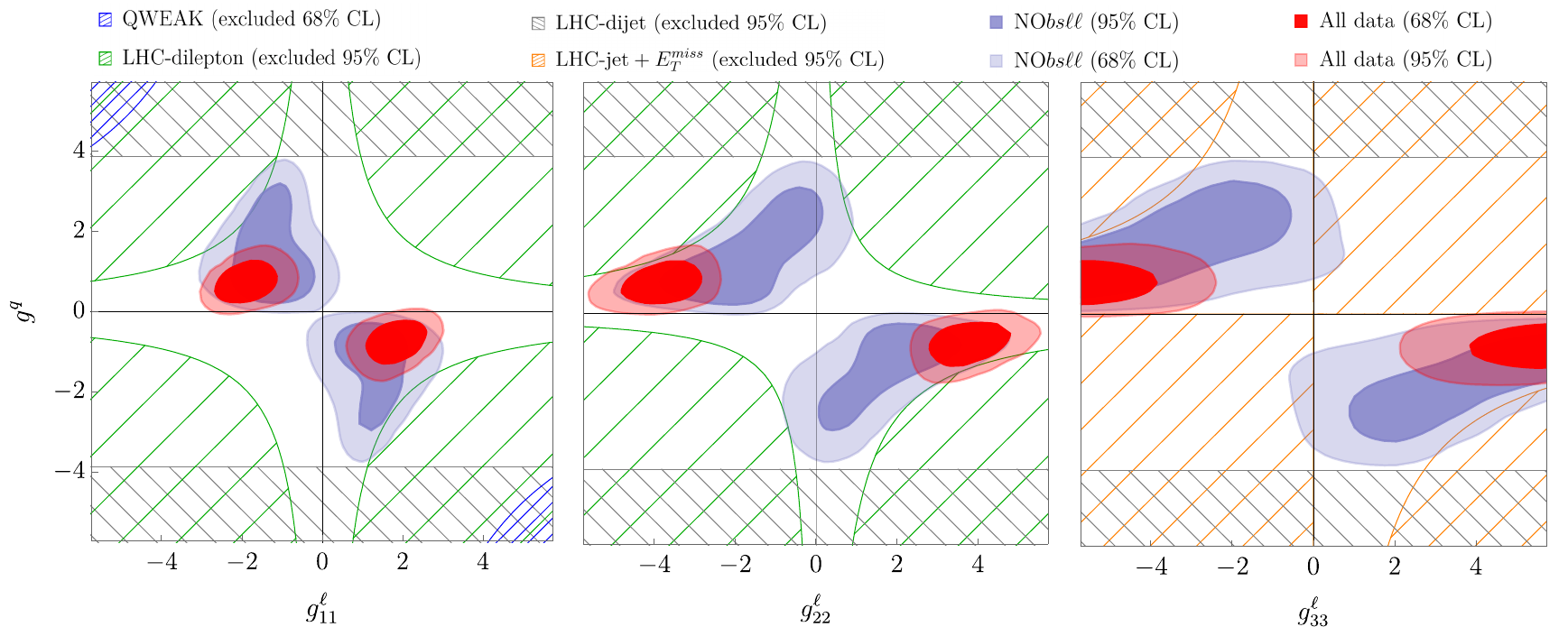}
	\caption{Global fit in the $g^\ell_{11}-g^q$, $g^\ell_{22}-g^q$ and $g^\ell_{33}-g^q$ planes for $M_X=10\,$TeV. Even though we included the LHC measurements into our global fit, we display them as well as hatched regions to show their constraining power and to verify that treating them as a hard cut would not change our results.}
	\label{fig.global}
\end{figure*}

\section{Observables}
\label{Observables}

Here we review the most relevant observables which are modified (compared to the SM) within our model. {These can be grouped into nine categories: 1) Charged Lepton Flavour Violation, 2) Electroweak Precision Observables, (e.g. $Z$ pole measurements), 3) LFU tests of the charged current, 4) the CAA,  5) $b\to s\ell^+\ell^-$ data, 6) $B_s-\bar B_s$ mixing,  7) LHC searches, 8) bounds from parity violation in cesium and $e p \to e p$ scattering (APV experiment and QWEAK collaboration), 9) LEP-II bounds. For details the interested reader can refer to Appendix \ref{app:Observables} while in the following we provide a small summary of how these observables set bounds to our model:}

1) Flavor violating decays of charged leptons constrain the off-diagonal elements $g^\ell_{ji}$ to be small~\cite{Bertl:2006up,Aubert:2009ag,TheMEG:2016wtm,Amhis:2019ckw} . Furthermore, since these elements do not interfere with the SM contribution in flavor conserving observables (their effects there are suppressed by $1/M_{X}^4$), they will not be considered in the following. Therefore, we will only consider $g^\ell_{11}$, $g^\ell_{22}$ and $g^\ell_{33}$ in the following.

2) Among the electroweak precision observables the quantities $G_F$, $\alpha_{em}$ and $M_Z$ have been measured with the highest accuracy~\cite{ALEPH:2005ab,Tanabashi:2018oca}. Therefore, they are commonly taken as Lagrangian parameters (fixed to their experimental values) and used to calculate all other EW observables~\cite{ALEPH:2005ab,Tanabashi:2018oca,Aaltonen:2016nuy,Chatrchyan:2011ya,Chatrchyan:2011ya,Aaij:2015lka} within the SM. Beyond the SM, this method can still be used, but the relations between the Lagrangian parameters and the measurements are changed. In particular, in our model the relations for the Fermi constant $G_F$, $M_Z$ and $M_W$ are modified. 
\begin{align}
\label{eq:GF}
G_F &= G_F^{\mathcal L} + \frac{{g_{11}^\ell g_{22}^\ell }}{{4 \sqrt{2} M_{X}^2}} \,,
\end{align}
where $G_F^{\mathcal L}$ is the Fermi constant in the Lagrangian which only equals the measured one within the pure SM. Furthermore, according to \eq{eq:MassRatios} the relation between the SM gauge bosons masses and the measured ones (which can be calculated as a function of $\alpha$, $M_Z$ and $G_F$ as well as the top and Higgs mass) is modified. The related list of observables, given in Appendix~\ref{app:Observables}, is calculated by and implemented in HEPfit~\cite{deBlas:2019okz} to which we added the modifications within our model. We also include as inputs in the EW fit the  Higgs~\cite{Aaboud:2018wps, CMS:2019drq} and top~\cite{TevatronElectroweakWorkingGroup:2016lid,  Aaboud:2018zbu,Sirunyan:2018mlv} masses together with $\alpha_s$~\cite{Tanabashi:2018oca} which enter the fit via loop effects. 

3) Deviations from LFU in the charged current can be measured in ratios like ${\pi\rightarrow\mu\nu}/{\pi\rightarrow e\nu}$ or ${\tau\rightarrow\mu\nu\bar{\nu}}/{\tau\rightarrow e\nu\bar{\nu}}$. Here we have modifications from tree-level $W$ exchange leading to 
\begin{eqnarray}
R\left[\frac{{{\tau  \to \mu \nu \nu }}}{{{\tau  \to e\nu \nu}}}\right] &=& \Big( {1 + \frac{{g_{33}^\ell \left( {g_{22}^\ell  - g_{11}^\ell } \right)}}{{g^2_2}}\frac{{M_W^2}}{{M_{W'}^2}}} \Big) \,. \nonumber \\
R\left[\frac{{{\pi  \to \mu \nu }}}{{{\pi  \to e\nu }}}\right] &=&\Big( {1 + \frac{{g^{q}_{}\left( {g_{22}^\ell  - g_{11}^\ell } \right)}}{{g^2_2}}\frac{{M_W^2}}{{M_{W'}^2}}} \Big) \,,
\label{LFUratios}
\end{eqnarray}
defined at the amplitude level and similarly for the other observables given in Appendix~\ref{app:Observables}. In addition, the modified $W$ couplings induced by $W$-$W^\prime$ mixing in \eq{eq:mixing} simply rescale the SM amplitudes. We implemented these modifications of the results given in Refs.~\cite{Lazzeroni:2012cx,Ambrosino:2009aa,Cirigliano:2007xi,Pich:2013lsa, Czapek:1993kc,Britton:1992pg,Bryman:1982em,Cirigliano:2007xi,Aguilar-Arevalo:2015cdf,Tanabashi:2018oca,Amhis:2019ckw,Tanabashi:2018oca,Antonelli:2010yf,Cirigliano:2011ny,Pich:2013lsa,Amhis:2019ckw,Tanabashi:2018oca} into HEPfit (see the Appendix for details). 

4) As outlined in the Introduction, the Lagrangian parameter $V_{us}^{\mathcal L}$ of the (unitary) CKM differs from the one extracted from the experiment assuming only the SM, which can be determined from kaon~\cite{Tanabashi:2018oca}, tau~\cite{Amhis:2019ckw} or beta decays; in particular superallowed beta decays~\cite{Czarnecki:2019mwq,Seng:2018yzq, Seng:2020wjq}. Note that only within the SM $V_{us}^{\mathcal L}=V_{us}$. This situation is illustrated in Fig.~\ref{VusPlot} where the different determinations of $V_{us}$ are compared. In particular, one can see that the $V_{us}$ from kaon and tau decays is significantly smaller than the $V_{us}$ obtained from superallowed beta decays via CKM unitarity. In our model, the $V_{us}^{\mathcal L}$ determination from superallowed beta decays receives, in addition to the direct modification of transition $d\to u\e \nu$ through tree-level $W^\prime$ exchange, an indirect one from the modification of $G_F$ from \eqref{eq:GF}. Therefore, the element of the unitary CKM matrix in the Lagrangian $V_{us}^{\mathcal L}$ is given in terms of the one extracted from the experiment with the SM $V_{us}^\beta$ as
\begin{equation}
V_{us}^{\mathcal L} \approx V_{us}^\beta \left( {1 + \frac{{{{\left| {V_{ud}^{\mathcal L}} \right|}^2}}}{{{{\left| {V_{us}^{\mathcal L}} \right|}^2}}}\frac{{g_{11}^\ell \left( {g^{q} - g_{22}^\ell } \right)}}{{g_2^2}}\frac{{M_W^2}}{{M_{X}^2}}} \right)\,.
\end{equation}
Similarly, the $V_{us}$ from $K\to\pi\ell\nu$ is affected, even though there is no enhancement factor of ${{{{\left| {V_{ud}^{\mathcal L}}/ \right|}^2}}}{{{{\left| {V_{us}^{\mathcal L}} \right|}^2}}}$ while $K\to\mu\nu/\pi\to\mu\nu$ and tau decays remain unchanged (see Appendix~\ref{app:Observables}).

5) For $b\to s\ell^+\ell^-$ our $Z^\prime$ contribution is purely left-handed but supplemented by a lepton flavor universal effect with an axial vector current on the lepton side resulting from the $Z-Z^\prime$ mixing. Such a scenario is known for being capable of providing a good fit to data~\cite{Crivellin:2019dun} including $R(K^{(*)})$.  In fact, from the analysis of all available $b\to s\ell^+\ell^-$ data (using the method and program of Ref.~\cite{Descotes-Genon:2015uva} with the data set given in Ref.~\cite{Alguero:2019ptt}) we find that our scenario yields a pull of $6.2\,\sigma$ with respect to the SM hypothesis.

6) The most important constraint on $Z^\prime-b-s$ couplings, i.e. $g_{23}^d$ in Eq.~\eqref{eq:ZpWpQuarkCouplings}, comes from $B_s-\bar B_s$ mixing. Here the SM agrees with experiment at the $1\,\sigma$ level but there is still space for NP of the order of 20\% compared to the SM~\cite{Bona:2007vi}. 

7) In our phenomenological analysis we will consider very heavy $Z^\prime$ and $W^\prime$ bosons, which cannot be produced on-shell at the LHC. In this case bounds from the tails of dijet~\cite{Sirunyan:2017ygf} and dilepton~\cite{Aaboud:2017buh} distributions apply. This allows us to put bounds on the $Z^\prime$ couplings directly from four-fermion operators which have the same scaling in coupling vs mass than flavor bounds and can thus be directly compared.

{8) } Further bounds on electron-quark interactions can also be extracted from atomic parity violation in cesium from the APV experiment~\cite{Wood:1997zq,Bennett:1999pd} and from parity violation in electron proton scattering from QWEAK~\cite{Androic:2018kni}.

{9) LEP-II sets bounds on the interactions of four charged leptons which in our model translate into bounds on the product of $g_{11} g_{\ell \ell}/M_X^2$.
}

\begin{figure}
	\centering
	\includegraphics[width=0.92\linewidth]{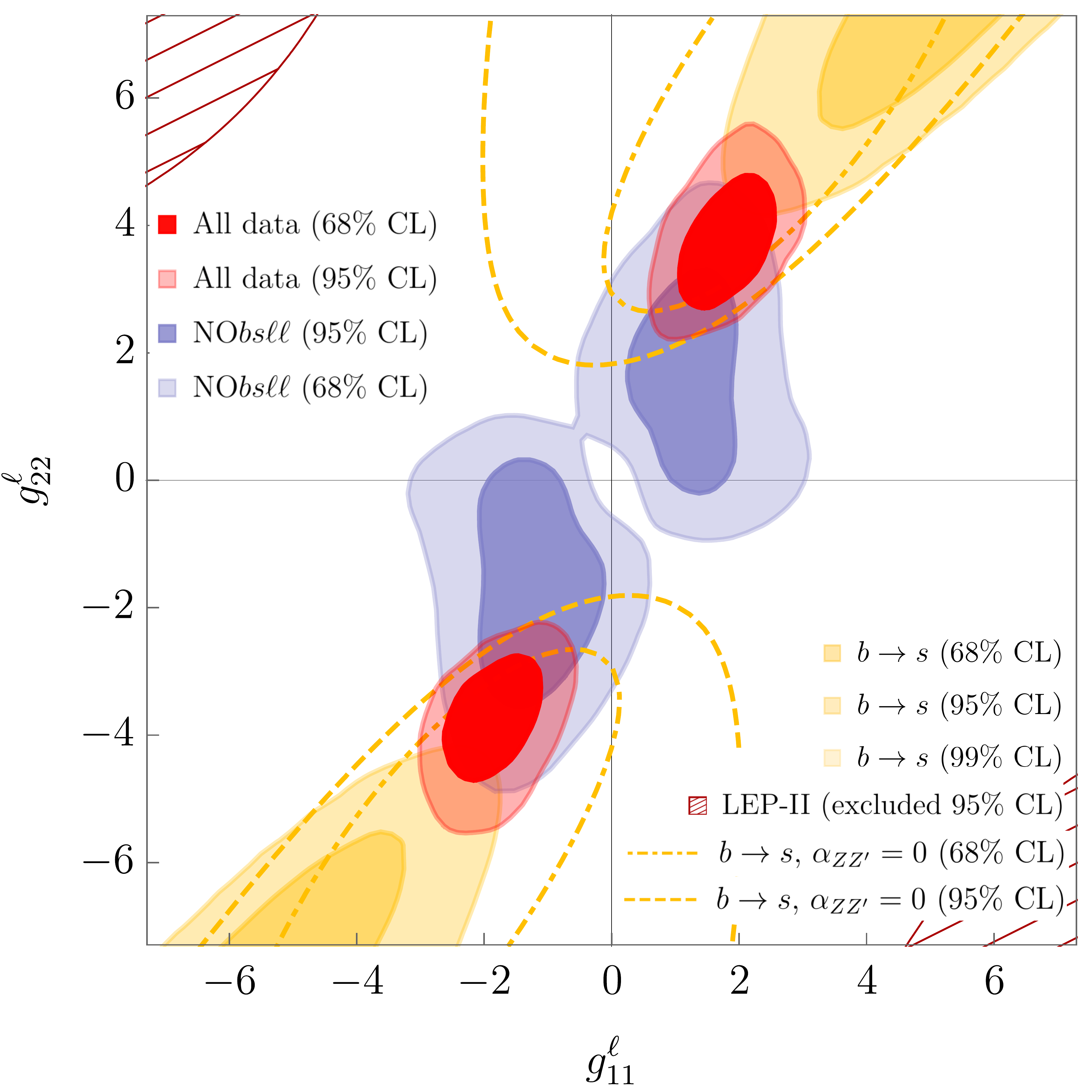}
	\caption{Global fit in the $g^\ell_{11}-g^\ell_{22}$ plane for $M_X=10\,$TeV. One can see that the blue region from the NO$bs\ell\ell$ data overlaps with the yellow one from the $b\to s\ell^+\ell^-$ data and $B_s-\bar B_s$ mixing at the 95\% C.L. Note that the overlap between the NO$bs\ell\ell$ region (which only mildly depends on $\sin\alpha_{ZZ^\prime}$) and the one from $b \to s \ell \ell$ is smaller when mixing is included. However, this does not mean that the agreement with the data is reduced. It is rather due to the fact that the three-dimensional scenario (including mixing) agrees better with data and its best fit point is further away from the SM hypothesis.}
	\label{fig.global2}
\end{figure}

\section{Phenomenological Analysis}\label{Analysis}

Let us now combine the observables discussed in the previous sections by performing a global fit. For this purpose we implemented in HEPfit~\cite{deBlas:2019okz} all of the observables testing LFU in the charged current, the standard EW observables (see Appendix~\ref{app:Observables}) and the CAA (encoded in the measurement of $V_{us}$)\footnote{In this subset we also included the LHC bounds for which we assumed a Gaussian distribution. However, we checked that in case a hard cut is implemented (which we will show in addition in the figures) the results only change marginally.}, called "NO$bs\ell\ell$" in the following, and performed a global fit with HEPfit. Furthermore, we translated the output for $b\to s\ell^+\ell^-$ data obtained with the code of Ref.~\cite{Descotes-Genon:2015uva} into a likelihood profile and included this, as well as the bound from $B_s-\bar B_s$ mixing, into HEPfit. With this setup we can now perform a Bayesian statistical analysis whose Markov Chain Monte Carlo (MCMC) determination of posteriors is powered by the Bayesian Analysis Toolkit (\texttt{BAT})~\cite{Caldwell:2008fw}.

As discussed in Sec.~\ref{Setup} we require that the $Z^\prime$, $W^\prime$ couplings to quarks respect an (approximate) $U(2)^3$ flavor symmetry such that potentially dangerous effects in $K^0-\bar K^0$ or $D^0-\bar D^0$ mixing are suppressed. Furthermore, we assume that the $Z^\prime$, $W^\prime$ mass is above the LHC production threshold such that the previously discussed bounds apply. For concreteness, we fix the common $Z^\prime$, $W^\prime$ mass $M_X$ to 10 TeV. In addition, we assume the couplings to leptons to be flavor diagonal, due to the stringent bounds from LFV observables (see Sec.~II  in Appendix~\ref{app:Observables}) and taking into account that such contributions do not interfere with the SM ones for the observables considered here. Therefore, the free parameters in our fit are $g^\ell_{11}$, $g^\ell_{22}$, $g^\ell_{33}$, $g^q$, $\sin\alpha_{ZZ^{\prime}}$ and $g^d_{23}$ for which we used a generously large prior of $[-10, 10]$.

Let us start with the combined fit NO$bs\ell\ell$ where $g^d_{23}$ does not enter. For $M_X = 10$ TeV, we find that $\sin \alpha_{ZZ'} = 2.9 \, 10^{-6} \pm 4.8 \, 10^{-5}$ and $g_{11}^\ell = 1.3 \pm 0.5$. The marginalized PDF's for the rest of the parameters are highly non-gaussian and therefore we give  the 68\% C.L. ranges as follows: $g_{22}^\ell \in [-2.7,0.6]\cup [1.5,2.3]$, $g_{33} \in [-8,1]\cup [1,6.5]$,	$g^q \in [-1.6,-0.] \cup [0.3,2.2]$.
The corresponding projections in the $g^\ell_{11}$--$g^q$, $g^\ell_{22}$--$g^q$ and $g^\ell_{33}$--$g^q$ planes are shown in blue in Fig.~\ref{fig.global} and the $g^\ell_{11}$--$g^\ell_{22}$ plane is shown in Fig.~\ref{fig.global2}. Here one can see that the SM point lies outside the 95\% C.L. regions, indicating a significantly better NP fit compared to the SM hypothesis. Furthermore, in Fig.~\ref{fig.global}  the bounds from LHC searches and parity violation experiments (hatched regions) are respected by the preferred regions.

Next we include the effect of $g^d_{23}$ which enters $b\to s\ell^+\ell^-$ transitions and $B_s-\bar B_s$ mixing. Here we combined both classes of observables in the $g^\ell_{11}$-$g^\ell_{22}$ plane by marginalizing over $g^d_{23}$ and $\sin \alpha_{ZZ^\prime}$, resulting in the yellow region in Fig.~\ref{fig.global2}. Interestingly one can see that this region overlaps significantly with the one favored by NO$bs\ell\ell$ data. Therefore, we can combine all data, $b\to s$ transitions and NO$bs\ell\ell$ observables, into one fit, resulting in the red regions of Figs.~\ref{fig.global} and ~\ref{fig.global2}.
The corresponding best fit point (note that there is an almost mirrored solution, though with a less significance) is at $\sin\alpha_{ZZ^{\prime}} = (-0.7 \pm 0.3)\times 10^{-4}$, $g_{11}^\ell = 1.8 \pm 0.5$, $g_{22}^\ell = 3.7 \pm 0.7$, $g_{33}^\ell = 6.3 \pm 1.7$,	$g^q = -0.8 \pm 0.4$, where, for the NP scenario, we find an Information Criterion~\cite{Kass:1995} value of $\approx 115$ compared to $\approx 167$ within the SM. This clearly shows that our dynamical model describes data significantly better than the SM hypothesis. In particular, we find that our global fit improves the agreement with $b\to s\ell^+\ell^-$ data by $\approx 5 \sigma$ compared to the SM, and that the CAA is alleviated. The tension in $V_{us}$ from $\beta$-decays is reduced by $2.6 \sigma$ and the one from semileptonic kaon decays into muons by $0.9\,\sigma$ (see Fig.~\ref{VusPlot}).
\section{Conclusions}
\label{Conclusions}
In this article we studied a simplified model with massive vector bosons transforming as a $SU(2)_L$ triplet in the context of the CAA anomaly and the hints for NP in $b\to s\ell^+\ell^-$ data. Within our setup, these anomalies clearly cannot be addressed without affecting other observables, in particular EW precision data, ratios testing LFU in the charged current, LHC bounds, parity violation experiments and $B_s-\bar B_s$ mixing. Therefore, assuming a $U(2)^3$ flavor symmetry in the quark sector, we performed a combined fit to six free parameters finding that the global fit for the flavor conserving observables is significantly improved. Furthermore, the preferred region of this fit overlaps with the one favoured by $b\to s\ell^+\ell^-$ data and the LHC bounds as well as $B_s-\bar B_s$ mixing. In particular, the $b\to s\ell^+\ell^-$ fit is improved by $5-6\,\sigma$ with respect to the SM while at the same time the CAA is reduced by more than 2$\,\sigma$. This shows that our model describes data significantly better than the SM hypothesis, testified by an IC value of $\approx 113$ compared to the SM value of $\approx 167$.

Looking towards the future, our model can be tested by improved measurements of LFU ratios (like $\pi\to\mu\nu/\pi\to e\nu$ at PEN~\cite{Glaser:2018aat} or $\tau\to\mu\nu\nu/\tau\to e\nu\nu$ at BELLE II~\cite{Kou:2018nap}), by additional data and modes for $b\to s\ell^+\ell^-$ transitions to be obtained by BELLE II~\cite{Kou:2018nap,Cerri:2018ypt} and the LHC~\cite{Cerri:2018ypt}, by LHC searches with increased luminosity~\cite{ApollinariG.:2017ojx} and by $Z$-pole measurements at future colliders such as CLIC~\cite{Aicheler:2012bya}, ILC~\cite{Behnke:2013lya} or FCC-ee~\cite{Abada:2019lih,Abada:2019zxq}. This, together with the accurate description of current data by our dynamical model clearly motivates the construction of UV compete realizations as a very promising direction for future research.

\begin{acknowledgements}
{We thank Javier Fuentes-Martin and Joaquim Matias for useful discussions and Antonio Coutinho for help with HEPfit. The work of A.C., C.A.M. and M.M. is supported by a Professorship Grant (No. PP00P2\_176884) of the Swiss National Science Foundation. B.C. is supported by the Italian Ministry of Research (MIUR) under the Grant No. PRIN 20172LNEEZ.}
\end{acknowledgements}

\appendix

\section{Observables}
\label{app:Observables}

In this section we provide more details on the relevant observables presented in the main text, and show how they are affected by our NP contributions. We will only give the explicit formulas for the direct $Z^\prime$ and $W^\prime$ contributions since one can easily recover the mixing induced effects by replacing the SM $W$ and $Z$ couplings with their modified versions.

\subsection{{Charged} Lepton Flavour Violation}
\label{secLFU}
The loop effects giving rise to $\ell\to\ell^\prime\gamma$ can be calculated in the unitary gauge since a finite result is obtained in our simplified model setup, which includes unavoidable Goldstone effects present in a UV complete model. Using the expressions given in Ref.~\cite{Crivellin:2018qmi} we obtain
\begin{equation}
{\rm Br}[\ell_i \to \ell_j \gamma]=\frac{m_{\ell_i}^3}{4\pi \, \Gamma_{\ell_i}} \big(|c^{ji}_{R} |^{2}+ |c^{ij}_{R} |^{2}\big),
\end{equation}
with
\begin{equation}
c_R^{ji} \approx -\frac{e m_{{\ell _i}}}{{16{\pi ^2}}}\frac{{g_{jk}^\ell g_{ki}^\ell }}{8} \dfrac{1}{{M_{X}^2}}\,.
\end{equation}
The current experimental limits for lepton flavor violation processes are shown in Refs.~\cite{Bertl:2006up,Aubert:2009ag,TheMEG:2016wtm} and yield the 90\% C.L. bounds
\begin{align}
\begin{split}
{\rm Br}\!\left[\mu\rightarrow e\gamma\right] &\leq 4.2\times10^{-13}\,, \quad  |{g_{ek}^\ell g_{k\mu}^\ell }| \leq  0.06\,, \\
{\rm Br}\!\left[\tau\rightarrow \mu\gamma\right] &\leq 4.4\times10^{-8} \,, \quad  |{g_{\mu k}^\ell g_{k\tau}^\ell }| \leq  112 \,,  \\
{\rm Br}\!\left[\tau\rightarrow e\gamma\right] &\leq 3.3\times10^{-8} \,, \quad  |{g_{e k}^\ell g_{k\tau}^\ell }| \leq  96 \,,
\end{split}
\end{align}
where we used $M_X=10\,$TeV as a reference point and the sum over $k$ is implied.

Three body decays to charged leptons are already mediated at tree-level but are phase space suppressed. In our model we find that
\begin{eqnarray}
\operatorname{Br}(\mu \rightarrow 3 e)&=&\frac{m_{\mu}^{5}}{768 \pi^{3} M_{X}^{4}, \Gamma_{\mu}} \frac{\left|g_{e \mu}^{\ell} g_{e e}^{\ell}\right|^{2}}{16} \,, \\
\operatorname{Br}(\tau \rightarrow e \mu \mu)&=&\frac{m_{\tau}^{5}}{1536 \pi^{3} M_{X}^{4} \Gamma_{\tau}} \frac{\left|g_{e \tau}^{\ell} g_{\mu \mu}^{\ell}\right|^{2}}{16} \,
\end{eqnarray}
where we neglected contributions involving two flavor changing couplings. Together with the experimental results~\cite{Amhis:2019ckw} this yields the following 90\% C.L. bounds (for $M_X=10\,$TeV)
\begin{align}
\begin{split}
\operatorname{Br}(\mu \rightarrow e e e)&\leq 1.0 \times 10^{-12}, \quad\left|g_{e \mu}^{\ell} g_{e e}^{\ell}\right| \leq 0.002 \,,\\
\operatorname{Br}(\tau \rightarrow \mu \mu \mu) &\leq 1.2 \times 10^{-8}, \quad\left|g_{\mu \tau}^{\ell} g_{\mu \mu}^{\ell}\right| \leq 0.85 \,,\\
\operatorname{Br}(\tau \rightarrow e e e)&\leq 1.4 \times 10^{-8}, \quad\left|g_{e \tau}^{\ell} g_{e e}^{\ell}\right| \leq 0.925\,, \\
\operatorname{Br}(\tau \rightarrow e \mu \mu) &\leq 1.6 \times 10^{-8}, \quad\left|g_{e \tau}^{\ell} g_{\mu \mu}^{\ell}\right| \leq 4.0 \,,\\
\operatorname{Br}(\tau \rightarrow \mu e e) &\leq 1.1 \times 10^{-8}, \quad\left|g_{\mu \tau}^{\ell} g_{e e}^{\ell}\right| \leq 3.3 \,.
\end{split}
\end{align}

Finally, following the conventions of Refs.~\cite{Crivellin:2014cta,Crivellin:2017rmk} we have that for $\mu\to e$ conversion in nuclei
\begin{align}
\Gamma _{\mu  \to e}^N = \frac{{m_\mu ^5{{| {g_{e\mu }^\ell g_{}^q} |}^2}}}{{M_{X}^2}}{\big| {\big( {V_N^{( n )} - V_N^{( p )}} \big)} \big|^2}\,,
\end{align}
which has to be normalized to the capture rate $\Gamma^{\rm capture}_N$ for gold~\cite{Egorov:2006bb}
\begin{align}
V_{Au}^{\left( n \right)} - V_{Au}^{\left( p \right)} =  - 0.0486,\;\; \Gamma^{\rm capture}_{Au}=8.7\times 10^{-15}\,{\rm MeV}.\nonumber
\end{align}
The current 90\% C.L. experimental limits are~\cite{Bertl:2006up}
\begin{align}
\begin{split}
\mathrm{Br}_{\mu \rightarrow e}^{\mathrm{Au}} &\leq 7.0 \times 10^{-13}\,,  \quad \left| {g_{e\mu }^\ell g_{}^q} \right|\leq 5.8\times 10^{-8}\,,
\end{split}
\end{align}
again for $M_X=10\,$TeV.

\begin{boldmath}
	\subsection{Electroweak Precision Observables}
	\label{sec:InputParameters}
\end{boldmath}

The high accuracy achieved in the determination of the EW observables $G_F$, $\alpha_{em}$ and $M_Z $ motivates the identification of their experimental values as Lagrangian parameters, so that they can be used to calculate all other EW observables within the SM. This strategy still applies in the presence of NP, but the relations between measurements and Lagrangian parameters must be adjusted. For our particular model, the Fermi-constant $G_F=1.16637(1)\times 10^{-5}{\rm GeV}^{-2}$, as measured from muon decays, is rewritten in terms of the one in the Lagrangian as
\begin{equation}
G_F = G_F^{\mathcal L} + \frac{{g_{11}^\ell g_{22}^\ell }}{{4 \sqrt{2} M_{X}^2}} \,.
\label{eq:GF}
\end{equation}
Likewise the measured $M_Z= 91.1875\pm0.0021$~\cite{ALEPH:2005ab} mass is given by
\begin{equation}
M_Z^2 = \left(M_{Z}^{\mathcal L}\right)^2  \Big( 1 - \sin^2 \alpha_{ZZ'}\frac{M_X^2}{(M_Z^{\mathcal L })^2} \Big)\,,
\end{equation}
where $G_F^{\mathcal L} = 1/(\sqrt{2} v^2)$ and $M_Z^{\mathcal L}$ is the measured $Z$ mass within the SM. In addition, the Higgs mass $M_H = 125.16\pm0.13$~\cite{Aaboud:2018wps, CMS:2019drq}, the top mass $m_t = 172.80 \pm 0.40$~\cite{TevatronElectroweakWorkingGroup:2016lid,  Aaboud:2018zbu,Sirunyan:2018mlv} and the strong coupling constant $\alpha_s = 0.1181\pm0.0011$~\cite{Tanabashi:2018oca} need to be included as fit parameters, since they enter EW observables indirectly via loop effects. Finally, the modification of the $W$ and $Z$ couplings to fermions due to gauge bosons mixing affects the relevant $W$ and $Z$ decays listed in Table~\ref{tab:EWCC}. We implemented these observables within our model into HEPfit~\cite{deBlas:2019okz} in order to perform our phenomenological analysis (see Appendix~\ref{app:Observables}).

\begin{table}[t!]
	\centering
	\begin{tabular}{c c c } 
		\hline
		Observable & Ref. & Measurement  \\
		\hline
		$M_W\,[\text{GeV}]$ & ~\cite{Tanabashi:2018oca} & $80.379(12)$  \\
		$\Gamma_W\,[\text{GeV}]$ & ~\cite{Tanabashi:2018oca} & $2.085(42)$  \\
		$\text{BR}(W\to \text{had})$ & ~\cite{Tanabashi:2018oca} & $0.6741(27)$  \\
		$\text{sin}^2\theta_{\rm eff(CDF)}^{\rm e}$ & ~\cite{Aaltonen:2016nuy}  & $0.23248(52)$  \\
		$\text{sin}^2\theta_{\rm eff(D0)}^{\rm e}$ & ~\cite{Abazov:2014jti}  & $0.23146(47)$ \\
		$\text{sin}^2\theta_{\rm eff(CDF)}^{\rm \mu}$ & ~\cite{Aaltonen:2014loa}  & $0.2315(20)$ \\
		$\text{sin}^2\theta_{\rm eff(CMS)}^{\rm \mu}$ & ~\cite{Chatrchyan:2011ya}  & $0.2287(32)$ \\
		$\text{sin}^2\theta_{\rm eff(LHCb)}^{\rm \mu}$ & ~\cite{Aaij:2015lka}  & $0.2314(11)$ \\
		$P_{\tau}^{\rm pol}$ &~\cite{ALEPH:2005ab} &$0.1465(33)$ \\
		$A_{e}$ &~\cite{ALEPH:2005ab} &$0.1516(21)$  \\
		$A_{\mu}$ &~\cite{ALEPH:2005ab} &$0.142(15)$  \\
		$A_{\tau}$ &~\cite{ALEPH:2005ab} &$0.136(15)$  \\
		$\Gamma_Z\,[\text{GeV}]$ &~\cite{ALEPH:2005ab} &$2.4952(23)$ \\
		$\sigma_h^{0}\,[\text{nb}]$ &~\cite{ALEPH:2005ab} &$41.541(37)$ \\
		$R^0_{\e}$ &~\cite{ALEPH:2005ab} &$20.804(50)$ \\
		$R^0_{\mu}$ &~\cite{ALEPH:2005ab} &$20.785(33)$  \\
		$R^0_{\tau}$ &~\cite{ALEPH:2005ab} &$20.764(45)$  \\
		$A_{\rm FB}^{0, e}$&~\cite{ALEPH:2005ab} &$0.0145(25)$   \\
		$A_{\rm FB}^{0, \mu}$&~\cite{ALEPH:2005ab} &$0.0169(13)$  \\
		$A_{\rm FB}^{0, \tau}$&~\cite{ALEPH:2005ab} &$0.0188(17)$ \\
		$R_{b}^{0}$ &~\cite{ALEPH:2005ab} &$0.21629(66)$\\
		$R_{c}^{0}$ &~\cite{ALEPH:2005ab} &$0.1721(30)$ \\
		$A_{\rm FB}^{0,b}$ &~\cite{ALEPH:2005ab} &$0.0992(16)$\\ 
		$A_{\rm FB}^{0,c}$ &~\cite{ALEPH:2005ab} &$0.0707(35)$ \\
		$A_{b}$ &~\cite{ALEPH:2005ab} &$0.923(20)$ \\
		$A_{c}$ &~\cite{ALEPH:2005ab} &$0.670(27)$ \\
	\end{tabular}
	\caption{ Electroweak observables used in our fit which are calculated (as a function of $M_Z^{\mathcal L}$, $\alpha$ and $G_F^{\mathcal L}$) by HEPfit~\cite{deBlas:2019okz}.}
	\label{tab:EWCC}
\end{table}

\bigskip

\begin{boldmath}
	\subsection{ $V_{us}$ and the CAA}
\end{boldmath}

{A modification of the Fermi-constant also affects the $V_{ud}$ from beta decays. However, we found that such an effect is too tightly constrained from EW precision data to account for the CAA.} The parameter $V_{us}^{\mathcal L}$ of the unitary CKM matrix of the Lagrangian within our model can be determined from kaon, tau or beta decays (particularly superallowed beta decays). The master formula for the latter reads~\cite{Hardy:2018zsb}
\begin{align}
|V_{ud}^\beta|^2=\frac{2984.432(3)s}{\Ft(1+\Delta_R^V)},
\end{align}
with $\Ft$-value $\Ft=3072.07(63)s$~\cite{Hardy:2018zsb}. The two different sets of radiative corrections
\begin{align}
\Delta_R^V\big|_\text{SGPR}&=0.02467(22)~\text{\cite{Seng:2018yzq}},\\
\Delta_R^V\big|_\text{CMS}&=0.02426(32)~\text{\cite{Czarnecki:2019mwq}},
\end{align}
lead to
\begin{align}
\left. V_{us}^\beta \right|_{\text{SGPR}} &=0.22782(62),\;\left. V_{us}^\beta\right|_{\text{CMS}}=0.22699(78),
\label{Vusbeta}
\end{align}
where we used unitarity with $|V_{ub}|=0.003683$ from~\cite{CKMfitter:2019,Charles:2004jd}, even though the precise value of $|V_{ub}|$ is unimportant here. This has to be compared to the average of the PDG value~\cite{Tanabashi:2018oca} for the $V_{us}$ from kaon and the HFLAV value~\cite{Amhis:2019ckw} from inclusive tau decays, $V_{us}^K=0.2243\pm0.0005$ and $V_{us}^\tau=0.2195 \pm 0.0019$, to get
\begin{equation}
V_{us}^{K+\tau}=0.2240\pm0.0005\,.
\end{equation}
When comparing $V_{us}^{K+\tau}$ with $V_{us}^\beta$ we notice a $\approx 3-5\,\sigma$ discrepancy which is the origin of the CAA. Note that we do not include $V_{us}$ exclusive tau decays here as these modes will be included in the rations testing LFU of the charged current.

Turning to NP corrections to these determinations, we have for $V_{us}$ from semileptonic kaon decays with muons
\begin{align}
|V_{us}^{{\mathcal L}}| = |V_{us}^{K_{\mu 3}}|\bigg(1-\frac{(g^{q}-g^{\ell}_{11})g^{\ell}_{22}}{g_2^2}\frac{M_W^2}{M_{X}^2}\bigg),
\label{eq:VusKtau}
\end{align}
and the determination of  $V_{us}/V_{ud}$ from $\B_{K^{\pm}\to\mu\nu}/\B_{\pi^{\pm}\to\mu\nu}$ is not modified. The same is true for its determination from $\tau\to K\nu/\tau\to\pi\nu$.
For $V_{us}^\beta$ there is, in addition to the direct modification of the transition $d\to u\e \nu$ an indirect one from the modification of $G_F$. Hence, the element of the unitary CKM matrix in the Lagrangian $V_{us}^{\mathcal L}$ can be expressed in terms of $V_{us}^\beta$ which is extracted from the experiment as follows
\begin{equation}
V_{us}^{\mathcal L} \approx V_{us}^\beta \left( {1 + \frac{{{{\left| {V_{ud}^{\mathcal L}} \right|}^2}}}{{{{\left| {V_{us}^{\mathcal L}} \right|}^2}}}\frac{{g_{11}^\ell \left( {g^{q} - g_{22}^\ell } \right)}}{{g_2^2}}\frac{{M_W^2}}{{M_{X}^2}}} \right).
\end{equation}
Note the important enhancement of $\left|V_{ud}^{\mathcal L}\right|^2/\left| V_{us}^{\mathcal L}\right|^2 \approx 20$~\cite{Crivellin:2020lzu}. This enhancement is not present in the modifications to the $V_{us}$ determination from kaon and tau decays. Therefore, the difference between Eq.~\eqref{Vusbeta} and Eq.~\eqref{eq:VusKtau} amounts to
\begin{align}
(85\pm 17)\times10^{-4} \approx\frac{{g_{11}^\ell\; (g_{22}^\ell-g^q)}}{{g_2^2}}\frac{{M_W^2}}{{M_{X}^2}}\;({\rm SGPR})\,,\\
(66\pm 20)\times10^{-4}\approx\frac{{g_{11}^\ell (g_{22}^\ell-g^q)}}{{g_2^2}}\frac{{M_W^2}}{{M_{X}^2}}\;({\rm CMS})\,,
\end{align}
by naively averaging the errors.
In the phenomenological section, we consider the SGPR determination due to its smaller error, which was recently confirmed by Ref.~\cite{Seng:2020wjq}, in our numerical analysis. Nonetheless, we found that choosing the CMS determination instead has only a marginal impact on the global fit.

\begin{table}[t]
	\begin{tabular}{l c l} \toprule
		Observable & Ref. & Measurement\\
		\colrule
		$R\left[\frac{K\rightarrow\mu\nu}{K\rightarrow e\nu}\right]$ &~\cite{Lazzeroni:2012cx,Ambrosino:2009aa,Cirigliano:2007xi,Pich:2013lsa} &$0.9978 \pm 0.0020$  \\
		$R\left[\frac{\pi\rightarrow\mu\nu}{\pi\rightarrow e\nu}\right]$&~\cite{Czapek:1993kc,Britton:1992pg,Bryman:1982em, Cirigliano:2007xi,Aguilar-Arevalo:2015cdf,Tanabashi:2018oca} & $1.0010 \pm 0.0009$  \\
		$R\left[\frac{\tau\rightarrow\mu\nu\bar{\nu}}{\tau\rightarrow e\nu\bar{\nu}}\right]$&~\cite{Amhis:2019ckw,Tanabashi:2018oca} & $1.0018 \pm 0.0014$ \\
		$R\left[\frac{K\rightarrow\pi\mu\bar{\nu}}{K\rightarrow\pi e\bar{\nu}}\right]$&~\cite{Antonelli:2010yf,Cirigliano:2011ny,Pich:2013lsa} & $1.0010 \pm 0.0025$  \\
		$R\left[\frac{\tau\rightarrow e\nu\bar{\nu}}{\mu\rightarrow e\bar{\nu}\nu}\right]$&~\cite{Amhis:2019ckw,Tanabashi:2018oca} & $1.0010 \pm 0.0014$ \\
		$R\left[\frac{\tau\rightarrow \pi\nu}{\pi\rightarrow \mu\bar{\nu}}\right]$&~\cite{Amhis:2019ckw}& $0.9961 \pm 0.0027$ \\
		$R\left[\frac{\tau\rightarrow K\nu}{K\rightarrow \mu\bar{\nu}}\right]$&~\cite{Amhis:2019ckw} & $0.9860 \pm 0.0070$ \\
		$R\left[\frac{\tau\rightarrow \mu\nu\bar{\nu}}{\mu\rightarrow e\nu\bar{\nu}}\right]$&~\cite{Amhis:2019ckw,Tanabashi:2018oca} & $1.0029 \pm 0.0014$
	\end{tabular}
	\caption{Measurements of the ratios testing LFU defined in \eq{LFUratios}. The correlations for the ratios involving tau decays are given in Ref.~\cite{Amhis:2019ckw}  while the value for the second row was determined by the weighted average of the measurements shown in  Refs.~\cite{Aguilar-Arevalo:2015cdf,Pich:2013lsa}.}
	\label{LFUobs}
\end{table}

\medskip

\subsection{Tests of LFU in the charged current}

In order to assess directly the modifications with respect to the SM we define the ratios
\begin{equation}
R(X)={\cal A}[X]/{\cal A}[X]_{\rm SM}\,,
\label{LFUratiosDef}
\end{equation}
where ${\cal A}$ is the amplitude, such that in the limit without NP they are unity. These ratios are modified as
\begin{eqnarray}
R\left[\frac{{{\tau  \to \mu \nu \nu }}}{{{\tau  \to e\nu \nu}}}\right] &=& \Big( {1 + \frac{{g_{33}^\ell \left( {g_{22}^\ell  - g_{11}^\ell } \right)}}{{g^2_2}}\frac{{M_W^2}}{{M_{W'}^2}}} \Big) \,. \nonumber \\
R\left[\frac{{{\tau  \to e \nu  \nu }}}{{{\mu  \to e \nu \nu}}}\right] &=& \Big( {1 + \frac{{g_{11}^\ell \left( {g_{33}^\ell  - g_{22}^\ell } \right)}}{{g^2_2}}\frac{{M_W^2}}{{M_{W'}^2}}} \Big) \,. \nonumber \\
R\left[\frac{{{\tau  \to \mu \nu  \nu }}}{{{\mu  \to e \nu \nu}}}\right] &=& \Big( {1 + \frac{{g_{22}^\ell \left( {g_{33}^\ell  - g_{11}^\ell } \right)}}{{g^2_2}}\frac{{M_W^2}}{{M_{W'}^2}}} \Big) \,. \nonumber \\
R\left[\frac{{{\pi  \to \mu \nu }}}{{{\pi  \to e\nu }}}\right] &=&\Big( {1 + \frac{{g^{q}_{}\left( {g_{22}^\ell  - g_{11}^\ell } \right)}}{{g^2_2}}\frac{{M_W^2}}{{M_{W'}^2}}} \Big) \,, \nonumber\\
R\left[\frac{{{K  \to \mu \nu }}}{{{K  \to e\nu }}}\right] &=&  R\left[\frac{{{K  \to \pi \mu \nu }}}{{{K  \to \pi e \nu }}}\right] = R\left[\frac{{{\pi  \to \mu \nu }}}{{{\pi  \to e\nu }}}\right]  \,, \nonumber\\
R\left[\frac{{{\tau  \to \pi \nu}}}{{{\pi  \to \mu \nu}}}\right] &=& \Big( {1 + \frac{{g^q \left( {g_{33}^\ell  - g_{22}^\ell } \right)}}{{g^2_2}}\frac{{M_W^2}}{{M_{W'}^2}}} \Big) \,. \nonumber \\
R\left[\frac{{{\tau  \to K \nu}}}{{{K  \to \mu \nu}}}\right] &=& R\left[\frac{{{\tau  \to \pi \nu}}}{{{\pi  \to \mu \nu}}}\right] \,,
\label{LFUratios}
\end{eqnarray}
by the tree-level $W^\prime$ effects. The corresponding experimental values are given in Table~\ref{LFUobs} with the correlations given in Ref.~\cite{Amhis:2019ckw}.

\begin{boldmath}
	\subsection{$B_s-\bar B_s$ Mixing}
\end{boldmath}

The most important constraint on $Z^\prime-b-s$ couplings, i.e. $g_{23}^d$, comes from $B_s-\bar B_s$ mixing where the contribution to the Hamiltonian $H_{\rm eff}=C_1 O_1$ with $O_1=\bar s \gamma^\mu P_L b\times \bar s \gamma_\mu P_L b$ is given by
\begin{equation}{C_1} = \frac{1}{{2M_{X}^2}}{\left( {\frac{g_{23}^d}{2}} \right)^2}\left( {1 + \frac{{{\alpha _s}}}{{4\pi }}\frac{{11}}{3}} \right)\,,\end{equation}
including the NLO matching corrections of Ref.~\cite{Buras:2012fs}. Note that the mixing induced effect generating $s-b-Z$ couplings can be neglected as it is a dim-8 effect. Employing the 2-loop RGE~\cite{Ciuchini:1997bw,Buras:2000if}, this leads to an effect, normalized to the SM one, of
\begin{equation}
\left(\frac{g^d_{23}}{0.26}\frac{M_{X}}{10 {\rm TeV}} \right)^2=0.110\pm0.090\label{Bsmixing}
\end{equation}
with the bag factor of Ref.~\cite{Aoki:2019cca} and the global fit to NP in $\delta F=2$ observables of Ref.~\cite{Bona:2007vi}.

\begin{boldmath}
	\subsection{$b\to s\ell^+\ell^-$}
\end{boldmath}

\begin{figure}[t!]
	\centering
	\includegraphics[width=0.95\linewidth]{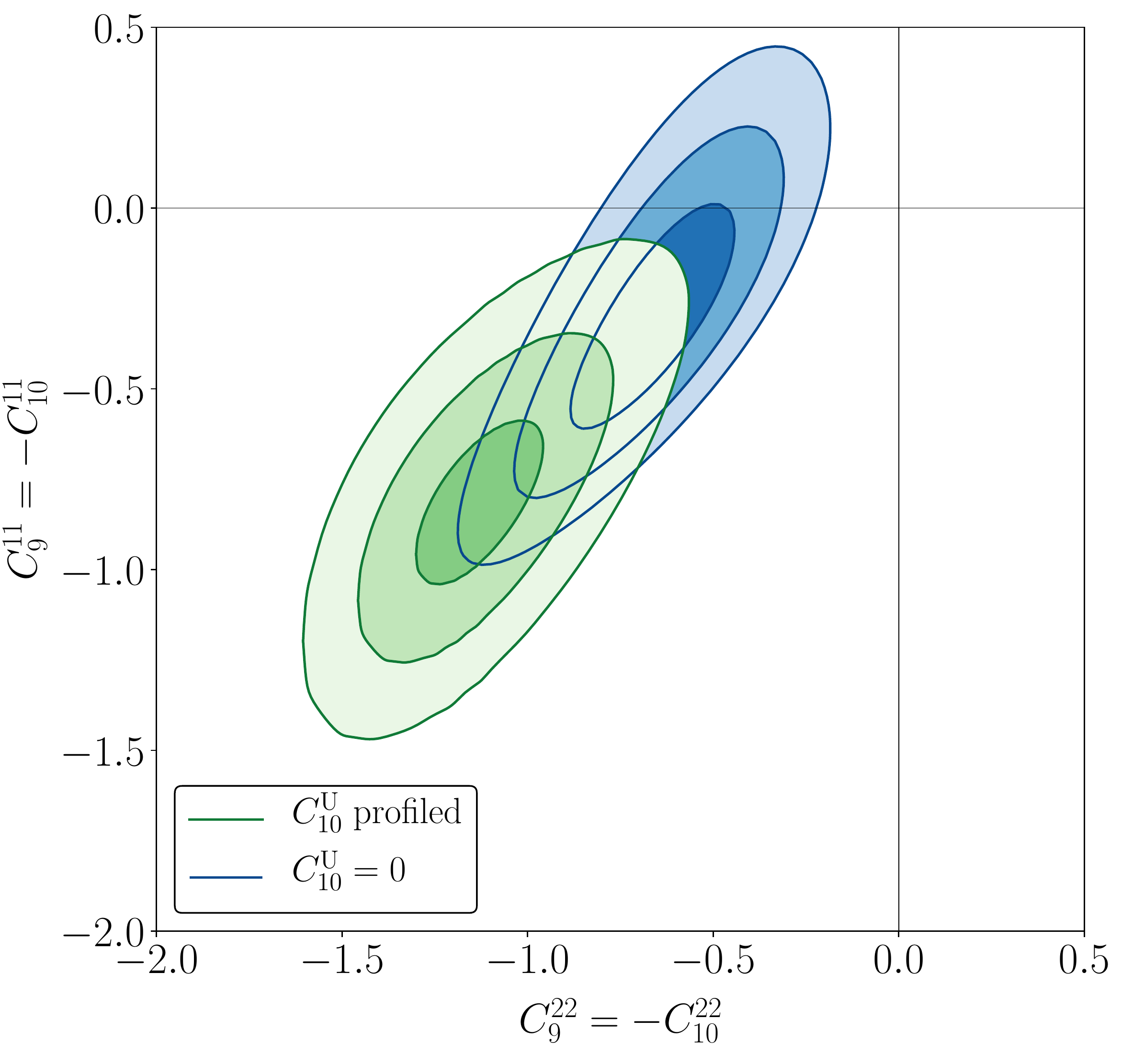}
	\caption{Preferred regions (68\%, 95\%, 99\% C.L.) of the two-dimensional ($C_{9}^{11} = - C_{10}^{11}$, $C_{9}^{22} = - C_{10}^{22}$) scenario (blue), and the three-dimensional scenario which includes $C_{10}^U$ (green).}
	\label{fig:C9mC10}
\end{figure}

For $b\to s\ell^+\ell^-$ our $Z^\prime$ contribution is purely left-handed and given by
\begin{equation}
C_9^{jj{}} =  - C_{10}^{jj{}} =  - \frac{{{\pi ^2}}}{{{e^2}}}\frac{{g_{23}^dg_{jj}^\ell }}{{\sqrt 2 {G_F}M_{X}^2{V_{tb}}V_{ts}^*}}\,,
\end{equation}
where $C_{9(10)}^{11}$ and $C_{9(10)}^{22}$  correspond to  $\mathcal{C}_{9 (10) e}^{\mathrm{NP}}$  and  $\mathcal{C}_{9(10) \mu}^{\mathrm{NP}}$  in the language of Refs.~\cite{Capdevila:2017bsm, Alguero:2019ptt}.
For the analysis of all available $b\to s\ell^+\ell^-$ data we use the method and program of Ref.~\cite{Descotes-Genon:2015uva} with the data set given in Ref.~\cite{Alguero:2019ptt}. Since our (two-dimensional) scenario with $C_9^{22}=-C_{10}^{22}$, $C_9^{11}=-C_{10}^{11}$, with a pull of 5.6 $\sigma$ with respect to the SM, was not explicitly given in Refs.~\cite{Capdevila:2017bsm,Alguero:2018nvb,Alguero:2019ptt} we show the corresponding preferred regions in Fig.~\ref{fig:C9mC10}.

{Since $C_9^U\approx 0$, the three-dimensional scenario $(C_9^{22}=-C_{10}^{22},C_9^{11}=-C_{10}^{11},\,C_{10}^U)$, is the most general scenario for the simplified model that we can explore. A global $b\to s\ell^+\ell^-$ fit to this structure yields a pull of $6.2\,\sigma$, with a best fit point and 68\% C.L. intervals of $( - 1.13, - 0.78, - 0.82)$  and $([ - 1.3, - 0.96],[ - 0.99, - 0.55],[ - 1.04, - 0.59])$, respectively.}

Gauge boson mixing induces extra LFU effects
\begin{align}
\begin{aligned}
C_9^{{\rm U}} &= -
\dfrac{{{\pi ^2}}}{{{e^2}}}\frac{{{g_2}\sin\alpha_{ZZ^\prime}\left( {1 - 4s_W^2} \right)}}{{\sqrt 2c_W {G_F}M_Z^2{V_{tb}}V_{ts}^*}}\,,\\
C_{10}^{{\rm U}} &=  \dfrac{{{\pi ^2}}}{{{e^2}}}\dfrac{{{{{g_2}\sin \alpha_{ZZ^\prime} g_{23}^d}}}}{{\sqrt 2c_W {G_F}M_Z^2{V_{tb}}V_{ts}^*}}\,.
\end{aligned}
\end{align}
Since $C_9^{{\rm U}}\approx 0$, the three-dimensional scenario
\begin{equation}
\left({{C}_9^{{\rm{22}}}=-{C}_{10}^{{\rm{22}}},{C}_9^{{\rm{11}}}=-{C}_{10}^{{\rm{11}}},{C}_{10}^{{\rm{U}}}} \right)\,,
\end{equation}
is the most general scenario for the simplified model that we can explore.

\subsection{LHC searches}

For 2-quark-2-lepton operators the 95\% C.L. bounds related to muons (electrons)~\cite{Aad:2020otl} and tau leptons are~\cite{Cirigliano:2018dyk}
\begin{eqnarray}
\begin{aligned}
- \dfrac{{4\pi }}{{{{\left( {{22(24)} \, {\rm TeV}} \right)}^2}}} &\le \dfrac{{g_{22}^\ell(g_{11}^\ell) g^q}}{{{2}M_{X}^2}} \le \dfrac{{4\pi }}{{{{\left( {{33 (26)} \, {\rm TeV}} \right)}^2}}}\,,\\
- 10.5\frac{{M_{X}^2}}{{{{\left( {10 \, {\rm TeV}} \right)}^2}}} &< g_{33}^\ell g^q < 0\,,
\end{aligned}
\label{eq:dijetbound}
\end{eqnarray}
while, from 2-jet events we obtain the approximate bound~\cite{Sirunyan:2017ygf}
\begin{equation}
{\left| {g^q} \right|^2} \lesssim 15\frac{{M_{X}^2}}{{{{\left( {10\,{\rm TeV}} \right)}^2}}}\,.
\end{equation}
Since Ref.~\cite{Sirunyan:2017ygf} did not distinguish between charged and neutral current contributions we estimated this bound by matching our $Z'$ on their EFT.

\subsection{Parity Violation}

Atomic parity violation in atoms, in particular cesium, and parity violation in electron proton scattering place limits on electron-quark interactions. Here the APV experiment~\cite{Wood:1997zq,Bennett:1999pd} and the QWEAK collaboration~\cite{Androic:2018kni} report
\begin{align}
\begin{aligned}
-2\left(2 C_{1 u}+C_{1 d}\right)=0.0719 \pm 0.0045\,,\\
-2\left(188 C_{1 u}+211 C_{1 d}\right)=-72.62 \pm 0.43\,,
\end{aligned}
\end{align}
respectively, with
\begin{align}
{C_{1d}} &= 0.3419 + \frac{{\sqrt 2 }}{{{G_F}}}\frac{{g_{}^qg_{11}^\ell }}{{16M_{X}^2}}\,,
\end{align}
\begin{align}
C_{1u} &=  - 0.1887 - \frac{{\sqrt 2 }}{{{G_F}}}\frac{{g_{}^qg_{11}^\ell }}{{16M_{X}^2}}\,.
\end{align}
Note that our NP contribution to ${C_{1d}}$ and $C_{1u}$ are of equal strength but have opposite signs. This nearly avoids the APV bound and significantly weakens the QWEAK one.

\subsection{{LEP-II bounds}}
\label{app:LEP2}

The bounds on four-lepton operators from LEP-II \cite{Schael:2013ita} impose the following bounds at 95\% C.L. on our model parameters (assuming that $Z-Z^\prime$ mixing is small):
\begin{eqnarray}
\begin{aligned}
- \dfrac{{4\pi }}{{{{\left( {8.8 \, {\rm TeV}} \right)}^2}}} &\le \dfrac{{g_{11}^\ell g_{11}^\ell }}{{2 M_{X}^2}} \le \dfrac{{4\pi }}{{{{\left( {8.0 \, {\rm TeV}} \right)}^2}}}\,,\\
- \dfrac{{4\pi }}{{{{\left( {12.2 \, {\rm TeV}} \right)}^2}}} &\le \dfrac{{g_{11}^\ell g_{22}^\ell }}{{4M_{X}^2}} \le \dfrac{{4\pi }}{{{{\left( {9.6 \, {\rm TeV}} \right)}^2}}}\,,\\
- \dfrac{{4\pi }}{{{{\left( {9.2 \, {\rm TeV}} \right)}^2}}} &\le \dfrac{{g_{11}^\ell g_{33}^\ell  }}{{4M_{X}^2}} \le \dfrac{{4\pi }}{{{{\left( {9.0 \, {\rm TeV}} \right)}^2}}}\,,
\end{aligned}
\label{eq:LEP2bound}
\end{eqnarray}

\bibliography{bibliography}

\begin{thebibliography}{169}%
\makeatletter
\providecommand \@ifxundefined [1]{%
 \@ifx{#1\undefined}
}%
\providecommand \@ifnum [1]{%
 \ifnum #1\expandafter \@firstoftwo
 \else \expandafter \@secondoftwo
 \fi
}%
\providecommand \@ifx [1]{%
 \ifx #1\expandafter \@firstoftwo
 \else \expandafter \@secondoftwo
 \fi
}%
\providecommand \natexlab [1]{#1}%
\providecommand \enquote  [1]{``#1''}%
\providecommand \bibnamefont  [1]{#1}%
\providecommand \bibfnamefont [1]{#1}%
\providecommand \citenamefont [1]{#1}%
\providecommand \href@noop [0]{\@secondoftwo}%
\providecommand \href [0]{\begingroup \@sanitize@url \@href}%
\providecommand \@href[1]{\@@startlink{#1}\@@href}%
\providecommand \@@href[1]{\endgroup#1\@@endlink}%
\providecommand \@sanitize@url [0]{\catcode `\\12\catcode `\$12\catcode
  `\&12\catcode `\#12\catcode `\^12\catcode `\_12\catcode `\%12\relax}%
\providecommand \@@startlink[1]{}%
\providecommand \@@endlink[0]{}%
\providecommand \url  [0]{\begingroup\@sanitize@url \@url }%
\providecommand \@url [1]{\endgroup\@href {#1}{\urlprefix }}%
\providecommand \urlprefix  [0]{URL }%
\providecommand \Eprint [0]{\href }%
\providecommand \doibase [0]{http://dx.doi.org/}%
\providecommand \selectlanguage [0]{\@gobble}%
\providecommand \bibinfo  [0]{\@secondoftwo}%
\providecommand \bibfield  [0]{\@secondoftwo}%
\providecommand \translation [1]{[#1]}%
\providecommand \BibitemOpen [0]{}%
\providecommand \bibitemStop [0]{}%
\providecommand \bibitemNoStop [0]{.\EOS\space}%
\providecommand \EOS [0]{\spacefactor3000\relax}%
\providecommand \BibitemShut  [1]{\csname bibitem#1\endcsname}%
\let\auto@bib@innerbib\@empty
\bibitem [{\citenamefont {Aad}\ \emph {et~al.}(2012)\citenamefont {Aad} \emph
  {et~al.}}]{Aad:2012tfa}%
  \BibitemOpen
  \bibfield  {author} {\bibinfo {author} {\bibfnamefont {G.}~\bibnamefont
  {Aad}} \emph {et~al.} (\bibinfo {collaboration} {ATLAS}),\ }\href {\doibase
  10.1016/j.physletb.2012.08.020} {\bibfield  {journal} {\bibinfo  {journal}
  {Phys. Lett.}\ }\textbf {\bibinfo {volume} {B716}},\ \bibinfo {pages} {1}
  (\bibinfo {year} {2012})},\ \Eprint {http://arxiv.org/abs/1207.7214}
  {arXiv:1207.7214 [hep-ex]} \BibitemShut {NoStop}%
\bibitem [{\citenamefont {Chatrchyan}\ \emph {et~al.}(2012)\citenamefont
  {Chatrchyan} \emph {et~al.}}]{Chatrchyan:2012xdj}%
  \BibitemOpen
  \bibfield  {author} {\bibinfo {author} {\bibfnamefont {S.}~\bibnamefont
  {Chatrchyan}} \emph {et~al.} (\bibinfo {collaboration} {CMS}),\ }\href
  {\doibase 10.1016/j.physletb.2012.08.021} {\bibfield  {journal} {\bibinfo
  {journal} {Phys. Lett.}\ }\textbf {\bibinfo {volume} {B716}},\ \bibinfo
  {pages} {30} (\bibinfo {year} {2012})},\ \Eprint
  {http://arxiv.org/abs/1207.7235} {arXiv:1207.7235 [hep-ex]} \BibitemShut
  {NoStop}%
\bibitem [{\citenamefont {Aaij}\ \emph {et~al.}(2017)\citenamefont {Aaij} \emph
  {et~al.}}]{Aaij:2017vbb}%
  \BibitemOpen
  \bibfield  {author} {\bibinfo {author} {\bibfnamefont {R.}~\bibnamefont
  {Aaij}} \emph {et~al.} (\bibinfo {collaboration} {LHCb}),\ }\href {\doibase
  10.1007/JHEP08(2017)055} {\bibfield  {journal} {\bibinfo  {journal} {JHEP}\
  }\textbf {\bibinfo {volume} {08}},\ \bibinfo {pages} {055} (\bibinfo {year}
  {2017})},\ \Eprint {http://arxiv.org/abs/1705.05802} {arXiv:1705.05802
  [hep-ex]} \BibitemShut {NoStop}%
\bibitem [{\citenamefont {Aaij}\ \emph {et~al.}(2019)\citenamefont {Aaij} \emph
  {et~al.}}]{Aaij:2019wad}%
  \BibitemOpen
  \bibfield  {author} {\bibinfo {author} {\bibfnamefont {R.}~\bibnamefont
  {Aaij}} \emph {et~al.} (\bibinfo {collaboration} {LHCb}),\ }\href {\doibase
  10.1103/PhysRevLett.122.191801} {\bibfield  {journal} {\bibinfo  {journal}
  {Phys. Rev. Lett.}\ }\textbf {\bibinfo {volume} {122}},\ \bibinfo {pages}
  {191801} (\bibinfo {year} {2019})},\ \Eprint
  {http://arxiv.org/abs/1903.09252} {arXiv:1903.09252 [hep-ex]} \BibitemShut
  {NoStop}%
\bibitem [{\citenamefont {Capdevila}\ \emph {et~al.}(2018)\citenamefont
  {Capdevila}, \citenamefont {Crivellin}, \citenamefont {Descotes-Genon},
  \citenamefont {Matias},\ and\ \citenamefont {Virto}}]{Capdevila:2017bsm}%
  \BibitemOpen
  \bibfield  {author} {\bibinfo {author} {\bibfnamefont {B.}~\bibnamefont
  {Capdevila}}, \bibinfo {author} {\bibfnamefont {A.}~\bibnamefont
  {Crivellin}}, \bibinfo {author} {\bibfnamefont {S.}~\bibnamefont
  {Descotes-Genon}}, \bibinfo {author} {\bibfnamefont {J.}~\bibnamefont
  {Matias}}, \ and\ \bibinfo {author} {\bibfnamefont {J.}~\bibnamefont
  {Virto}},\ }\href {\doibase 10.1007/JHEP01(2018)093} {\bibfield  {journal}
  {\bibinfo  {journal} {JHEP}\ }\textbf {\bibinfo {volume} {01}},\ \bibinfo
  {pages} {093} (\bibinfo {year} {2018})},\ \Eprint
  {http://arxiv.org/abs/1704.05340} {arXiv:1704.05340 [hep-ph]} \BibitemShut
  {NoStop}%
\bibitem [{\citenamefont {Altmannshofer}\ \emph {et~al.}(2017)\citenamefont
  {Altmannshofer}, \citenamefont {Stangl},\ and\ \citenamefont
  {Straub}}]{Altmannshofer:2017yso}%
  \BibitemOpen
  \bibfield  {author} {\bibinfo {author} {\bibfnamefont {W.}~\bibnamefont
  {Altmannshofer}}, \bibinfo {author} {\bibfnamefont {P.}~\bibnamefont
  {Stangl}}, \ and\ \bibinfo {author} {\bibfnamefont {D.~M.}\ \bibnamefont
  {Straub}},\ }\href {\doibase 10.1103/PhysRevD.96.055008} {\bibfield
  {journal} {\bibinfo  {journal} {Phys. Rev.}\ }\textbf {\bibinfo {volume}
  {D96}},\ \bibinfo {pages} {055008} (\bibinfo {year} {2017})},\ \Eprint
  {http://arxiv.org/abs/1704.05435} {arXiv:1704.05435 [hep-ph]} \BibitemShut
  {NoStop}%
\bibitem [{\citenamefont {D'Amico}\ \emph {et~al.}(2017)\citenamefont
  {D'Amico}, \citenamefont {Nardecchia}, \citenamefont {Panci}, \citenamefont
  {Sannino}, \citenamefont {Strumia}, \citenamefont {Torre},\ and\
  \citenamefont {Urbano}}]{DAmico:2017mtc}%
  \BibitemOpen
  \bibfield  {author} {\bibinfo {author} {\bibfnamefont {G.}~\bibnamefont
  {D'Amico}}, \bibinfo {author} {\bibfnamefont {M.}~\bibnamefont {Nardecchia}},
  \bibinfo {author} {\bibfnamefont {P.}~\bibnamefont {Panci}}, \bibinfo
  {author} {\bibfnamefont {F.}~\bibnamefont {Sannino}}, \bibinfo {author}
  {\bibfnamefont {A.}~\bibnamefont {Strumia}}, \bibinfo {author} {\bibfnamefont
  {R.}~\bibnamefont {Torre}}, \ and\ \bibinfo {author} {\bibfnamefont
  {A.}~\bibnamefont {Urbano}},\ }\href {\doibase 10.1007/JHEP09(2017)010}
  {\bibfield  {journal} {\bibinfo  {journal} {JHEP}\ }\textbf {\bibinfo
  {volume} {09}},\ \bibinfo {pages} {010} (\bibinfo {year} {2017})},\ \Eprint
  {http://arxiv.org/abs/1704.05438} {arXiv:1704.05438 [hep-ph]} \BibitemShut
  {NoStop}%
\bibitem [{\citenamefont {Ciuchini}\ \emph {et~al.}(2017)\citenamefont
  {Ciuchini}, \citenamefont {Coutinho}, \citenamefont {Fedele}, \citenamefont
  {Franco}, \citenamefont {Paul}, \citenamefont {Silvestrini},\ and\
  \citenamefont {Valli}}]{Ciuchini:2017mik}%
  \BibitemOpen
  \bibfield  {author} {\bibinfo {author} {\bibfnamefont {M.}~\bibnamefont
  {Ciuchini}}, \bibinfo {author} {\bibfnamefont {A.~M.}\ \bibnamefont
  {Coutinho}}, \bibinfo {author} {\bibfnamefont {M.}~\bibnamefont {Fedele}},
  \bibinfo {author} {\bibfnamefont {E.}~\bibnamefont {Franco}}, \bibinfo
  {author} {\bibfnamefont {A.}~\bibnamefont {Paul}}, \bibinfo {author}
  {\bibfnamefont {L.}~\bibnamefont {Silvestrini}}, \ and\ \bibinfo {author}
  {\bibfnamefont {M.}~\bibnamefont {Valli}},\ }\href {\doibase
  10.1140/epjc/s10052-017-5270-2} {\bibfield  {journal} {\bibinfo  {journal}
  {Eur. Phys. J.}\ }\textbf {\bibinfo {volume} {C77}},\ \bibinfo {pages} {688}
  (\bibinfo {year} {2017})},\ \Eprint {http://arxiv.org/abs/1704.05447}
  {arXiv:1704.05447 [hep-ph]} \BibitemShut {NoStop}%
\bibitem [{\citenamefont {Hiller}\ and\ \citenamefont
  {Nisandzic}(2017)}]{Hiller:2017bzc}%
  \BibitemOpen
  \bibfield  {author} {\bibinfo {author} {\bibfnamefont {G.}~\bibnamefont
  {Hiller}}\ and\ \bibinfo {author} {\bibfnamefont {I.}~\bibnamefont
  {Nisandzic}},\ }\href {\doibase 10.1103/PhysRevD.96.035003} {\bibfield
  {journal} {\bibinfo  {journal} {Phys. Rev.}\ }\textbf {\bibinfo {volume}
  {D96}},\ \bibinfo {pages} {035003} (\bibinfo {year} {2017})},\ \Eprint
  {http://arxiv.org/abs/1704.05444} {arXiv:1704.05444 [hep-ph]} \BibitemShut
  {NoStop}%
\bibitem [{\citenamefont {Geng}\ \emph {et~al.}(2017)\citenamefont {Geng},
  \citenamefont {Grinstein}, \citenamefont {Jäger}, \citenamefont
  {Martin~Camalich}, \citenamefont {Ren},\ and\ \citenamefont
  {Shi}}]{Geng:2017svp}%
  \BibitemOpen
  \bibfield  {author} {\bibinfo {author} {\bibfnamefont {L.-S.}\ \bibnamefont
  {Geng}}, \bibinfo {author} {\bibfnamefont {B.}~\bibnamefont {Grinstein}},
  \bibinfo {author} {\bibfnamefont {S.}~\bibnamefont {Jäger}}, \bibinfo
  {author} {\bibfnamefont {J.}~\bibnamefont {Martin~Camalich}}, \bibinfo
  {author} {\bibfnamefont {X.-L.}\ \bibnamefont {Ren}}, \ and\ \bibinfo
  {author} {\bibfnamefont {R.-X.}\ \bibnamefont {Shi}},\ }\href {\doibase
  10.1103/PhysRevD.96.093006} {\bibfield  {journal} {\bibinfo  {journal} {Phys.
  Rev.}\ }\textbf {\bibinfo {volume} {D96}},\ \bibinfo {pages} {093006}
  (\bibinfo {year} {2017})},\ \Eprint {http://arxiv.org/abs/1704.05446}
  {arXiv:1704.05446 [hep-ph]} \BibitemShut {NoStop}%
\bibitem [{\citenamefont {Hurth}\ \emph {et~al.}(2017)\citenamefont {Hurth},
  \citenamefont {Mahmoudi}, \citenamefont {Martinez~Santos},\ and\
  \citenamefont {Neshatpour}}]{Hurth:2017hxg}%
  \BibitemOpen
  \bibfield  {author} {\bibinfo {author} {\bibfnamefont {T.}~\bibnamefont
  {Hurth}}, \bibinfo {author} {\bibfnamefont {F.}~\bibnamefont {Mahmoudi}},
  \bibinfo {author} {\bibfnamefont {D.}~\bibnamefont {Martinez~Santos}}, \ and\
  \bibinfo {author} {\bibfnamefont {S.}~\bibnamefont {Neshatpour}},\ }\href
  {\doibase 10.1103/PhysRevD.96.095034} {\bibfield  {journal} {\bibinfo
  {journal} {Phys. Rev.}\ }\textbf {\bibinfo {volume} {D96}},\ \bibinfo {pages}
  {095034} (\bibinfo {year} {2017})},\ \Eprint
  {http://arxiv.org/abs/1705.06274} {arXiv:1705.06274 [hep-ph]} \BibitemShut
  {NoStop}%
\bibitem [{\citenamefont {Algueró}\ \emph
  {et~al.}(2019{\natexlab{a}})\citenamefont {Algueró}, \citenamefont
  {Capdevila}, \citenamefont {Crivellin}, \citenamefont {Descotes-Genon},
  \citenamefont {Masjuan}, \citenamefont {Matias}, \citenamefont
  {Novoa~Brunet},\ and\ \citenamefont {Virto}}]{Alguero:2019ptt}%
  \BibitemOpen
  \bibfield  {author} {\bibinfo {author} {\bibfnamefont {M.}~\bibnamefont
  {Algueró}}, \bibinfo {author} {\bibfnamefont {B.}~\bibnamefont {Capdevila}},
  \bibinfo {author} {\bibfnamefont {A.}~\bibnamefont {Crivellin}}, \bibinfo
  {author} {\bibfnamefont {S.}~\bibnamefont {Descotes-Genon}}, \bibinfo
  {author} {\bibfnamefont {P.}~\bibnamefont {Masjuan}}, \bibinfo {author}
  {\bibfnamefont {J.}~\bibnamefont {Matias}}, \bibinfo {author} {\bibfnamefont
  {M.}~\bibnamefont {Novoa~Brunet}}, \ and\ \bibinfo {author} {\bibfnamefont
  {J.}~\bibnamefont {Virto}},\ }\href {\doibase 10.1140/epjc/s10052-019-7216-3}
  {\bibfield  {journal} {\bibinfo  {journal} {Eur. Phys. J.}\ }\textbf
  {\bibinfo {volume} {C79}},\ \bibinfo {pages} {714} (\bibinfo {year}
  {2019}{\natexlab{a}})},\ \Eprint {http://arxiv.org/abs/1903.09578}
  {arXiv:1903.09578 [hep-ph]} \BibitemShut {NoStop}%
\bibitem [{\citenamefont {Aebischer}\ \emph
  {et~al.}(2020{\natexlab{a}})\citenamefont {Aebischer}, \citenamefont
  {Altmannshofer}, \citenamefont {Guadagnoli}, \citenamefont {Reboud},
  \citenamefont {Stangl},\ and\ \citenamefont {Straub}}]{Aebischer:2019mlg}%
  \BibitemOpen
  \bibfield  {author} {\bibinfo {author} {\bibfnamefont {J.}~\bibnamefont
  {Aebischer}}, \bibinfo {author} {\bibfnamefont {W.}~\bibnamefont
  {Altmannshofer}}, \bibinfo {author} {\bibfnamefont {D.}~\bibnamefont
  {Guadagnoli}}, \bibinfo {author} {\bibfnamefont {M.}~\bibnamefont {Reboud}},
  \bibinfo {author} {\bibfnamefont {P.}~\bibnamefont {Stangl}}, \ and\ \bibinfo
  {author} {\bibfnamefont {D.~M.}\ \bibnamefont {Straub}},\ }\href {\doibase
  10.1140/epjc/s10052-020-7817-x} {\bibfield  {journal} {\bibinfo  {journal}
  {Eur. Phys. J.}\ }\textbf {\bibinfo {volume} {C80}},\ \bibinfo {pages} {252}
  (\bibinfo {year} {2020}{\natexlab{a}})},\ \Eprint
  {http://arxiv.org/abs/1903.10434} {arXiv:1903.10434 [hep-ph]} \BibitemShut
  {NoStop}%
\bibitem [{\citenamefont {Ciuchini}\ \emph {et~al.}(2019)\citenamefont
  {Ciuchini}, \citenamefont {Coutinho}, \citenamefont {Fedele}, \citenamefont
  {Franco}, \citenamefont {Paul}, \citenamefont {Silvestrini},\ and\
  \citenamefont {Valli}}]{Ciuchini:2019usw}%
  \BibitemOpen
  \bibfield  {author} {\bibinfo {author} {\bibfnamefont {M.}~\bibnamefont
  {Ciuchini}}, \bibinfo {author} {\bibfnamefont {A.~M.}\ \bibnamefont
  {Coutinho}}, \bibinfo {author} {\bibfnamefont {M.}~\bibnamefont {Fedele}},
  \bibinfo {author} {\bibfnamefont {E.}~\bibnamefont {Franco}}, \bibinfo
  {author} {\bibfnamefont {A.}~\bibnamefont {Paul}}, \bibinfo {author}
  {\bibfnamefont {L.}~\bibnamefont {Silvestrini}}, \ and\ \bibinfo {author}
  {\bibfnamefont {M.}~\bibnamefont {Valli}},\ }\href {\doibase
  10.1140/epjc/s10052-019-7210-9} {\bibfield  {journal} {\bibinfo  {journal}
  {Eur. Phys. J.}\ }\textbf {\bibinfo {volume} {C79}},\ \bibinfo {pages} {719}
  (\bibinfo {year} {2019})},\ \Eprint {http://arxiv.org/abs/1903.09632}
  {arXiv:1903.09632 [hep-ph]} \BibitemShut {NoStop}%
\bibitem [{\citenamefont {Arbey}\ \emph {et~al.}(2019)\citenamefont {Arbey},
  \citenamefont {Hurth}, \citenamefont {Mahmoudi}, \citenamefont {Santos},\
  and\ \citenamefont {Neshatpour}}]{Arbey:2019duh}%
  \BibitemOpen
  \bibfield  {author} {\bibinfo {author} {\bibfnamefont {A.}~\bibnamefont
  {Arbey}}, \bibinfo {author} {\bibfnamefont {T.}~\bibnamefont {Hurth}},
  \bibinfo {author} {\bibfnamefont {F.}~\bibnamefont {Mahmoudi}}, \bibinfo
  {author} {\bibfnamefont {D.~M.}\ \bibnamefont {Santos}}, \ and\ \bibinfo
  {author} {\bibfnamefont {S.}~\bibnamefont {Neshatpour}},\ }\href {\doibase
  10.1103/PhysRevD.100.015045} {\bibfield  {journal} {\bibinfo  {journal}
  {Phys. Rev.}\ }\textbf {\bibinfo {volume} {D100}},\ \bibinfo {pages} {015045}
  (\bibinfo {year} {2019})},\ \Eprint {http://arxiv.org/abs/1904.08399}
  {arXiv:1904.08399 [hep-ph]} \BibitemShut {NoStop}%
\bibitem [{\citenamefont {Amhis}\ \emph {et~al.}(2019)\citenamefont {Amhis}
  \emph {et~al.}}]{Amhis:2019ckw}%
  \BibitemOpen
  \bibfield  {author} {\bibinfo {author} {\bibfnamefont {Y.~S.}\ \bibnamefont
  {Amhis}} \emph {et~al.} (\bibinfo {collaboration} {HFLAV}),\ }\href@noop {}
  {\  (\bibinfo {year} {2019})},\ \bibinfo {note} {{updated results and plots
  available at
  \href{https://hflav.web.cern.ch/}{{\texttt{https://hflav.web.cern.ch/}}}}},\
  \Eprint {http://arxiv.org/abs/1909.12524} {arXiv:1909.12524 [hep-ex]}
  \BibitemShut {NoStop}%
\bibitem [{\citenamefont {Matias}\ \emph {et~al.}(2012)\citenamefont {Matias},
  \citenamefont {Mescia}, \citenamefont {Ramon},\ and\ \citenamefont
  {Virto}}]{Matias:2012xw}%
  \BibitemOpen
  \bibfield  {author} {\bibinfo {author} {\bibfnamefont {J.}~\bibnamefont
  {Matias}}, \bibinfo {author} {\bibfnamefont {F.}~\bibnamefont {Mescia}},
  \bibinfo {author} {\bibfnamefont {M.}~\bibnamefont {Ramon}}, \ and\ \bibinfo
  {author} {\bibfnamefont {J.}~\bibnamefont {Virto}},\ }\href {\doibase
  10.1007/JHEP04(2012)104} {\bibfield  {journal} {\bibinfo  {journal} {JHEP}\
  }\textbf {\bibinfo {volume} {04}},\ \bibinfo {pages} {104} (\bibinfo {year}
  {2012})},\ \Eprint {http://arxiv.org/abs/1202.4266} {arXiv:1202.4266
  [hep-ph]} \BibitemShut {NoStop}%
\bibitem [{\citenamefont {Descotes-Genon}\ \emph {et~al.}(2013)\citenamefont
  {Descotes-Genon}, \citenamefont {Hurth}, \citenamefont {Matias},\ and\
  \citenamefont {Virto}}]{Descotes-Genon:2013vna}%
  \BibitemOpen
  \bibfield  {author} {\bibinfo {author} {\bibfnamefont {S.}~\bibnamefont
  {Descotes-Genon}}, \bibinfo {author} {\bibfnamefont {T.}~\bibnamefont
  {Hurth}}, \bibinfo {author} {\bibfnamefont {J.}~\bibnamefont {Matias}}, \
  and\ \bibinfo {author} {\bibfnamefont {J.}~\bibnamefont {Virto}},\ }\href
  {\doibase 10.1007/JHEP05(2013)137} {\bibfield  {journal} {\bibinfo  {journal}
  {JHEP}\ }\textbf {\bibinfo {volume} {05}},\ \bibinfo {pages} {137} (\bibinfo
  {year} {2013})},\ \Eprint {http://arxiv.org/abs/1303.5794} {arXiv:1303.5794
  [hep-ph]} \BibitemShut {NoStop}%
\bibitem [{\citenamefont {Aaij}\ \emph {et~al.}(2016)\citenamefont {Aaij} \emph
  {et~al.}}]{Aaij:2015oid}%
  \BibitemOpen
  \bibfield  {author} {\bibinfo {author} {\bibfnamefont {R.}~\bibnamefont
  {Aaij}} \emph {et~al.} (\bibinfo {collaboration} {LHCb}),\ }\href {\doibase
  10.1007/JHEP02(2016)104} {\bibfield  {journal} {\bibinfo  {journal} {JHEP}\
  }\textbf {\bibinfo {volume} {02}},\ \bibinfo {pages} {104} (\bibinfo {year}
  {2016})},\ \Eprint {http://arxiv.org/abs/1512.04442} {arXiv:1512.04442
  [hep-ex]} \BibitemShut {NoStop}%
\bibitem [{\citenamefont {Aaij}\ \emph {et~al.}(2020)\citenamefont {Aaij} \emph
  {et~al.}}]{Aaij:2020nrf}%
  \BibitemOpen
  \bibfield  {author} {\bibinfo {author} {\bibfnamefont {R.}~\bibnamefont
  {Aaij}} \emph {et~al.} (\bibinfo {collaboration} {LHCb}),\ }\href@noop {} {\
  (\bibinfo {year} {2020})},\ \Eprint {http://arxiv.org/abs/2003.04831}
  {arXiv:2003.04831 [hep-ex]} \BibitemShut {NoStop}%
\bibitem [{\citenamefont {Belfatto}\ \emph {et~al.}(2020)\citenamefont
  {Belfatto}, \citenamefont {Beradze},\ and\ \citenamefont
  {Berezhiani}}]{Belfatto:2019swo}%
  \BibitemOpen
  \bibfield  {author} {\bibinfo {author} {\bibfnamefont {B.}~\bibnamefont
  {Belfatto}}, \bibinfo {author} {\bibfnamefont {R.}~\bibnamefont {Beradze}}, \
  and\ \bibinfo {author} {\bibfnamefont {Z.}~\bibnamefont {Berezhiani}},\
  }\href {\doibase 10.1140/epjc/s10052-020-7691-6} {\bibfield  {journal}
  {\bibinfo  {journal} {Eur. Phys. J.}\ }\textbf {\bibinfo {volume} {C80}},\
  \bibinfo {pages} {149} (\bibinfo {year} {2020})},\ \Eprint
  {http://arxiv.org/abs/1906.02714} {arXiv:1906.02714 [hep-ph]} \BibitemShut
  {NoStop}%
\bibitem [{\citenamefont {Grossman}\ \emph {et~al.}(2019)\citenamefont
  {Grossman}, \citenamefont {Passemar},\ and\ \citenamefont
  {Schacht}}]{Grossman:2019bzp}%
  \BibitemOpen
  \bibfield  {author} {\bibinfo {author} {\bibfnamefont {Y.}~\bibnamefont
  {Grossman}}, \bibinfo {author} {\bibfnamefont {E.}~\bibnamefont {Passemar}},
  \ and\ \bibinfo {author} {\bibfnamefont {S.}~\bibnamefont {Schacht}},\
  }\href@noop {} {\  (\bibinfo {year} {2019})},\ \Eprint
  {http://arxiv.org/abs/1911.07821} {arXiv:1911.07821 [hep-ph]} \BibitemShut
  {NoStop}%
\bibitem [{\citenamefont {Shiells}\ \emph {et~al.}(2020)\citenamefont
  {Shiells}, \citenamefont {Blunden},\ and\ \citenamefont
  {Melnitchouk}}]{Shiells:2020fqp}%
  \BibitemOpen
  \bibfield  {author} {\bibinfo {author} {\bibfnamefont {K.}~\bibnamefont
  {Shiells}}, \bibinfo {author} {\bibfnamefont {P.}~\bibnamefont {Blunden}}, \
  and\ \bibinfo {author} {\bibfnamefont {W.}~\bibnamefont {Melnitchouk}},\
  }\href@noop {} {\  (\bibinfo {year} {2020})},\ \Eprint
  {http://arxiv.org/abs/2012.01580} {arXiv:2012.01580 [hep-ph]} \BibitemShut
  {NoStop}%
\bibitem [{\citenamefont {Coutinho}\ \emph {et~al.}(2019)\citenamefont
  {Coutinho}, \citenamefont {Crivellin},\ and\ \citenamefont
  {Manzari}}]{Coutinho:2019aiy}%
  \BibitemOpen
  \bibfield  {author} {\bibinfo {author} {\bibfnamefont {A.~M.}\ \bibnamefont
  {Coutinho}}, \bibinfo {author} {\bibfnamefont {A.}~\bibnamefont {Crivellin}},
  \ and\ \bibinfo {author} {\bibfnamefont {C.~A.}\ \bibnamefont {Manzari}},\
  }\href@noop {} {\  (\bibinfo {year} {2019})},\ \Eprint
  {http://arxiv.org/abs/1912.08823} {arXiv:1912.08823 [hep-ph]} \BibitemShut
  {NoStop}%
\bibitem [{\citenamefont {Crivellin}\ and\ \citenamefont
  {Hoferichter}(2020)}]{Crivellin:2020lzu}%
  \BibitemOpen
  \bibfield  {author} {\bibinfo {author} {\bibfnamefont {A.}~\bibnamefont
  {Crivellin}}\ and\ \bibinfo {author} {\bibfnamefont {M.}~\bibnamefont
  {Hoferichter}},\ }\href@noop {} {\  (\bibinfo {year} {2020})},\ \Eprint
  {http://arxiv.org/abs/2002.07184} {arXiv:2002.07184 [hep-ph]} \BibitemShut
  {NoStop}%
\bibitem [{\citenamefont {Endo}\ and\ \citenamefont
  {Mishima}(2020)}]{Endo:2020tkb}%
  \BibitemOpen
  \bibfield  {author} {\bibinfo {author} {\bibfnamefont {M.}~\bibnamefont
  {Endo}}\ and\ \bibinfo {author} {\bibfnamefont {S.}~\bibnamefont {Mishima}},\
  }\href@noop {} {\  (\bibinfo {year} {2020})},\ \Eprint
  {http://arxiv.org/abs/2005.03933} {arXiv:2005.03933 [hep-ph]} \BibitemShut
  {NoStop}%
\bibitem [{\citenamefont {Cheung}\ \emph {et~al.}(2020)\citenamefont {Cheung},
  \citenamefont {Keung}, \citenamefont {Lu},\ and\ \citenamefont
  {Tseng}}]{Cheung:2020vqm}%
  \BibitemOpen
  \bibfield  {author} {\bibinfo {author} {\bibfnamefont {K.}~\bibnamefont
  {Cheung}}, \bibinfo {author} {\bibfnamefont {W.-Y.}\ \bibnamefont {Keung}},
  \bibinfo {author} {\bibfnamefont {C.-T.}\ \bibnamefont {Lu}}, \ and\ \bibinfo
  {author} {\bibfnamefont {P.-Y.}\ \bibnamefont {Tseng}},\ }\href@noop {} {\
  (\bibinfo {year} {2020})},\ \Eprint {http://arxiv.org/abs/2001.02853}
  {arXiv:2001.02853 [hep-ph]} \BibitemShut {NoStop}%
\bibitem [{\citenamefont {Bobeth}\ \emph {et~al.}(2017)\citenamefont {Bobeth},
  \citenamefont {Buras}, \citenamefont {Celis},\ and\ \citenamefont
  {Jung}}]{Bobeth:2016llm}%
  \BibitemOpen
  \bibfield  {author} {\bibinfo {author} {\bibfnamefont {C.}~\bibnamefont
  {Bobeth}}, \bibinfo {author} {\bibfnamefont {A.~J.}\ \bibnamefont {Buras}},
  \bibinfo {author} {\bibfnamefont {A.}~\bibnamefont {Celis}}, \ and\ \bibinfo
  {author} {\bibfnamefont {M.}~\bibnamefont {Jung}},\ }\href {\doibase
  10.1007/JHEP04(2017)079} {\bibfield  {journal} {\bibinfo  {journal} {JHEP}\
  }\textbf {\bibinfo {volume} {04}},\ \bibinfo {pages} {079} (\bibinfo {year}
  {2017})},\ \Eprint {http://arxiv.org/abs/1609.04783} {arXiv:1609.04783
  [hep-ph]} \BibitemShut {NoStop}%
\bibitem [{\citenamefont {Bernard}\ \emph {et~al.}(2008)\citenamefont
  {Bernard}, \citenamefont {Oertel}, \citenamefont {Passemar},\ and\
  \citenamefont {Stern}}]{Bernard:2007cf}%
  \BibitemOpen
  \bibfield  {author} {\bibinfo {author} {\bibfnamefont {V.}~\bibnamefont
  {Bernard}}, \bibinfo {author} {\bibfnamefont {M.}~\bibnamefont {Oertel}},
  \bibinfo {author} {\bibfnamefont {E.}~\bibnamefont {Passemar}}, \ and\
  \bibinfo {author} {\bibfnamefont {J.}~\bibnamefont {Stern}},\ }\href
  {\doibase 10.1088/1126-6708/2008/01/015} {\bibfield  {journal} {\bibinfo
  {journal} {JHEP}\ }\textbf {\bibinfo {volume} {01}},\ \bibinfo {pages} {015}
  (\bibinfo {year} {2008})},\ \Eprint {http://arxiv.org/abs/0707.4194}
  {arXiv:0707.4194 [hep-ph]} \BibitemShut {NoStop}%
\bibitem [{\citenamefont {Crivellin}(2010)}]{Crivellin:2009sd}%
  \BibitemOpen
  \bibfield  {author} {\bibinfo {author} {\bibfnamefont {A.}~\bibnamefont
  {Crivellin}},\ }\href {\doibase 10.1103/PhysRevD.81.031301} {\bibfield
  {journal} {\bibinfo  {journal} {Phys. Rev.}\ }\textbf {\bibinfo {volume}
  {D81}},\ \bibinfo {pages} {031301} (\bibinfo {year} {2010})},\ \Eprint
  {http://arxiv.org/abs/0907.2461} {arXiv:0907.2461 [hep-ph]} \BibitemShut
  {NoStop}%
\bibitem [{\citenamefont {Perez-Victoria}(2011)}]{PerezVictoria:2011uj}%
  \BibitemOpen
  \bibfield  {author} {\bibinfo {author} {\bibfnamefont {M.}~\bibnamefont
  {Perez-Victoria}},\ }in\ \href@noop {} {\emph {\bibinfo {booktitle}
  {{Proceedings, 46th Rencontres de Moriond on Electroweak Interactions and
  Unified Theories: La Thuile, Italy, March 13-20, 2011}}}}\ (\bibinfo {year}
  {2011})\ pp.\ \bibinfo {pages} {95--100},\ \Eprint
  {http://arxiv.org/abs/1107.0851} {arXiv:1107.0851 [hep-ph]} \BibitemShut
  {NoStop}%
\bibitem [{\citenamefont {Buras}\ and\ \citenamefont
  {Girrbach}(2013{\natexlab{a}})}]{Buras:2013qja}%
  \BibitemOpen
  \bibfield  {author} {\bibinfo {author} {\bibfnamefont {A.~J.}\ \bibnamefont
  {Buras}}\ and\ \bibinfo {author} {\bibfnamefont {J.}~\bibnamefont
  {Girrbach}},\ }\href {\doibase 10.1007/JHEP12(2013)009} {\bibfield  {journal}
  {\bibinfo  {journal} {JHEP}\ }\textbf {\bibinfo {volume} {12}},\ \bibinfo
  {pages} {009} (\bibinfo {year} {2013}{\natexlab{a}})},\ \Eprint
  {http://arxiv.org/abs/1309.2466} {arXiv:1309.2466 [hep-ph]} \BibitemShut
  {NoStop}%
\bibitem [{\citenamefont {Gauld}\ \emph
  {et~al.}(2014{\natexlab{a}})\citenamefont {Gauld}, \citenamefont {Goertz},\
  and\ \citenamefont {Haisch}}]{Gauld:2013qba}%
  \BibitemOpen
  \bibfield  {author} {\bibinfo {author} {\bibfnamefont {R.}~\bibnamefont
  {Gauld}}, \bibinfo {author} {\bibfnamefont {F.}~\bibnamefont {Goertz}}, \
  and\ \bibinfo {author} {\bibfnamefont {U.}~\bibnamefont {Haisch}},\ }\href
  {\doibase 10.1103/PhysRevD.89.015005} {\bibfield  {journal} {\bibinfo
  {journal} {Phys. Rev.}\ }\textbf {\bibinfo {volume} {D89}},\ \bibinfo {pages}
  {015005} (\bibinfo {year} {2014}{\natexlab{a}})},\ \Eprint
  {http://arxiv.org/abs/1308.1959} {arXiv:1308.1959 [hep-ph]} \BibitemShut
  {NoStop}%
\bibitem [{\citenamefont {Gauld}\ \emph
  {et~al.}(2014{\natexlab{b}})\citenamefont {Gauld}, \citenamefont {Goertz},\
  and\ \citenamefont {Haisch}}]{Gauld:2013qja}%
  \BibitemOpen
  \bibfield  {author} {\bibinfo {author} {\bibfnamefont {R.}~\bibnamefont
  {Gauld}}, \bibinfo {author} {\bibfnamefont {F.}~\bibnamefont {Goertz}}, \
  and\ \bibinfo {author} {\bibfnamefont {U.}~\bibnamefont {Haisch}},\ }\href
  {\doibase 10.1007/JHEP01(2014)069} {\bibfield  {journal} {\bibinfo  {journal}
  {JHEP}\ }\textbf {\bibinfo {volume} {01}},\ \bibinfo {pages} {069} (\bibinfo
  {year} {2014}{\natexlab{b}})},\ \Eprint {http://arxiv.org/abs/1310.1082}
  {arXiv:1310.1082 [hep-ph]} \BibitemShut {NoStop}%
\bibitem [{\citenamefont {Altmannshofer}\ \emph {et~al.}(2014)\citenamefont
  {Altmannshofer}, \citenamefont {Gori}, \citenamefont {Pospelov},\ and\
  \citenamefont {Yavin}}]{Altmannshofer:2014cfa}%
  \BibitemOpen
  \bibfield  {author} {\bibinfo {author} {\bibfnamefont {W.}~\bibnamefont
  {Altmannshofer}}, \bibinfo {author} {\bibfnamefont {S.}~\bibnamefont {Gori}},
  \bibinfo {author} {\bibfnamefont {M.}~\bibnamefont {Pospelov}}, \ and\
  \bibinfo {author} {\bibfnamefont {I.}~\bibnamefont {Yavin}},\ }\href
  {\doibase 10.1103/PhysRevD.89.095033} {\bibfield  {journal} {\bibinfo
  {journal} {Phys. Rev.}\ }\textbf {\bibinfo {volume} {D89}},\ \bibinfo {pages}
  {095033} (\bibinfo {year} {2014})},\ \Eprint {http://arxiv.org/abs/1403.1269}
  {arXiv:1403.1269 [hep-ph]} \BibitemShut {NoStop}%
\bibitem [{\citenamefont {Crivellin}\ \emph
  {et~al.}(2015{\natexlab{a}})\citenamefont {Crivellin}, \citenamefont
  {D'Ambrosio},\ and\ \citenamefont {Heeck}}]{Crivellin:2015mga}%
  \BibitemOpen
  \bibfield  {author} {\bibinfo {author} {\bibfnamefont {A.}~\bibnamefont
  {Crivellin}}, \bibinfo {author} {\bibfnamefont {G.}~\bibnamefont
  {D'Ambrosio}}, \ and\ \bibinfo {author} {\bibfnamefont {J.}~\bibnamefont
  {Heeck}},\ }\href {\doibase 10.1103/PhysRevLett.114.151801} {\bibfield
  {journal} {\bibinfo  {journal} {Phys. Rev. Lett.}\ }\textbf {\bibinfo
  {volume} {114}},\ \bibinfo {pages} {151801} (\bibinfo {year}
  {2015}{\natexlab{a}})},\ \Eprint {http://arxiv.org/abs/1501.00993}
  {arXiv:1501.00993 [hep-ph]} \BibitemShut {NoStop}%
\bibitem [{\citenamefont {Crivellin}\ \emph
  {et~al.}(2015{\natexlab{b}})\citenamefont {Crivellin}, \citenamefont
  {D'Ambrosio},\ and\ \citenamefont {Heeck}}]{Crivellin:2015lwa}%
  \BibitemOpen
  \bibfield  {author} {\bibinfo {author} {\bibfnamefont {A.}~\bibnamefont
  {Crivellin}}, \bibinfo {author} {\bibfnamefont {G.}~\bibnamefont
  {D'Ambrosio}}, \ and\ \bibinfo {author} {\bibfnamefont {J.}~\bibnamefont
  {Heeck}},\ }\href {\doibase 10.1103/PhysRevD.91.075006} {\bibfield  {journal}
  {\bibinfo  {journal} {Phys. Rev.}\ }\textbf {\bibinfo {volume} {D91}},\
  \bibinfo {pages} {075006} (\bibinfo {year} {2015}{\natexlab{b}})},\ \Eprint
  {http://arxiv.org/abs/1503.03477} {arXiv:1503.03477 [hep-ph]} \BibitemShut
  {NoStop}%
\bibitem [{\citenamefont {Niehoff}\ \emph {et~al.}(2015)\citenamefont
  {Niehoff}, \citenamefont {Stangl},\ and\ \citenamefont
  {Straub}}]{Niehoff:2015bfa}%
  \BibitemOpen
  \bibfield  {author} {\bibinfo {author} {\bibfnamefont {C.}~\bibnamefont
  {Niehoff}}, \bibinfo {author} {\bibfnamefont {P.}~\bibnamefont {Stangl}}, \
  and\ \bibinfo {author} {\bibfnamefont {D.~M.}\ \bibnamefont {Straub}},\
  }\href {\doibase 10.1016/j.physletb.2015.05.063} {\bibfield  {journal}
  {\bibinfo  {journal} {Phys. Lett.}\ }\textbf {\bibinfo {volume} {B747}},\
  \bibinfo {pages} {182} (\bibinfo {year} {2015})},\ \Eprint
  {http://arxiv.org/abs/1503.03865} {arXiv:1503.03865 [hep-ph]} \BibitemShut
  {NoStop}%
\bibitem [{\citenamefont {Carmona}\ and\ \citenamefont
  {Goertz}(2016)}]{Carmona:2015ena}%
  \BibitemOpen
  \bibfield  {author} {\bibinfo {author} {\bibfnamefont {A.}~\bibnamefont
  {Carmona}}\ and\ \bibinfo {author} {\bibfnamefont {F.}~\bibnamefont
  {Goertz}},\ }\href {\doibase 10.1103/PhysRevLett.116.251801} {\bibfield
  {journal} {\bibinfo  {journal} {Phys. Rev. Lett.}\ }\textbf {\bibinfo
  {volume} {116}},\ \bibinfo {pages} {251801} (\bibinfo {year} {2016})},\
  \Eprint {http://arxiv.org/abs/1510.07658} {arXiv:1510.07658 [hep-ph]}
  \BibitemShut {NoStop}%
\bibitem [{\citenamefont {Falkowski}\ \emph {et~al.}(2015)\citenamefont
  {Falkowski}, \citenamefont {Nardecchia},\ and\ \citenamefont
  {Ziegler}}]{Falkowski:2015zwa}%
  \BibitemOpen
  \bibfield  {author} {\bibinfo {author} {\bibfnamefont {A.}~\bibnamefont
  {Falkowski}}, \bibinfo {author} {\bibfnamefont {M.}~\bibnamefont
  {Nardecchia}}, \ and\ \bibinfo {author} {\bibfnamefont {R.}~\bibnamefont
  {Ziegler}},\ }\href {\doibase 10.1007/JHEP11(2015)173} {\bibfield  {journal}
  {\bibinfo  {journal} {JHEP}\ }\textbf {\bibinfo {volume} {11}},\ \bibinfo
  {pages} {173} (\bibinfo {year} {2015})},\ \Eprint
  {http://arxiv.org/abs/1509.01249} {arXiv:1509.01249 [hep-ph]} \BibitemShut
  {NoStop}%
\bibitem [{\citenamefont {Celis}\ \emph {et~al.}(2016)\citenamefont {Celis},
  \citenamefont {Feng},\ and\ \citenamefont {Lüst}}]{Celis:2015eqs}%
  \BibitemOpen
  \bibfield  {author} {\bibinfo {author} {\bibfnamefont {A.}~\bibnamefont
  {Celis}}, \bibinfo {author} {\bibfnamefont {W.-Z.}\ \bibnamefont {Feng}}, \
  and\ \bibinfo {author} {\bibfnamefont {D.}~\bibnamefont {Lüst}},\ }\href
  {\doibase 10.1007/JHEP02(2016)007} {\bibfield  {journal} {\bibinfo  {journal}
  {JHEP}\ }\textbf {\bibinfo {volume} {02}},\ \bibinfo {pages} {007} (\bibinfo
  {year} {2016})},\ \Eprint {http://arxiv.org/abs/1512.02218} {arXiv:1512.02218
  [hep-ph]} \BibitemShut {NoStop}%
\bibitem [{\citenamefont {Celis}\ \emph {et~al.}(2015)\citenamefont {Celis},
  \citenamefont {Fuentes-Martin}, \citenamefont {Jung},\ and\ \citenamefont
  {Serodio}}]{Celis:2015ara}%
  \BibitemOpen
  \bibfield  {author} {\bibinfo {author} {\bibfnamefont {A.}~\bibnamefont
  {Celis}}, \bibinfo {author} {\bibfnamefont {J.}~\bibnamefont
  {Fuentes-Martin}}, \bibinfo {author} {\bibfnamefont {M.}~\bibnamefont
  {Jung}}, \ and\ \bibinfo {author} {\bibfnamefont {H.}~\bibnamefont
  {Serodio}},\ }\href {\doibase 10.1103/PhysRevD.92.015007} {\bibfield
  {journal} {\bibinfo  {journal} {Phys. Rev.}\ }\textbf {\bibinfo {volume}
  {D92}},\ \bibinfo {pages} {015007} (\bibinfo {year} {2015})},\ \Eprint
  {http://arxiv.org/abs/1505.03079} {arXiv:1505.03079 [hep-ph]} \BibitemShut
  {NoStop}%
\bibitem [{\citenamefont {Crivellin}\ \emph
  {et~al.}(2015{\natexlab{c}})\citenamefont {Crivellin}, \citenamefont {Hofer},
  \citenamefont {Matias}, \citenamefont {Nierste}, \citenamefont {Pokorski},\
  and\ \citenamefont {Rosiek}}]{Crivellin:2015era}%
  \BibitemOpen
  \bibfield  {author} {\bibinfo {author} {\bibfnamefont {A.}~\bibnamefont
  {Crivellin}}, \bibinfo {author} {\bibfnamefont {L.}~\bibnamefont {Hofer}},
  \bibinfo {author} {\bibfnamefont {J.}~\bibnamefont {Matias}}, \bibinfo
  {author} {\bibfnamefont {U.}~\bibnamefont {Nierste}}, \bibinfo {author}
  {\bibfnamefont {S.}~\bibnamefont {Pokorski}}, \ and\ \bibinfo {author}
  {\bibfnamefont {J.}~\bibnamefont {Rosiek}},\ }\href {\doibase
  10.1103/PhysRevD.92.054013} {\bibfield  {journal} {\bibinfo  {journal} {Phys.
  Rev.}\ }\textbf {\bibinfo {volume} {D92}},\ \bibinfo {pages} {054013}
  (\bibinfo {year} {2015}{\natexlab{c}})},\ \Eprint
  {http://arxiv.org/abs/1504.07928} {arXiv:1504.07928 [hep-ph]} \BibitemShut
  {NoStop}%
\bibitem [{\citenamefont {Crivellin}\ \emph
  {et~al.}(2017{\natexlab{a}})\citenamefont {Crivellin}, \citenamefont
  {Fuentes-Martin}, \citenamefont {Greljo},\ and\ \citenamefont
  {Isidori}}]{Crivellin:2016ejn}%
  \BibitemOpen
  \bibfield  {author} {\bibinfo {author} {\bibfnamefont {A.}~\bibnamefont
  {Crivellin}}, \bibinfo {author} {\bibfnamefont {J.}~\bibnamefont
  {Fuentes-Martin}}, \bibinfo {author} {\bibfnamefont {A.}~\bibnamefont
  {Greljo}}, \ and\ \bibinfo {author} {\bibfnamefont {G.}~\bibnamefont
  {Isidori}},\ }\href {\doibase 10.1016/j.physletb.2016.12.057} {\bibfield
  {journal} {\bibinfo  {journal} {Phys. Lett.}\ }\textbf {\bibinfo {volume}
  {B766}},\ \bibinfo {pages} {77} (\bibinfo {year} {2017}{\natexlab{a}})},\
  \Eprint {http://arxiv.org/abs/1611.02703} {arXiv:1611.02703 [hep-ph]}
  \BibitemShut {NoStop}%
\bibitem [{\citenamefont {Garcia~Garcia}(2017)}]{GarciaGarcia:2016nvr}%
  \BibitemOpen
  \bibfield  {author} {\bibinfo {author} {\bibfnamefont {I.}~\bibnamefont
  {Garcia~Garcia}},\ }\href {\doibase 10.1007/JHEP03(2017)040} {\bibfield
  {journal} {\bibinfo  {journal} {JHEP}\ }\textbf {\bibinfo {volume} {03}},\
  \bibinfo {pages} {040} (\bibinfo {year} {2017})},\ \Eprint
  {http://arxiv.org/abs/1611.03507} {arXiv:1611.03507 [hep-ph]} \BibitemShut
  {NoStop}%
\bibitem [{\citenamefont {Altmannshofer}\ \emph {et~al.}(2016)\citenamefont
  {Altmannshofer}, \citenamefont {Carena},\ and\ \citenamefont
  {Crivellin}}]{Altmannshofer:2016oaq}%
  \BibitemOpen
  \bibfield  {author} {\bibinfo {author} {\bibfnamefont {W.}~\bibnamefont
  {Altmannshofer}}, \bibinfo {author} {\bibfnamefont {M.}~\bibnamefont
  {Carena}}, \ and\ \bibinfo {author} {\bibfnamefont {A.}~\bibnamefont
  {Crivellin}},\ }\href {\doibase 10.1103/PhysRevD.94.095026} {\bibfield
  {journal} {\bibinfo  {journal} {Phys. Rev.}\ }\textbf {\bibinfo {volume}
  {D94}},\ \bibinfo {pages} {095026} (\bibinfo {year} {2016})},\ \Eprint
  {http://arxiv.org/abs/1604.08221} {arXiv:1604.08221 [hep-ph]} \BibitemShut
  {NoStop}%
\bibitem [{\citenamefont {Faisel}\ and\ \citenamefont
  {Tandean}(2018)}]{Faisel:2017glo}%
  \BibitemOpen
  \bibfield  {author} {\bibinfo {author} {\bibfnamefont {G.}~\bibnamefont
  {Faisel}}\ and\ \bibinfo {author} {\bibfnamefont {J.}~\bibnamefont
  {Tandean}},\ }\href {\doibase 10.1007/JHEP02(2018)074} {\bibfield  {journal}
  {\bibinfo  {journal} {JHEP}\ }\textbf {\bibinfo {volume} {02}},\ \bibinfo
  {pages} {074} (\bibinfo {year} {2018})},\ \Eprint
  {http://arxiv.org/abs/1710.11102} {arXiv:1710.11102 [hep-ph]} \BibitemShut
  {NoStop}%
\bibitem [{\citenamefont {King}(2017)}]{King:2017anf}%
  \BibitemOpen
  \bibfield  {author} {\bibinfo {author} {\bibfnamefont {S.~F.}\ \bibnamefont
  {King}},\ }\href {\doibase 10.1007/JHEP08(2017)019} {\bibfield  {journal}
  {\bibinfo  {journal} {JHEP}\ }\textbf {\bibinfo {volume} {08}},\ \bibinfo
  {pages} {019} (\bibinfo {year} {2017})},\ \Eprint
  {http://arxiv.org/abs/1706.06100} {arXiv:1706.06100 [hep-ph]} \BibitemShut
  {NoStop}%
\bibitem [{\citenamefont {Chiang}\ \emph {et~al.}(2017)\citenamefont {Chiang},
  \citenamefont {He}, \citenamefont {Tandean},\ and\ \citenamefont
  {Yuan}}]{Chiang:2017hlj}%
  \BibitemOpen
  \bibfield  {author} {\bibinfo {author} {\bibfnamefont {C.-W.}\ \bibnamefont
  {Chiang}}, \bibinfo {author} {\bibfnamefont {X.-G.}\ \bibnamefont {He}},
  \bibinfo {author} {\bibfnamefont {J.}~\bibnamefont {Tandean}}, \ and\
  \bibinfo {author} {\bibfnamefont {X.-B.}\ \bibnamefont {Yuan}},\ }\href
  {\doibase 10.1103/PhysRevD.96.115022} {\bibfield  {journal} {\bibinfo
  {journal} {Phys. Rev.}\ }\textbf {\bibinfo {volume} {D96}},\ \bibinfo {pages}
  {115022} (\bibinfo {year} {2017})},\ \Eprint
  {http://arxiv.org/abs/1706.02696} {arXiv:1706.02696 [hep-ph]} \BibitemShut
  {NoStop}%
\bibitem [{\citenamefont {Di~Chiara}\ \emph {et~al.}(2017)\citenamefont
  {Di~Chiara}, \citenamefont {Fowlie}, \citenamefont {Fraser}, \citenamefont
  {Marzo}, \citenamefont {Marzola}, \citenamefont {Raidal},\ and\ \citenamefont
  {Spethmann}}]{DiChiara:2017cjq}%
  \BibitemOpen
  \bibfield  {author} {\bibinfo {author} {\bibfnamefont {S.}~\bibnamefont
  {Di~Chiara}}, \bibinfo {author} {\bibfnamefont {A.}~\bibnamefont {Fowlie}},
  \bibinfo {author} {\bibfnamefont {S.}~\bibnamefont {Fraser}}, \bibinfo
  {author} {\bibfnamefont {C.}~\bibnamefont {Marzo}}, \bibinfo {author}
  {\bibfnamefont {L.}~\bibnamefont {Marzola}}, \bibinfo {author} {\bibfnamefont
  {M.}~\bibnamefont {Raidal}}, \ and\ \bibinfo {author} {\bibfnamefont
  {C.}~\bibnamefont {Spethmann}},\ }\href {\doibase
  10.1016/j.nuclphysb.2017.08.003} {\bibfield  {journal} {\bibinfo  {journal}
  {Nucl. Phys.}\ }\textbf {\bibinfo {volume} {B923}},\ \bibinfo {pages} {245}
  (\bibinfo {year} {2017})},\ \Eprint {http://arxiv.org/abs/1704.06200}
  {arXiv:1704.06200 [hep-ph]} \BibitemShut {NoStop}%
\bibitem [{\citenamefont {Ko}\ \emph {et~al.}(2017)\citenamefont {Ko},
  \citenamefont {Omura}, \citenamefont {Shigekami},\ and\ \citenamefont
  {Yu}}]{Ko:2017lzd}%
  \BibitemOpen
  \bibfield  {author} {\bibinfo {author} {\bibfnamefont {P.}~\bibnamefont
  {Ko}}, \bibinfo {author} {\bibfnamefont {Y.}~\bibnamefont {Omura}}, \bibinfo
  {author} {\bibfnamefont {Y.}~\bibnamefont {Shigekami}}, \ and\ \bibinfo
  {author} {\bibfnamefont {C.}~\bibnamefont {Yu}},\ }\href {\doibase
  10.1103/PhysRevD.95.115040} {\bibfield  {journal} {\bibinfo  {journal} {Phys.
  Rev.}\ }\textbf {\bibinfo {volume} {D95}},\ \bibinfo {pages} {115040}
  (\bibinfo {year} {2017})},\ \Eprint {http://arxiv.org/abs/1702.08666}
  {arXiv:1702.08666 [hep-ph]} \BibitemShut {NoStop}%
\bibitem [{\citenamefont {Sannino}\ \emph {et~al.}(2018)\citenamefont
  {Sannino}, \citenamefont {Stangl}, \citenamefont {Straub},\ and\
  \citenamefont {Thomsen}}]{Sannino:2017utc}%
  \BibitemOpen
  \bibfield  {author} {\bibinfo {author} {\bibfnamefont {F.}~\bibnamefont
  {Sannino}}, \bibinfo {author} {\bibfnamefont {P.}~\bibnamefont {Stangl}},
  \bibinfo {author} {\bibfnamefont {D.~M.}\ \bibnamefont {Straub}}, \ and\
  \bibinfo {author} {\bibfnamefont {A.~E.}\ \bibnamefont {Thomsen}},\ }\href
  {\doibase 10.1103/PhysRevD.97.115046} {\bibfield  {journal} {\bibinfo
  {journal} {Phys. Rev.}\ }\textbf {\bibinfo {volume} {D97}},\ \bibinfo {pages}
  {115046} (\bibinfo {year} {2018})},\ \Eprint
  {http://arxiv.org/abs/1712.07646} {arXiv:1712.07646 [hep-ph]} \BibitemShut
  {NoStop}%
\bibitem [{\citenamefont {Falkowski}\ \emph {et~al.}(2018)\citenamefont
  {Falkowski}, \citenamefont {King}, \citenamefont {Perdomo},\ and\
  \citenamefont {Pierre}}]{Falkowski:2018dsl}%
  \BibitemOpen
  \bibfield  {author} {\bibinfo {author} {\bibfnamefont {A.}~\bibnamefont
  {Falkowski}}, \bibinfo {author} {\bibfnamefont {S.~F.}\ \bibnamefont {King}},
  \bibinfo {author} {\bibfnamefont {E.}~\bibnamefont {Perdomo}}, \ and\
  \bibinfo {author} {\bibfnamefont {M.}~\bibnamefont {Pierre}},\ }\href
  {\doibase 10.1007/JHEP08(2018)061} {\bibfield  {journal} {\bibinfo  {journal}
  {JHEP}\ }\textbf {\bibinfo {volume} {08}},\ \bibinfo {pages} {061} (\bibinfo
  {year} {2018})},\ \Eprint {http://arxiv.org/abs/1803.04430} {arXiv:1803.04430
  [hep-ph]} \BibitemShut {NoStop}%
\bibitem [{\citenamefont {Benavides}\ \emph {et~al.}(2018)\citenamefont
  {Benavides}, \citenamefont {Muñoz}, \citenamefont {Ponce}, \citenamefont
  {Rodríguez},\ and\ \citenamefont {Rojas}}]{Benavides:2018rgh}%
  \BibitemOpen
  \bibfield  {author} {\bibinfo {author} {\bibfnamefont {R.~H.}\ \bibnamefont
  {Benavides}}, \bibinfo {author} {\bibfnamefont {L.}~\bibnamefont {Muñoz}},
  \bibinfo {author} {\bibfnamefont {W.~A.}\ \bibnamefont {Ponce}}, \bibinfo
  {author} {\bibfnamefont {O.}~\bibnamefont {Rodríguez}}, \ and\ \bibinfo
  {author} {\bibfnamefont {E.}~\bibnamefont {Rojas}},\ }\href@noop {} {\
  (\bibinfo {year} {2018})},\ \Eprint {http://arxiv.org/abs/1812.05077}
  {arXiv:1812.05077 [hep-ph]} \BibitemShut {NoStop}%
\bibitem [{\citenamefont {Maji}\ \emph {et~al.}(2019)\citenamefont {Maji},
  \citenamefont {Nayek},\ and\ \citenamefont {Sahoo}}]{Maji:2018gvz}%
  \BibitemOpen
  \bibfield  {author} {\bibinfo {author} {\bibfnamefont {P.}~\bibnamefont
  {Maji}}, \bibinfo {author} {\bibfnamefont {P.}~\bibnamefont {Nayek}}, \ and\
  \bibinfo {author} {\bibfnamefont {S.}~\bibnamefont {Sahoo}},\ }\href
  {\doibase 10.1093/ptep/ptz010} {\bibfield  {journal} {\bibinfo  {journal}
  {PTEP}\ }\textbf {\bibinfo {volume} {2019}},\ \bibinfo {pages} {033B06}
  (\bibinfo {year} {2019})},\ \Eprint {http://arxiv.org/abs/1811.03869}
  {arXiv:1811.03869 [hep-ph]} \BibitemShut {NoStop}%
\bibitem [{\citenamefont {Singirala}\ \emph {et~al.}(2019)\citenamefont
  {Singirala}, \citenamefont {Sahoo},\ and\ \citenamefont
  {Mohanta}}]{Singirala:2018mio}%
  \BibitemOpen
  \bibfield  {author} {\bibinfo {author} {\bibfnamefont {S.}~\bibnamefont
  {Singirala}}, \bibinfo {author} {\bibfnamefont {S.}~\bibnamefont {Sahoo}}, \
  and\ \bibinfo {author} {\bibfnamefont {R.}~\bibnamefont {Mohanta}},\ }\href
  {\doibase 10.1103/PhysRevD.99.035042} {\bibfield  {journal} {\bibinfo
  {journal} {Phys. Rev.}\ }\textbf {\bibinfo {volume} {D99}},\ \bibinfo {pages}
  {035042} (\bibinfo {year} {2019})},\ \Eprint
  {http://arxiv.org/abs/1809.03213} {arXiv:1809.03213 [hep-ph]} \BibitemShut
  {NoStop}%
\bibitem [{\citenamefont {Guadagnoli}\ \emph {et~al.}(2018)\citenamefont
  {Guadagnoli}, \citenamefont {Reboud},\ and\ \citenamefont
  {Sumensari}}]{Guadagnoli:2018ojc}%
  \BibitemOpen
  \bibfield  {author} {\bibinfo {author} {\bibfnamefont {D.}~\bibnamefont
  {Guadagnoli}}, \bibinfo {author} {\bibfnamefont {M.}~\bibnamefont {Reboud}},
  \ and\ \bibinfo {author} {\bibfnamefont {O.}~\bibnamefont {Sumensari}},\
  }\href {\doibase 10.1007/JHEP11(2018)163} {\bibfield  {journal} {\bibinfo
  {journal} {JHEP}\ }\textbf {\bibinfo {volume} {11}},\ \bibinfo {pages} {163}
  (\bibinfo {year} {2018})},\ \Eprint {http://arxiv.org/abs/1807.03285}
  {arXiv:1807.03285 [hep-ph]} \BibitemShut {NoStop}%
\bibitem [{\citenamefont {Allanach}\ and\ \citenamefont
  {Davighi}(2018)}]{Allanach:2018lvl}%
  \BibitemOpen
  \bibfield  {author} {\bibinfo {author} {\bibfnamefont {B.~C.}\ \bibnamefont
  {Allanach}}\ and\ \bibinfo {author} {\bibfnamefont {J.}~\bibnamefont
  {Davighi}},\ }\href {\doibase 10.1007/JHEP12(2018)075} {\bibfield  {journal}
  {\bibinfo  {journal} {JHEP}\ }\textbf {\bibinfo {volume} {12}},\ \bibinfo
  {pages} {075} (\bibinfo {year} {2018})},\ \Eprint
  {http://arxiv.org/abs/1809.01158} {arXiv:1809.01158 [hep-ph]} \BibitemShut
  {NoStop}%
\bibitem [{\citenamefont {Duan}\ \emph {et~al.}(2019)\citenamefont {Duan},
  \citenamefont {Fan}, \citenamefont {Frank}, \citenamefont {Han},\ and\
  \citenamefont {Yang}}]{Duan:2018akc}%
  \BibitemOpen
  \bibfield  {author} {\bibinfo {author} {\bibfnamefont {G.~H.}\ \bibnamefont
  {Duan}}, \bibinfo {author} {\bibfnamefont {X.}~\bibnamefont {Fan}}, \bibinfo
  {author} {\bibfnamefont {M.}~\bibnamefont {Frank}}, \bibinfo {author}
  {\bibfnamefont {C.}~\bibnamefont {Han}}, \ and\ \bibinfo {author}
  {\bibfnamefont {J.~M.}\ \bibnamefont {Yang}},\ }\href {\doibase
  10.1016/j.physletb.2018.12.005} {\bibfield  {journal} {\bibinfo  {journal}
  {Phys. Lett.}\ }\textbf {\bibinfo {volume} {B789}},\ \bibinfo {pages} {54}
  (\bibinfo {year} {2019})},\ \Eprint {http://arxiv.org/abs/1808.04116}
  {arXiv:1808.04116 [hep-ph]} \BibitemShut {NoStop}%
\bibitem [{\citenamefont {King}(2018)}]{King:2018fcg}%
  \BibitemOpen
  \bibfield  {author} {\bibinfo {author} {\bibfnamefont {S.~F.}\ \bibnamefont
  {King}},\ }\href {\doibase 10.1007/JHEP09(2018)069} {\bibfield  {journal}
  {\bibinfo  {journal} {JHEP}\ }\textbf {\bibinfo {volume} {09}},\ \bibinfo
  {pages} {069} (\bibinfo {year} {2018})},\ \Eprint
  {http://arxiv.org/abs/1806.06780} {arXiv:1806.06780 [hep-ph]} \BibitemShut
  {NoStop}%
\bibitem [{\citenamefont {Kohda}\ \emph {et~al.}(2018)\citenamefont {Kohda},
  \citenamefont {Modak},\ and\ \citenamefont {Soffer}}]{Kohda:2018xbc}%
  \BibitemOpen
  \bibfield  {author} {\bibinfo {author} {\bibfnamefont {M.}~\bibnamefont
  {Kohda}}, \bibinfo {author} {\bibfnamefont {T.}~\bibnamefont {Modak}}, \ and\
  \bibinfo {author} {\bibfnamefont {A.}~\bibnamefont {Soffer}},\ }\href
  {\doibase 10.1103/PhysRevD.97.115019} {\bibfield  {journal} {\bibinfo
  {journal} {Phys. Rev.}\ }\textbf {\bibinfo {volume} {D97}},\ \bibinfo {pages}
  {115019} (\bibinfo {year} {2018})},\ \Eprint
  {http://arxiv.org/abs/1803.07492} {arXiv:1803.07492 [hep-ph]} \BibitemShut
  {NoStop}%
\bibitem [{\citenamefont {Dwivedi}\ \emph {et~al.}(2020)\citenamefont
  {Dwivedi}, \citenamefont {Falkowski}, \citenamefont {Kumar~Ghosh},\ and\
  \citenamefont {Ghosh}}]{Dwivedi:2019uqd}%
  \BibitemOpen
  \bibfield  {author} {\bibinfo {author} {\bibfnamefont {S.}~\bibnamefont
  {Dwivedi}}, \bibinfo {author} {\bibfnamefont {A.}~\bibnamefont {Falkowski}},
  \bibinfo {author} {\bibfnamefont {D.}~\bibnamefont {Kumar~Ghosh}}, \ and\
  \bibinfo {author} {\bibfnamefont {N.}~\bibnamefont {Ghosh}},\ }\href
  {\doibase 10.1140/epjc/s10052-020-7810-4} {\bibfield  {journal} {\bibinfo
  {journal} {Eur. Phys. J.}\ }\textbf {\bibinfo {volume} {C80}},\ \bibinfo
  {pages} {263} (\bibinfo {year} {2020})},\ \Eprint
  {http://arxiv.org/abs/1908.03031} {arXiv:1908.03031 [hep-ph]} \BibitemShut
  {NoStop}%
\bibitem [{\citenamefont {Foldenauer}(7 03)}]{Foldenauer:2019vgn}%
  \BibitemOpen
  \bibfield  {author} {\bibinfo {author} {\bibfnamefont {P.}~\bibnamefont
  {Foldenauer}},\ }\emph {\bibinfo {title} {{Phenomenology of Extra Abelian
  Gauge Symmetries}}},\ \href {\doibase 10.11588/heidok.00026777} {Ph.D.
  thesis},\ \bibinfo  {school} {U. Heidelberg (main)} (\bibinfo {year}
  {2019-07-03})\BibitemShut {NoStop}%
\bibitem [{\citenamefont {Ko}\ \emph {et~al.}(2019)\citenamefont {Ko},
  \citenamefont {Nomura},\ and\ \citenamefont {Yu}}]{Ko:2019tts}%
  \BibitemOpen
  \bibfield  {author} {\bibinfo {author} {\bibfnamefont {P.}~\bibnamefont
  {Ko}}, \bibinfo {author} {\bibfnamefont {T.}~\bibnamefont {Nomura}}, \ and\
  \bibinfo {author} {\bibfnamefont {C.}~\bibnamefont {Yu}},\ }\href {\doibase
  10.1007/JHEP04(2019)102} {\bibfield  {journal} {\bibinfo  {journal} {JHEP}\
  }\textbf {\bibinfo {volume} {04}},\ \bibinfo {pages} {102} (\bibinfo {year}
  {2019})},\ \Eprint {http://arxiv.org/abs/1902.06107} {arXiv:1902.06107
  [hep-ph]} \BibitemShut {NoStop}%
\bibitem [{\citenamefont {Allanach}\ and\ \citenamefont
  {Davighi}(2019)}]{Allanach:2019iiy}%
  \BibitemOpen
  \bibfield  {author} {\bibinfo {author} {\bibfnamefont {B.~C.}\ \bibnamefont
  {Allanach}}\ and\ \bibinfo {author} {\bibfnamefont {J.}~\bibnamefont
  {Davighi}},\ }\href {\doibase 10.1140/epjc/s10052-019-7414-z} {\bibfield
  {journal} {\bibinfo  {journal} {Eur. Phys. J.}\ }\textbf {\bibinfo {volume}
  {C79}},\ \bibinfo {pages} {908} (\bibinfo {year} {2019})},\ \Eprint
  {http://arxiv.org/abs/1905.10327} {arXiv:1905.10327 [hep-ph]} \BibitemShut
  {NoStop}%
\bibitem [{\citenamefont {Altmannshofer}\ \emph {et~al.}(2020)\citenamefont
  {Altmannshofer}, \citenamefont {Davighi},\ and\ \citenamefont
  {Nardecchia}}]{Altmannshofer:2019xda}%
  \BibitemOpen
  \bibfield  {author} {\bibinfo {author} {\bibfnamefont {W.}~\bibnamefont
  {Altmannshofer}}, \bibinfo {author} {\bibfnamefont {J.}~\bibnamefont
  {Davighi}}, \ and\ \bibinfo {author} {\bibfnamefont {M.}~\bibnamefont
  {Nardecchia}},\ }\href {\doibase 10.1103/PhysRevD.101.015004} {\bibfield
  {journal} {\bibinfo  {journal} {Phys. Rev.}\ }\textbf {\bibinfo {volume}
  {D101}},\ \bibinfo {pages} {015004} (\bibinfo {year} {2020})},\ \Eprint
  {http://arxiv.org/abs/1909.02021} {arXiv:1909.02021 [hep-ph]} \BibitemShut
  {NoStop}%
\bibitem [{\citenamefont {Calibbi}\ \emph {et~al.}(2020)\citenamefont
  {Calibbi}, \citenamefont {Crivellin}, \citenamefont {Kirk}, \citenamefont
  {Manzari},\ and\ \citenamefont {Vernazza}}]{Calibbi:2019lvs}%
  \BibitemOpen
  \bibfield  {author} {\bibinfo {author} {\bibfnamefont {L.}~\bibnamefont
  {Calibbi}}, \bibinfo {author} {\bibfnamefont {A.}~\bibnamefont {Crivellin}},
  \bibinfo {author} {\bibfnamefont {F.}~\bibnamefont {Kirk}}, \bibinfo {author}
  {\bibfnamefont {C.~A.}\ \bibnamefont {Manzari}}, \ and\ \bibinfo {author}
  {\bibfnamefont {L.}~\bibnamefont {Vernazza}},\ }\href {\doibase
  10.1103/PhysRevD.101.095003} {\bibfield  {journal} {\bibinfo  {journal}
  {Phys. Rev.}\ }\textbf {\bibinfo {volume} {D101}},\ \bibinfo {pages} {095003}
  (\bibinfo {year} {2020})},\ \Eprint {http://arxiv.org/abs/1910.00014}
  {arXiv:1910.00014 [hep-ph]} \BibitemShut {NoStop}%
\bibitem [{\citenamefont {Aebischer}\ \emph
  {et~al.}(2020{\natexlab{b}})\citenamefont {Aebischer}, \citenamefont {Buras},
  \citenamefont {Cerdà-Sevilla},\ and\ \citenamefont
  {De~Fazio}}]{Aebischer:2019blw}%
  \BibitemOpen
  \bibfield  {author} {\bibinfo {author} {\bibfnamefont {J.}~\bibnamefont
  {Aebischer}}, \bibinfo {author} {\bibfnamefont {A.~J.}\ \bibnamefont
  {Buras}}, \bibinfo {author} {\bibfnamefont {M.}~\bibnamefont
  {Cerdà-Sevilla}}, \ and\ \bibinfo {author} {\bibfnamefont {F.}~\bibnamefont
  {De~Fazio}},\ }\href {\doibase 10.1007/JHEP02(2020)183} {\bibfield  {journal}
  {\bibinfo  {journal} {JHEP}\ }\textbf {\bibinfo {volume} {02}},\ \bibinfo
  {pages} {183} (\bibinfo {year} {2020}{\natexlab{b}})},\ \Eprint
  {http://arxiv.org/abs/1912.09308} {arXiv:1912.09308 [hep-ph]} \BibitemShut
  {NoStop}%
\bibitem [{\citenamefont {del Aguila}\ \emph {et~al.}(2010)\citenamefont {del
  Aguila}, \citenamefont {de~Blas},\ and\ \citenamefont
  {Perez-Victoria}}]{delAguila:2010mx}%
  \BibitemOpen
  \bibfield  {author} {\bibinfo {author} {\bibfnamefont {F.}~\bibnamefont {del
  Aguila}}, \bibinfo {author} {\bibfnamefont {J.}~\bibnamefont {de~Blas}}, \
  and\ \bibinfo {author} {\bibfnamefont {M.}~\bibnamefont {Perez-Victoria}},\
  }\href {\doibase 10.1007/JHEP09(2010)033} {\bibfield  {journal} {\bibinfo
  {journal} {JHEP}\ }\textbf {\bibinfo {volume} {09}},\ \bibinfo {pages} {033}
  (\bibinfo {year} {2010})},\ \Eprint {http://arxiv.org/abs/1005.3998}
  {arXiv:1005.3998 [hep-ph]} \BibitemShut {NoStop}%
\bibitem [{\citenamefont {de~Blas}\ \emph {et~al.}(2013)\citenamefont
  {de~Blas}, \citenamefont {Lizana},\ and\ \citenamefont
  {Perez-Victoria}}]{deBlas:2012qp}%
  \BibitemOpen
  \bibfield  {author} {\bibinfo {author} {\bibfnamefont {J.}~\bibnamefont
  {de~Blas}}, \bibinfo {author} {\bibfnamefont {J.~M.}\ \bibnamefont {Lizana}},
  \ and\ \bibinfo {author} {\bibfnamefont {M.}~\bibnamefont {Perez-Victoria}},\
  }\href {\doibase 10.1007/JHEP01(2013)166} {\bibfield  {journal} {\bibinfo
  {journal} {JHEP}\ }\textbf {\bibinfo {volume} {01}},\ \bibinfo {pages} {166}
  (\bibinfo {year} {2013})},\ \Eprint {http://arxiv.org/abs/1211.2229}
  {arXiv:1211.2229 [hep-ph]} \BibitemShut {NoStop}%
\bibitem [{\citenamefont {Pappadopulo}\ \emph {et~al.}(2014)\citenamefont
  {Pappadopulo}, \citenamefont {Thamm}, \citenamefont {Torre},\ and\
  \citenamefont {Wulzer}}]{Pappadopulo:2014qza}%
  \BibitemOpen
  \bibfield  {author} {\bibinfo {author} {\bibfnamefont {D.}~\bibnamefont
  {Pappadopulo}}, \bibinfo {author} {\bibfnamefont {A.}~\bibnamefont {Thamm}},
  \bibinfo {author} {\bibfnamefont {R.}~\bibnamefont {Torre}}, \ and\ \bibinfo
  {author} {\bibfnamefont {A.}~\bibnamefont {Wulzer}},\ }\href {\doibase
  10.1007/JHEP09(2014)060} {\bibfield  {journal} {\bibinfo  {journal} {JHEP}\
  }\textbf {\bibinfo {volume} {09}},\ \bibinfo {pages} {060} (\bibinfo {year}
  {2014})},\ \Eprint {http://arxiv.org/abs/1402.4431} {arXiv:1402.4431
  [hep-ph]} \BibitemShut {NoStop}%
\bibitem [{\citenamefont {Arkani-Hamed}\ \emph {et~al.}(1998)\citenamefont
  {Arkani-Hamed}, \citenamefont {Dimopoulos},\ and\ \citenamefont
  {Dvali}}]{ArkaniHamed:1998rs}%
  \BibitemOpen
  \bibfield  {author} {\bibinfo {author} {\bibfnamefont {N.}~\bibnamefont
  {Arkani-Hamed}}, \bibinfo {author} {\bibfnamefont {S.}~\bibnamefont
  {Dimopoulos}}, \ and\ \bibinfo {author} {\bibfnamefont {G.~R.}\ \bibnamefont
  {Dvali}},\ }\href {\doibase 10.1016/S0370-2693(98)00466-3} {\bibfield
  {journal} {\bibinfo  {journal} {Phys. Lett.}\ }\textbf {\bibinfo {volume}
  {B429}},\ \bibinfo {pages} {263} (\bibinfo {year} {1998})},\ \Eprint
  {http://arxiv.org/abs/hep-ph/9803315} {arXiv:hep-ph/9803315 [hep-ph]}
  \BibitemShut {NoStop}%
\bibitem [{\citenamefont {Rizzo}(2000)}]{Rizzo:1999en}%
  \BibitemOpen
  \bibfield  {author} {\bibinfo {author} {\bibfnamefont {T.~G.}\ \bibnamefont
  {Rizzo}},\ }\href {\doibase 10.1103/PhysRevD.61.055005} {\bibfield  {journal}
  {\bibinfo  {journal} {Phys. Rev.}\ }\textbf {\bibinfo {volume} {D61}},\
  \bibinfo {pages} {055005} (\bibinfo {year} {2000})},\ \Eprint
  {http://arxiv.org/abs/hep-ph/9909232} {arXiv:hep-ph/9909232 [hep-ph]}
  \BibitemShut {NoStop}%
\bibitem [{\citenamefont {Csaki}(2004)}]{Csaki:2004ay}%
  \BibitemOpen
  \bibfield  {author} {\bibinfo {author} {\bibfnamefont {C.}~\bibnamefont
  {Csaki}},\ }in\ \href@noop {} {\emph {\bibinfo {booktitle} {{From fields to
  strings: Circumnavigating theoretical physics. Ian Kogan memorial collection
  (3 volume set)}}}}\ (\bibinfo {year} {2004})\ pp.\ \bibinfo {pages}
  {605--698},\ \Eprint {http://arxiv.org/abs/hep-ph/0404096}
  {arXiv:hep-ph/0404096 [hep-ph]} \BibitemShut {NoStop}%
\bibitem [{\citenamefont {Rizzo}(2009)}]{Rizzo:2009pu}%
  \BibitemOpen
  \bibfield  {author} {\bibinfo {author} {\bibfnamefont {T.~G.}\ \bibnamefont
  {Rizzo}},\ }\href {\doibase 10.1088/1126-6708/2009/08/082} {\bibfield
  {journal} {\bibinfo  {journal} {JHEP}\ }\textbf {\bibinfo {volume} {08}},\
  \bibinfo {pages} {082} (\bibinfo {year} {2009})},\ \Eprint
  {http://arxiv.org/abs/0904.2534} {arXiv:0904.2534 [hep-ph]} \BibitemShut
  {NoStop}%
\bibitem [{\citenamefont {Bella}\ \emph
  {et~al.}(2010{\natexlab{a}})\citenamefont {Bella}, \citenamefont {Etzion},
  \citenamefont {Hod},\ and\ \citenamefont {Sutton}}]{Bella:2010vi}%
  \BibitemOpen
  \bibfield  {author} {\bibinfo {author} {\bibfnamefont {G.}~\bibnamefont
  {Bella}}, \bibinfo {author} {\bibfnamefont {E.}~\bibnamefont {Etzion}},
  \bibinfo {author} {\bibfnamefont {N.}~\bibnamefont {Hod}}, \ and\ \bibinfo
  {author} {\bibfnamefont {M.}~\bibnamefont {Sutton}},\ }\href@noop {} {\
  (\bibinfo {year} {2010}{\natexlab{a}})},\ \Eprint
  {http://arxiv.org/abs/1004.1649} {arXiv:1004.1649 [hep-ex]} \BibitemShut
  {NoStop}%
\bibitem [{\citenamefont {Bella}\ \emph
  {et~al.}(2010{\natexlab{b}})\citenamefont {Bella}, \citenamefont {Etzion},
  \citenamefont {Hod}, \citenamefont {Oz}, \citenamefont {Silver},\ and\
  \citenamefont {Sutton}}]{Bella:2010sc}%
  \BibitemOpen
  \bibfield  {author} {\bibinfo {author} {\bibfnamefont {G.}~\bibnamefont
  {Bella}}, \bibinfo {author} {\bibfnamefont {E.}~\bibnamefont {Etzion}},
  \bibinfo {author} {\bibfnamefont {N.}~\bibnamefont {Hod}}, \bibinfo {author}
  {\bibfnamefont {Y.}~\bibnamefont {Oz}}, \bibinfo {author} {\bibfnamefont
  {Y.}~\bibnamefont {Silver}}, \ and\ \bibinfo {author} {\bibfnamefont
  {M.}~\bibnamefont {Sutton}},\ }\href {\doibase 10.1007/JHEP09(2010)025}
  {\bibfield  {journal} {\bibinfo  {journal} {JHEP}\ }\textbf {\bibinfo
  {volume} {09}},\ \bibinfo {pages} {025} (\bibinfo {year}
  {2010}{\natexlab{b}})},\ \Eprint {http://arxiv.org/abs/1004.2432}
  {arXiv:1004.2432 [hep-ex]} \BibitemShut {NoStop}%
\bibitem [{\citenamefont {Contino}\ \emph {et~al.}(2007)\citenamefont
  {Contino}, \citenamefont {Kramer}, \citenamefont {Son},\ and\ \citenamefont
  {Sundrum}}]{Contino:2006nn}%
  \BibitemOpen
  \bibfield  {author} {\bibinfo {author} {\bibfnamefont {R.}~\bibnamefont
  {Contino}}, \bibinfo {author} {\bibfnamefont {T.}~\bibnamefont {Kramer}},
  \bibinfo {author} {\bibfnamefont {M.}~\bibnamefont {Son}}, \ and\ \bibinfo
  {author} {\bibfnamefont {R.}~\bibnamefont {Sundrum}},\ }\href {\doibase
  10.1088/1126-6708/2007/05/074} {\bibfield  {journal} {\bibinfo  {journal}
  {JHEP}\ }\textbf {\bibinfo {volume} {05}},\ \bibinfo {pages} {074} (\bibinfo
  {year} {2007})},\ \Eprint {http://arxiv.org/abs/hep-ph/0612180}
  {arXiv:hep-ph/0612180 [hep-ph]} \BibitemShut {NoStop}%
\bibitem [{\citenamefont {Contino}\ \emph {et~al.}(2011)\citenamefont
  {Contino}, \citenamefont {Marzocca}, \citenamefont {Pappadopulo},\ and\
  \citenamefont {Rattazzi}}]{Contino:2011np}%
  \BibitemOpen
  \bibfield  {author} {\bibinfo {author} {\bibfnamefont {R.}~\bibnamefont
  {Contino}}, \bibinfo {author} {\bibfnamefont {D.}~\bibnamefont {Marzocca}},
  \bibinfo {author} {\bibfnamefont {D.}~\bibnamefont {Pappadopulo}}, \ and\
  \bibinfo {author} {\bibfnamefont {R.}~\bibnamefont {Rattazzi}},\ }\href
  {\doibase 10.1007/JHEP10(2011)081} {\bibfield  {journal} {\bibinfo  {journal}
  {JHEP}\ }\textbf {\bibinfo {volume} {10}},\ \bibinfo {pages} {081} (\bibinfo
  {year} {2011})},\ \Eprint {http://arxiv.org/abs/1109.1570} {arXiv:1109.1570
  [hep-ph]} \BibitemShut {NoStop}%
\bibitem [{\citenamefont {Bellazzini}\ \emph {et~al.}(2012)\citenamefont
  {Bellazzini}, \citenamefont {Csaki}, \citenamefont {Hubisz}, \citenamefont
  {Serra},\ and\ \citenamefont {Terning}}]{Bellazzini:2012tv}%
  \BibitemOpen
  \bibfield  {author} {\bibinfo {author} {\bibfnamefont {B.}~\bibnamefont
  {Bellazzini}}, \bibinfo {author} {\bibfnamefont {C.}~\bibnamefont {Csaki}},
  \bibinfo {author} {\bibfnamefont {J.}~\bibnamefont {Hubisz}}, \bibinfo
  {author} {\bibfnamefont {J.}~\bibnamefont {Serra}}, \ and\ \bibinfo {author}
  {\bibfnamefont {J.}~\bibnamefont {Terning}},\ }\href {\doibase
  10.1007/JHEP11(2012)003} {\bibfield  {journal} {\bibinfo  {journal} {JHEP}\
  }\textbf {\bibinfo {volume} {11}},\ \bibinfo {pages} {003} (\bibinfo {year}
  {2012})},\ \Eprint {http://arxiv.org/abs/1205.4032} {arXiv:1205.4032
  [hep-ph]} \BibitemShut {NoStop}%
\bibitem [{\citenamefont {Li}\ and\ \citenamefont {Ma}(1981)}]{Li:1981nk}%
  \BibitemOpen
  \bibfield  {author} {\bibinfo {author} {\bibfnamefont {X.}~\bibnamefont
  {Li}}\ and\ \bibinfo {author} {\bibfnamefont {E.}~\bibnamefont {Ma}},\ }\href
  {\doibase 10.1103/PhysRevLett.47.1788} {\bibfield  {journal} {\bibinfo
  {journal} {Phys. Rev. Lett.}\ }\textbf {\bibinfo {volume} {47}},\ \bibinfo
  {pages} {1788} (\bibinfo {year} {1981})}\BibitemShut {NoStop}%
\bibitem [{\citenamefont {Muller}\ and\ \citenamefont
  {Nandi}(1996)}]{Muller:1996dj}%
  \BibitemOpen
  \bibfield  {author} {\bibinfo {author} {\bibfnamefont {D.~J.}\ \bibnamefont
  {Muller}}\ and\ \bibinfo {author} {\bibfnamefont {S.}~\bibnamefont {Nandi}},\
  }\href {\doibase 10.1016/0370-2693(96)00745-9} {\bibfield  {journal}
  {\bibinfo  {journal} {Phys. Lett.}\ }\textbf {\bibinfo {volume} {B383}},\
  \bibinfo {pages} {345} (\bibinfo {year} {1996})},\ \Eprint
  {http://arxiv.org/abs/hep-ph/9602390} {arXiv:hep-ph/9602390 [hep-ph]}
  \BibitemShut {NoStop}%
\bibitem [{\citenamefont {Morrissey}\ \emph {et~al.}(2005)\citenamefont
  {Morrissey}, \citenamefont {Tait},\ and\ \citenamefont
  {Wagner}}]{Morrissey:2005uza}%
  \BibitemOpen
  \bibfield  {author} {\bibinfo {author} {\bibfnamefont {D.~E.}\ \bibnamefont
  {Morrissey}}, \bibinfo {author} {\bibfnamefont {T.~M.~P.}\ \bibnamefont
  {Tait}}, \ and\ \bibinfo {author} {\bibfnamefont {C.~E.~M.}\ \bibnamefont
  {Wagner}},\ }\href {\doibase 10.1103/PhysRevD.72.095003} {\bibfield
  {journal} {\bibinfo  {journal} {Phys. Rev.}\ }\textbf {\bibinfo {volume}
  {D72}},\ \bibinfo {pages} {095003} (\bibinfo {year} {2005})},\ \Eprint
  {http://arxiv.org/abs/hep-ph/0508123} {arXiv:hep-ph/0508123 [hep-ph]}
  \BibitemShut {NoStop}%
\bibitem [{\citenamefont {Chiang}\ \emph {et~al.}(2010)\citenamefont {Chiang},
  \citenamefont {Deshpande}, \citenamefont {He},\ and\ \citenamefont
  {Jiang}}]{Chiang:2009kb}%
  \BibitemOpen
  \bibfield  {author} {\bibinfo {author} {\bibfnamefont {C.-W.}\ \bibnamefont
  {Chiang}}, \bibinfo {author} {\bibfnamefont {N.~G.}\ \bibnamefont
  {Deshpande}}, \bibinfo {author} {\bibfnamefont {X.-G.}\ \bibnamefont {He}}, \
  and\ \bibinfo {author} {\bibfnamefont {J.}~\bibnamefont {Jiang}},\ }\href
  {\doibase 10.1103/PhysRevD.81.015006} {\bibfield  {journal} {\bibinfo
  {journal} {Phys. Rev.}\ }\textbf {\bibinfo {volume} {D81}},\ \bibinfo {pages}
  {015006} (\bibinfo {year} {2010})},\ \Eprint {http://arxiv.org/abs/0911.1480}
  {arXiv:0911.1480 [hep-ph]} \BibitemShut {NoStop}%
\bibitem [{\citenamefont {Fuentes-Martin}\ \emph {et~al.}(2015)\citenamefont
  {Fuentes-Martin}, \citenamefont {Portoles},\ and\ \citenamefont
  {Ruiz-Femenia}}]{Fuentes-Martin:2014fxa}%
  \BibitemOpen
  \bibfield  {author} {\bibinfo {author} {\bibfnamefont {J.}~\bibnamefont
  {Fuentes-Martin}}, \bibinfo {author} {\bibfnamefont {J.}~\bibnamefont
  {Portoles}}, \ and\ \bibinfo {author} {\bibfnamefont {P.}~\bibnamefont
  {Ruiz-Femenia}},\ }\href {\doibase 10.1007/JHEP01(2015)134} {\bibfield
  {journal} {\bibinfo  {journal} {JHEP}\ }\textbf {\bibinfo {volume} {01}},\
  \bibinfo {pages} {134} (\bibinfo {year} {2015})},\ \Eprint
  {http://arxiv.org/abs/1411.2471} {arXiv:1411.2471 [hep-ph]} \BibitemShut
  {NoStop}%
\bibitem [{\citenamefont {He}\ and\ \citenamefont
  {Valencia}(2013)}]{He:2012zp}%
  \BibitemOpen
  \bibfield  {author} {\bibinfo {author} {\bibfnamefont {X.-G.}\ \bibnamefont
  {He}}\ and\ \bibinfo {author} {\bibfnamefont {G.}~\bibnamefont {Valencia}},\
  }\href {\doibase 10.1103/PhysRevD.87.014014} {\bibfield  {journal} {\bibinfo
  {journal} {Phys. Rev.}\ }\textbf {\bibinfo {volume} {D87}},\ \bibinfo {pages}
  {014014} (\bibinfo {year} {2013})},\ \Eprint {http://arxiv.org/abs/1211.0348}
  {arXiv:1211.0348 [hep-ph]} \BibitemShut {NoStop}%
\bibitem [{\citenamefont {Bhattacharya}\ \emph {et~al.}(2015)\citenamefont
  {Bhattacharya}, \citenamefont {Datta}, \citenamefont {London},\ and\
  \citenamefont {Shivashankara}}]{Bhattacharya:2014wla}%
  \BibitemOpen
  \bibfield  {author} {\bibinfo {author} {\bibfnamefont {B.}~\bibnamefont
  {Bhattacharya}}, \bibinfo {author} {\bibfnamefont {A.}~\bibnamefont {Datta}},
  \bibinfo {author} {\bibfnamefont {D.}~\bibnamefont {London}}, \ and\ \bibinfo
  {author} {\bibfnamefont {S.}~\bibnamefont {Shivashankara}},\ }\href {\doibase
  10.1016/j.physletb.2015.02.011} {\bibfield  {journal} {\bibinfo  {journal}
  {Phys. Lett.}\ }\textbf {\bibinfo {volume} {B742}},\ \bibinfo {pages} {370}
  (\bibinfo {year} {2015})},\ \Eprint {http://arxiv.org/abs/1412.7164}
  {arXiv:1412.7164 [hep-ph]} \BibitemShut {NoStop}%
\bibitem [{\citenamefont {Alonso}\ \emph {et~al.}(2014)\citenamefont {Alonso},
  \citenamefont {Grinstein},\ and\ \citenamefont
  {Martin~Camalich}}]{Alonso:2014csa}%
  \BibitemOpen
  \bibfield  {author} {\bibinfo {author} {\bibfnamefont {R.}~\bibnamefont
  {Alonso}}, \bibinfo {author} {\bibfnamefont {B.}~\bibnamefont {Grinstein}}, \
  and\ \bibinfo {author} {\bibfnamefont {J.}~\bibnamefont {Martin~Camalich}},\
  }\href {\doibase 10.1103/PhysRevLett.113.241802} {\bibfield  {journal}
  {\bibinfo  {journal} {Phys. Rev. Lett.}\ }\textbf {\bibinfo {volume} {113}},\
  \bibinfo {pages} {241802} (\bibinfo {year} {2014})},\ \Eprint
  {http://arxiv.org/abs/1407.7044} {arXiv:1407.7044 [hep-ph]} \BibitemShut
  {NoStop}%
\bibitem [{\citenamefont {Greljo}\ \emph {et~al.}(2015)\citenamefont {Greljo},
  \citenamefont {Isidori},\ and\ \citenamefont {Marzocca}}]{Greljo:2015mma}%
  \BibitemOpen
  \bibfield  {author} {\bibinfo {author} {\bibfnamefont {A.}~\bibnamefont
  {Greljo}}, \bibinfo {author} {\bibfnamefont {G.}~\bibnamefont {Isidori}}, \
  and\ \bibinfo {author} {\bibfnamefont {D.}~\bibnamefont {Marzocca}},\ }\href
  {\doibase 10.1007/JHEP07(2015)142} {\bibfield  {journal} {\bibinfo  {journal}
  {JHEP}\ }\textbf {\bibinfo {volume} {07}},\ \bibinfo {pages} {142} (\bibinfo
  {year} {2015})},\ \Eprint {http://arxiv.org/abs/1506.01705} {arXiv:1506.01705
  [hep-ph]} \BibitemShut {NoStop}%
\bibitem [{\citenamefont {Boucenna}\ \emph
  {et~al.}(2016{\natexlab{a}})\citenamefont {Boucenna}, \citenamefont {Celis},
  \citenamefont {Fuentes-Martin}, \citenamefont {Vicente},\ and\ \citenamefont
  {Virto}}]{Boucenna:2016wpr}%
  \BibitemOpen
  \bibfield  {author} {\bibinfo {author} {\bibfnamefont {S.~M.}\ \bibnamefont
  {Boucenna}}, \bibinfo {author} {\bibfnamefont {A.}~\bibnamefont {Celis}},
  \bibinfo {author} {\bibfnamefont {J.}~\bibnamefont {Fuentes-Martin}},
  \bibinfo {author} {\bibfnamefont {A.}~\bibnamefont {Vicente}}, \ and\
  \bibinfo {author} {\bibfnamefont {J.}~\bibnamefont {Virto}},\ }\href
  {\doibase 10.1016/j.physletb.2016.06.067} {\bibfield  {journal} {\bibinfo
  {journal} {Phys. Lett.}\ }\textbf {\bibinfo {volume} {B760}},\ \bibinfo
  {pages} {214} (\bibinfo {year} {2016}{\natexlab{a}})},\ \Eprint
  {http://arxiv.org/abs/1604.03088} {arXiv:1604.03088 [hep-ph]} \BibitemShut
  {NoStop}%
\bibitem [{\citenamefont {Boucenna}\ \emph
  {et~al.}(2016{\natexlab{b}})\citenamefont {Boucenna}, \citenamefont {Celis},
  \citenamefont {Fuentes-Martin}, \citenamefont {Vicente},\ and\ \citenamefont
  {Virto}}]{Boucenna:2016qad}%
  \BibitemOpen
  \bibfield  {author} {\bibinfo {author} {\bibfnamefont {S.~M.}\ \bibnamefont
  {Boucenna}}, \bibinfo {author} {\bibfnamefont {A.}~\bibnamefont {Celis}},
  \bibinfo {author} {\bibfnamefont {J.}~\bibnamefont {Fuentes-Martin}},
  \bibinfo {author} {\bibfnamefont {A.}~\bibnamefont {Vicente}}, \ and\
  \bibinfo {author} {\bibfnamefont {J.}~\bibnamefont {Virto}},\ }\href
  {\doibase 10.1007/JHEP12(2016)059} {\bibfield  {journal} {\bibinfo  {journal}
  {JHEP}\ }\textbf {\bibinfo {volume} {12}},\ \bibinfo {pages} {059} (\bibinfo
  {year} {2016}{\natexlab{b}})},\ \Eprint {http://arxiv.org/abs/1608.01349}
  {arXiv:1608.01349 [hep-ph]} \BibitemShut {NoStop}%
\bibitem [{\citenamefont {Bhattacharya}\ \emph {et~al.}(2017)\citenamefont
  {Bhattacharya}, \citenamefont {Datta}, \citenamefont {Guévin}, \citenamefont
  {London},\ and\ \citenamefont {Watanabe}}]{Bhattacharya:2016mcc}%
  \BibitemOpen
  \bibfield  {author} {\bibinfo {author} {\bibfnamefont {B.}~\bibnamefont
  {Bhattacharya}}, \bibinfo {author} {\bibfnamefont {A.}~\bibnamefont {Datta}},
  \bibinfo {author} {\bibfnamefont {J.-P.}\ \bibnamefont {Guévin}}, \bibinfo
  {author} {\bibfnamefont {D.}~\bibnamefont {London}}, \ and\ \bibinfo {author}
  {\bibfnamefont {R.}~\bibnamefont {Watanabe}},\ }\href {\doibase
  10.1007/JHEP01(2017)015} {\bibfield  {journal} {\bibinfo  {journal} {JHEP}\
  }\textbf {\bibinfo {volume} {01}},\ \bibinfo {pages} {015} (\bibinfo {year}
  {2017})},\ \Eprint {http://arxiv.org/abs/1609.09078} {arXiv:1609.09078
  [hep-ph]} \BibitemShut {NoStop}%
\bibitem [{\citenamefont {Buttazzo}\ \emph {et~al.}(2017)\citenamefont
  {Buttazzo}, \citenamefont {Greljo}, \citenamefont {Isidori},\ and\
  \citenamefont {Marzocca}}]{Buttazzo:2017ixm}%
  \BibitemOpen
  \bibfield  {author} {\bibinfo {author} {\bibfnamefont {D.}~\bibnamefont
  {Buttazzo}}, \bibinfo {author} {\bibfnamefont {A.}~\bibnamefont {Greljo}},
  \bibinfo {author} {\bibfnamefont {G.}~\bibnamefont {Isidori}}, \ and\
  \bibinfo {author} {\bibfnamefont {D.}~\bibnamefont {Marzocca}},\ }\href
  {\doibase 10.1007/JHEP11(2017)044} {\bibfield  {journal} {\bibinfo  {journal}
  {JHEP}\ }\textbf {\bibinfo {volume} {11}},\ \bibinfo {pages} {044} (\bibinfo
  {year} {2017})},\ \Eprint {http://arxiv.org/abs/1706.07808} {arXiv:1706.07808
  [hep-ph]} \BibitemShut {NoStop}%
\bibitem [{\citenamefont {Kumar}\ \emph {et~al.}(2019)\citenamefont {Kumar},
  \citenamefont {London},\ and\ \citenamefont {Watanabe}}]{Kumar:2018kmr}%
  \BibitemOpen
  \bibfield  {author} {\bibinfo {author} {\bibfnamefont {J.}~\bibnamefont
  {Kumar}}, \bibinfo {author} {\bibfnamefont {D.}~\bibnamefont {London}}, \
  and\ \bibinfo {author} {\bibfnamefont {R.}~\bibnamefont {Watanabe}},\ }\href
  {\doibase 10.1103/PhysRevD.99.015007} {\bibfield  {journal} {\bibinfo
  {journal} {Phys. Rev.}\ }\textbf {\bibinfo {volume} {D99}},\ \bibinfo {pages}
  {015007} (\bibinfo {year} {2019})},\ \Eprint
  {http://arxiv.org/abs/1806.07403} {arXiv:1806.07403 [hep-ph]} \BibitemShut
  {NoStop}%
\bibitem [{\citenamefont {Blanke}\ and\ \citenamefont
  {Crivellin}(2018)}]{Blanke:2018sro}%
  \BibitemOpen
  \bibfield  {author} {\bibinfo {author} {\bibfnamefont {M.}~\bibnamefont
  {Blanke}}\ and\ \bibinfo {author} {\bibfnamefont {A.}~\bibnamefont
  {Crivellin}},\ }\href {\doibase 10.1103/PhysRevLett.121.011801} {\bibfield
  {journal} {\bibinfo  {journal} {Phys. Rev. Lett.}\ }\textbf {\bibinfo
  {volume} {121}},\ \bibinfo {pages} {011801} (\bibinfo {year} {2018})},\
  \Eprint {http://arxiv.org/abs/1801.07256} {arXiv:1801.07256 [hep-ph]}
  \BibitemShut {NoStop}%
\bibitem [{\citenamefont {Tanabashi}\ \emph {et~al.}(2018)\citenamefont
  {Tanabashi} \emph {et~al.}}]{Tanabashi:2018oca}%
  \BibitemOpen
  \bibfield  {author} {\bibinfo {author} {\bibfnamefont {M.}~\bibnamefont
  {Tanabashi}} \emph {et~al.} (\bibinfo {collaboration} {Particle Data
  Group}),\ }\href {\doibase 10.1103/PhysRevD.98.030001} {\bibfield  {journal}
  {\bibinfo  {journal} {Phys. Rev.}\ }\textbf {\bibinfo {volume} {D98}},\
  \bibinfo {pages} {030001} (\bibinfo {year} {2018})}\BibitemShut {NoStop}%
\bibitem [{\citenamefont {Hardy}\ and\ \citenamefont
  {Towner}(2018)}]{Hardy:2018zsb}%
  \BibitemOpen
  \bibfield  {author} {\bibinfo {author} {\bibfnamefont {J.~C.}\ \bibnamefont
  {Hardy}}\ and\ \bibinfo {author} {\bibfnamefont {I.~S.}\ \bibnamefont
  {Towner}},\ }in\ \href@noop {} {\emph {\bibinfo {booktitle} {{13th Conference
  on the Intersections of Particle and Nuclear Physics (CIPANP 2018) Palm
  Springs, California, USA, May 29-June 3, 2018}}}}\ (\bibinfo {year} {2018})\
  \Eprint {http://arxiv.org/abs/1807.01146} {arXiv:1807.01146 [nucl-ex]}
  \BibitemShut {NoStop}%
\bibitem [{\citenamefont {Seng}\ \emph {et~al.}(2020)\citenamefont {Seng},
  \citenamefont {Feng}, \citenamefont {Gorchtein},\ and\ \citenamefont
  {Jin}}]{Seng:2020wjq}%
  \BibitemOpen
  \bibfield  {author} {\bibinfo {author} {\bibfnamefont {C.-Y.}\ \bibnamefont
  {Seng}}, \bibinfo {author} {\bibfnamefont {X.}~\bibnamefont {Feng}}, \bibinfo
  {author} {\bibfnamefont {M.}~\bibnamefont {Gorchtein}}, \ and\ \bibinfo
  {author} {\bibfnamefont {L.-C.}\ \bibnamefont {Jin}},\ }\href@noop {} {\
  (\bibinfo {year} {2020})},\ \Eprint {http://arxiv.org/abs/2003.11264}
  {arXiv:2003.11264 [hep-ph]} \BibitemShut {NoStop}%
\bibitem [{\citenamefont {Barbieri}\ \emph {et~al.}(1996)\citenamefont
  {Barbieri}, \citenamefont {Dvali},\ and\ \citenamefont
  {Hall}}]{Barbieri:1995uv}%
  \BibitemOpen
  \bibfield  {author} {\bibinfo {author} {\bibfnamefont {R.}~\bibnamefont
  {Barbieri}}, \bibinfo {author} {\bibfnamefont {G.~R.}\ \bibnamefont {Dvali}},
  \ and\ \bibinfo {author} {\bibfnamefont {L.~J.}\ \bibnamefont {Hall}},\
  }\href {\doibase 10.1016/0370-2693(96)00318-8} {\bibfield  {journal}
  {\bibinfo  {journal} {Phys. Lett.}\ }\textbf {\bibinfo {volume} {B377}},\
  \bibinfo {pages} {76} (\bibinfo {year} {1996})},\ \Eprint
  {http://arxiv.org/abs/hep-ph/9512388} {arXiv:hep-ph/9512388 [hep-ph]}
  \BibitemShut {NoStop}%
\bibitem [{\citenamefont {Barbieri}\ \emph {et~al.}(1997)\citenamefont
  {Barbieri}, \citenamefont {Hall},\ and\ \citenamefont
  {Romanino}}]{Barbieri:1997tu}%
  \BibitemOpen
  \bibfield  {author} {\bibinfo {author} {\bibfnamefont {R.}~\bibnamefont
  {Barbieri}}, \bibinfo {author} {\bibfnamefont {L.~J.}\ \bibnamefont {Hall}},
  \ and\ \bibinfo {author} {\bibfnamefont {A.}~\bibnamefont {Romanino}},\
  }\href {\doibase 10.1016/S0370-2693(97)00372-9} {\bibfield  {journal}
  {\bibinfo  {journal} {Phys. Lett.}\ }\textbf {\bibinfo {volume} {B401}},\
  \bibinfo {pages} {47} (\bibinfo {year} {1997})},\ \Eprint
  {http://arxiv.org/abs/hep-ph/9702315} {arXiv:hep-ph/9702315 [hep-ph]}
  \BibitemShut {NoStop}%
\bibitem [{\citenamefont {Barbieri}\ \emph
  {et~al.}(2011{\natexlab{a}})\citenamefont {Barbieri}, \citenamefont {Campli},
  \citenamefont {Isidori}, \citenamefont {Sala},\ and\ \citenamefont
  {Straub}}]{Barbieri:2011fc}%
  \BibitemOpen
  \bibfield  {author} {\bibinfo {author} {\bibfnamefont {R.}~\bibnamefont
  {Barbieri}}, \bibinfo {author} {\bibfnamefont {P.}~\bibnamefont {Campli}},
  \bibinfo {author} {\bibfnamefont {G.}~\bibnamefont {Isidori}}, \bibinfo
  {author} {\bibfnamefont {F.}~\bibnamefont {Sala}}, \ and\ \bibinfo {author}
  {\bibfnamefont {D.~M.}\ \bibnamefont {Straub}},\ }\href {\doibase
  10.1140/epjc/s10052-011-1812-1} {\bibfield  {journal} {\bibinfo  {journal}
  {Eur. Phys. J.}\ }\textbf {\bibinfo {volume} {C71}},\ \bibinfo {pages} {1812}
  (\bibinfo {year} {2011}{\natexlab{a}})},\ \Eprint
  {http://arxiv.org/abs/1108.5125} {arXiv:1108.5125 [hep-ph]} \BibitemShut
  {NoStop}%
\bibitem [{\citenamefont {Barbieri}\ \emph
  {et~al.}(2011{\natexlab{b}})\citenamefont {Barbieri}, \citenamefont
  {Isidori}, \citenamefont {Jones-Perez}, \citenamefont {Lodone},\ and\
  \citenamefont {Straub}}]{Barbieri:2011ci}%
  \BibitemOpen
  \bibfield  {author} {\bibinfo {author} {\bibfnamefont {R.}~\bibnamefont
  {Barbieri}}, \bibinfo {author} {\bibfnamefont {G.}~\bibnamefont {Isidori}},
  \bibinfo {author} {\bibfnamefont {J.}~\bibnamefont {Jones-Perez}}, \bibinfo
  {author} {\bibfnamefont {P.}~\bibnamefont {Lodone}}, \ and\ \bibinfo {author}
  {\bibfnamefont {D.~M.}\ \bibnamefont {Straub}},\ }\href {\doibase
  10.1140/epjc/s10052-011-1725-z} {\bibfield  {journal} {\bibinfo  {journal}
  {Eur. Phys. J.}\ }\textbf {\bibinfo {volume} {C71}},\ \bibinfo {pages} {1725}
  (\bibinfo {year} {2011}{\natexlab{b}})},\ \Eprint
  {http://arxiv.org/abs/1105.2296} {arXiv:1105.2296 [hep-ph]} \BibitemShut
  {NoStop}%
\bibitem [{\citenamefont {Crivellin}\ \emph {et~al.}(2011)\citenamefont
  {Crivellin}, \citenamefont {Hofer},\ and\ \citenamefont
  {Nierste}}]{Crivellin:2011fb}%
  \BibitemOpen
  \bibfield  {author} {\bibinfo {author} {\bibfnamefont {A.}~\bibnamefont
  {Crivellin}}, \bibinfo {author} {\bibfnamefont {L.}~\bibnamefont {Hofer}}, \
  and\ \bibinfo {author} {\bibfnamefont {U.}~\bibnamefont {Nierste}},\
  }\bibfield  {booktitle} {\emph {\bibinfo {booktitle} {{Proceedings, 21st
  International Europhysics Conference on High energy physics (EPS-HEP 2011):
  Grenoble, France, July 21-27, 2011}}},\ }\href {\doibase 10.22323/1.134.0145}
  {\bibfield  {journal} {\bibinfo  {journal} {PoS}\ }\textbf {\bibinfo {volume}
  {EPS-HEP2011}},\ \bibinfo {pages} {145} (\bibinfo {year} {2011})},\ \Eprint
  {http://arxiv.org/abs/1111.0246} {arXiv:1111.0246 [hep-ph]} \BibitemShut
  {NoStop}%
\bibitem [{\citenamefont {Barbieri}\ \emph
  {et~al.}(2012{\natexlab{a}})\citenamefont {Barbieri}, \citenamefont
  {Buttazzo}, \citenamefont {Sala},\ and\ \citenamefont
  {Straub}}]{Barbieri:2012uh}%
  \BibitemOpen
  \bibfield  {author} {\bibinfo {author} {\bibfnamefont {R.}~\bibnamefont
  {Barbieri}}, \bibinfo {author} {\bibfnamefont {D.}~\bibnamefont {Buttazzo}},
  \bibinfo {author} {\bibfnamefont {F.}~\bibnamefont {Sala}}, \ and\ \bibinfo
  {author} {\bibfnamefont {D.~M.}\ \bibnamefont {Straub}},\ }\href {\doibase
  10.1007/JHEP07(2012)181} {\bibfield  {journal} {\bibinfo  {journal} {JHEP}\
  }\textbf {\bibinfo {volume} {07}},\ \bibinfo {pages} {181} (\bibinfo {year}
  {2012}{\natexlab{a}})},\ \Eprint {http://arxiv.org/abs/1203.4218}
  {arXiv:1203.4218 [hep-ph]} \BibitemShut {NoStop}%
\bibitem [{\citenamefont {Barbieri}\ \emph
  {et~al.}(2012{\natexlab{b}})\citenamefont {Barbieri}, \citenamefont
  {Buttazzo}, \citenamefont {Sala},\ and\ \citenamefont
  {Straub}}]{Barbieri:2012bh}%
  \BibitemOpen
  \bibfield  {author} {\bibinfo {author} {\bibfnamefont {R.}~\bibnamefont
  {Barbieri}}, \bibinfo {author} {\bibfnamefont {D.}~\bibnamefont {Buttazzo}},
  \bibinfo {author} {\bibfnamefont {F.}~\bibnamefont {Sala}}, \ and\ \bibinfo
  {author} {\bibfnamefont {D.~M.}\ \bibnamefont {Straub}},\ }\href {\doibase
  10.1007/JHEP10(2012)040} {\bibfield  {journal} {\bibinfo  {journal} {JHEP}\
  }\textbf {\bibinfo {volume} {10}},\ \bibinfo {pages} {040} (\bibinfo {year}
  {2012}{\natexlab{b}})},\ \Eprint {http://arxiv.org/abs/1206.1327}
  {arXiv:1206.1327 [hep-ph]} \BibitemShut {NoStop}%
\bibitem [{\citenamefont {Buras}\ and\ \citenamefont
  {Girrbach}(2013{\natexlab{b}})}]{Buras:2012sd}%
  \BibitemOpen
  \bibfield  {author} {\bibinfo {author} {\bibfnamefont {A.~J.}\ \bibnamefont
  {Buras}}\ and\ \bibinfo {author} {\bibfnamefont {J.}~\bibnamefont
  {Girrbach}},\ }\href {\doibase 10.1007/JHEP01(2013)007} {\bibfield  {journal}
  {\bibinfo  {journal} {JHEP}\ }\textbf {\bibinfo {volume} {01}},\ \bibinfo
  {pages} {007} (\bibinfo {year} {2013}{\natexlab{b}})},\ \Eprint
  {http://arxiv.org/abs/1206.3878} {arXiv:1206.3878 [hep-ph]} \BibitemShut
  {NoStop}%
\bibitem [{\citenamefont {Chivukula}\ \emph {et~al.}(1987)\citenamefont
  {Chivukula}, \citenamefont {Georgi},\ and\ \citenamefont
  {Randall}}]{Chivukula:1987fw}%
  \BibitemOpen
  \bibfield  {author} {\bibinfo {author} {\bibfnamefont {R.~S.}\ \bibnamefont
  {Chivukula}}, \bibinfo {author} {\bibfnamefont {H.}~\bibnamefont {Georgi}}, \
  and\ \bibinfo {author} {\bibfnamefont {L.}~\bibnamefont {Randall}},\ }\href
  {\doibase 10.1016/0550-3213(87)90638-9} {\bibfield  {journal} {\bibinfo
  {journal} {Nucl. Phys.}\ }\textbf {\bibinfo {volume} {B292}},\ \bibinfo
  {pages} {93} (\bibinfo {year} {1987})}\BibitemShut {NoStop}%
\bibitem [{\citenamefont {Hall}\ and\ \citenamefont
  {Randall}(1990)}]{Hall:1990ac}%
  \BibitemOpen
  \bibfield  {author} {\bibinfo {author} {\bibfnamefont {L.~J.}\ \bibnamefont
  {Hall}}\ and\ \bibinfo {author} {\bibfnamefont {L.}~\bibnamefont {Randall}},\
  }\href {\doibase 10.1103/PhysRevLett.65.2939} {\bibfield  {journal} {\bibinfo
   {journal} {Phys. Rev. Lett.}\ }\textbf {\bibinfo {volume} {65}},\ \bibinfo
  {pages} {2939} (\bibinfo {year} {1990})}\BibitemShut {NoStop}%
\bibitem [{\citenamefont {Buras}\ \emph {et~al.}(2001)\citenamefont {Buras},
  \citenamefont {Gambino}, \citenamefont {Gorbahn}, \citenamefont {Jager},\
  and\ \citenamefont {Silvestrini}}]{Buras:2000dm}%
  \BibitemOpen
  \bibfield  {author} {\bibinfo {author} {\bibfnamefont {A.~J.}\ \bibnamefont
  {Buras}}, \bibinfo {author} {\bibfnamefont {P.}~\bibnamefont {Gambino}},
  \bibinfo {author} {\bibfnamefont {M.}~\bibnamefont {Gorbahn}}, \bibinfo
  {author} {\bibfnamefont {S.}~\bibnamefont {Jager}}, \ and\ \bibinfo {author}
  {\bibfnamefont {L.}~\bibnamefont {Silvestrini}},\ }\href {\doibase
  10.1016/S0370-2693(01)00061-2} {\bibfield  {journal} {\bibinfo  {journal}
  {Phys. Lett.}\ }\textbf {\bibinfo {volume} {B500}},\ \bibinfo {pages} {161}
  (\bibinfo {year} {2001})},\ \Eprint {http://arxiv.org/abs/hep-ph/0007085}
  {arXiv:hep-ph/0007085 [hep-ph]} \BibitemShut {NoStop}%
\bibitem [{\citenamefont {D'Ambrosio}\ \emph {et~al.}(2002)\citenamefont
  {D'Ambrosio}, \citenamefont {Giudice}, \citenamefont {Isidori},\ and\
  \citenamefont {Strumia}}]{DAmbrosio:2002vsn}%
  \BibitemOpen
  \bibfield  {author} {\bibinfo {author} {\bibfnamefont {G.}~\bibnamefont
  {D'Ambrosio}}, \bibinfo {author} {\bibfnamefont {G.~F.}\ \bibnamefont
  {Giudice}}, \bibinfo {author} {\bibfnamefont {G.}~\bibnamefont {Isidori}}, \
  and\ \bibinfo {author} {\bibfnamefont {A.}~\bibnamefont {Strumia}},\ }\href
  {\doibase 10.1016/S0550-3213(02)00836-2} {\bibfield  {journal} {\bibinfo
  {journal} {Nucl. Phys.}\ }\textbf {\bibinfo {volume} {B645}},\ \bibinfo
  {pages} {155} (\bibinfo {year} {2002})},\ \Eprint
  {http://arxiv.org/abs/hep-ph/0207036} {arXiv:hep-ph/0207036 [hep-ph]}
  \BibitemShut {NoStop}%
\bibitem [{\citenamefont {Bertl}\ \emph {et~al.}(2006)\citenamefont {Bertl}
  \emph {et~al.}}]{Bertl:2006up}%
  \BibitemOpen
  \bibfield  {author} {\bibinfo {author} {\bibfnamefont {W.~H.}\ \bibnamefont
  {Bertl}} \emph {et~al.} (\bibinfo {collaboration} {SINDRUM II}),\ }\href
  {\doibase 10.1140/epjc/s2006-02582-x} {\bibfield  {journal} {\bibinfo
  {journal} {Eur. Phys. J.}\ }\textbf {\bibinfo {volume} {C47}},\ \bibinfo
  {pages} {337} (\bibinfo {year} {2006})}\BibitemShut {NoStop}%
\bibitem [{\citenamefont {Aubert}\ \emph {et~al.}(2010)\citenamefont {Aubert}
  \emph {et~al.}}]{Aubert:2009ag}%
  \BibitemOpen
  \bibfield  {author} {\bibinfo {author} {\bibfnamefont {B.}~\bibnamefont
  {Aubert}} \emph {et~al.} (\bibinfo {collaboration} {BaBar}),\ }\href
  {\doibase 10.1103/PhysRevLett.104.021802} {\bibfield  {journal} {\bibinfo
  {journal} {Phys. Rev. Lett.}\ }\textbf {\bibinfo {volume} {104}},\ \bibinfo
  {pages} {021802} (\bibinfo {year} {2010})},\ \Eprint
  {http://arxiv.org/abs/0908.2381} {arXiv:0908.2381 [hep-ex]} \BibitemShut
  {NoStop}%
\bibitem [{\citenamefont {Baldini}\ \emph {et~al.}(2016)\citenamefont {Baldini}
  \emph {et~al.}}]{TheMEG:2016wtm}%
  \BibitemOpen
  \bibfield  {author} {\bibinfo {author} {\bibfnamefont {A.~M.}\ \bibnamefont
  {Baldini}} \emph {et~al.} (\bibinfo {collaboration} {MEG}),\ }\href {\doibase
  10.1140/epjc/s10052-016-4271-x} {\bibfield  {journal} {\bibinfo  {journal}
  {Eur. Phys. J.}\ }\textbf {\bibinfo {volume} {C76}},\ \bibinfo {pages} {434}
  (\bibinfo {year} {2016})},\ \Eprint {http://arxiv.org/abs/1605.05081}
  {arXiv:1605.05081 [hep-ex]} \BibitemShut {NoStop}%
\bibitem [{\citenamefont {Schael}\ \emph {et~al.}(2006)\citenamefont {Schael}
  \emph {et~al.}}]{ALEPH:2005ab}%
  \BibitemOpen
  \bibfield  {author} {\bibinfo {author} {\bibfnamefont {S.}~\bibnamefont
  {Schael}} \emph {et~al.} (\bibinfo {collaboration} {ALEPH, DELPHI, L3, OPAL,
  SLD, LEP Electroweak Working Group, SLD Electroweak Group, SLD Heavy Flavour
  Group}),\ }\href {\doibase 10.1016/j.physrep.2005.12.006} {\bibfield
  {journal} {\bibinfo  {journal} {Phys. Rept.}\ }\textbf {\bibinfo {volume}
  {427}},\ \bibinfo {pages} {257} (\bibinfo {year} {2006})},\ \Eprint
  {http://arxiv.org/abs/hep-ex/0509008} {arXiv:hep-ex/0509008 [hep-ex]}
  \BibitemShut {NoStop}%
\bibitem [{\citenamefont {Aaltonen}\ \emph {et~al.}(2016)\citenamefont
  {Aaltonen} \emph {et~al.}}]{Aaltonen:2016nuy}%
  \BibitemOpen
  \bibfield  {author} {\bibinfo {author} {\bibfnamefont {T.~A.}\ \bibnamefont
  {Aaltonen}} \emph {et~al.} (\bibinfo {collaboration} {CDF}),\ }\href
  {\doibase 10.1103/PhysRevD.93.112016} {\bibfield  {journal} {\bibinfo
  {journal} {Phys. Rev. D}\ }\textbf {\bibinfo {volume} {93}},\ \bibinfo
  {pages} {112016} (\bibinfo {year} {2016})},\ \bibinfo {note} {[Addendum:
  Phys.Rev.D 95, 119901 (2017)]},\ \Eprint {http://arxiv.org/abs/1605.02719}
  {arXiv:1605.02719 [hep-ex]} \BibitemShut {NoStop}%
\bibitem [{\citenamefont {Chatrchyan}\ \emph {et~al.}(2011)\citenamefont
  {Chatrchyan} \emph {et~al.}}]{Chatrchyan:2011ya}%
  \BibitemOpen
  \bibfield  {author} {\bibinfo {author} {\bibfnamefont {S.}~\bibnamefont
  {Chatrchyan}} \emph {et~al.} (\bibinfo {collaboration} {CMS}),\ }\href
  {\doibase 10.1103/PhysRevD.84.112002} {\bibfield  {journal} {\bibinfo
  {journal} {Phys. Rev. D}\ }\textbf {\bibinfo {volume} {84}},\ \bibinfo
  {pages} {112002} (\bibinfo {year} {2011})},\ \Eprint
  {http://arxiv.org/abs/1110.2682} {arXiv:1110.2682 [hep-ex]} \BibitemShut
  {NoStop}%
\bibitem [{\citenamefont {Aaij}\ \emph {et~al.}(2015)\citenamefont {Aaij} \emph
  {et~al.}}]{Aaij:2015lka}%
  \BibitemOpen
  \bibfield  {author} {\bibinfo {author} {\bibfnamefont {R.}~\bibnamefont
  {Aaij}} \emph {et~al.} (\bibinfo {collaboration} {LHCb}),\ }\href {\doibase
  10.1007/JHEP11(2015)190} {\bibfield  {journal} {\bibinfo  {journal} {JHEP}\
  }\textbf {\bibinfo {volume} {11}},\ \bibinfo {pages} {190} (\bibinfo {year}
  {2015})},\ \Eprint {http://arxiv.org/abs/1509.07645} {arXiv:1509.07645
  [hep-ex]} \BibitemShut {NoStop}%
\bibitem [{\citenamefont {De~Blas}\ \emph {et~al.}(2020)\citenamefont {De~Blas}
  \emph {et~al.}}]{deBlas:2019okz}%
  \BibitemOpen
  \bibfield  {author} {\bibinfo {author} {\bibfnamefont {J.}~\bibnamefont
  {De~Blas}} \emph {et~al.},\ }\href {\doibase 10.1140/epjc/s10052-020-7904-z}
  {\bibfield  {journal} {\bibinfo  {journal} {Eur. Phys. J.}\ }\textbf
  {\bibinfo {volume} {C80}},\ \bibinfo {pages} {456} (\bibinfo {year}
  {2020})},\ \Eprint {http://arxiv.org/abs/1910.14012} {arXiv:1910.14012
  [hep-ph]} \BibitemShut {NoStop}%
\bibitem [{\citenamefont {Aaboud}\ \emph {et~al.}(2018)\citenamefont {Aaboud}
  \emph {et~al.}}]{Aaboud:2018wps}%
  \BibitemOpen
  \bibfield  {author} {\bibinfo {author} {\bibfnamefont {M.}~\bibnamefont
  {Aaboud}} \emph {et~al.} (\bibinfo {collaboration} {ATLAS}),\ }\href
  {\doibase 10.1016/j.physletb.2018.07.050} {\bibfield  {journal} {\bibinfo
  {journal} {Phys. Lett. B}\ }\textbf {\bibinfo {volume} {784}},\ \bibinfo
  {pages} {345} (\bibinfo {year} {2018})},\ \Eprint
  {http://arxiv.org/abs/1806.00242} {arXiv:1806.00242 [hep-ex]} \BibitemShut
  {NoStop}%
\bibitem [{\citenamefont {CMS-collaboration}()}]{CMS:2019drq}%
  \BibitemOpen
  \bibfield  {author} {\bibinfo {author} {\bibnamefont {CMS-collaboration}},\
  }\href@noop {} {\ }\bibinfo {note} {CMS-PAS-HIG-19-004}\BibitemShut {NoStop}%
\bibitem [{\citenamefont
  {Group}(2016)}]{TevatronElectroweakWorkingGroup:2016lid}%
  \BibitemOpen
  \bibfield  {author} {\bibinfo {author} {\bibfnamefont {T.~E.~W.}\
  \bibnamefont {Group}} (\bibinfo {collaboration} {CDF, D0}),\ }\href@noop {}
  {\  (\bibinfo {year} {2016})},\ \Eprint {http://arxiv.org/abs/1608.01881}
  {arXiv:1608.01881 [hep-ex]} \BibitemShut {NoStop}%
\bibitem [{\citenamefont {Aaboud}\ \emph {et~al.}(2019)\citenamefont {Aaboud}
  \emph {et~al.}}]{Aaboud:2018zbu}%
  \BibitemOpen
  \bibfield  {author} {\bibinfo {author} {\bibfnamefont {M.}~\bibnamefont
  {Aaboud}} \emph {et~al.} (\bibinfo {collaboration} {ATLAS}),\ }\href
  {\doibase 10.1140/epjc/s10052-019-6757-9} {\bibfield  {journal} {\bibinfo
  {journal} {Eur. Phys. J. C}\ }\textbf {\bibinfo {volume} {79}},\ \bibinfo
  {pages} {290} (\bibinfo {year} {2019})},\ \Eprint
  {http://arxiv.org/abs/1810.01772} {arXiv:1810.01772 [hep-ex]} \BibitemShut
  {NoStop}%
\bibitem [{\citenamefont {Sirunyan}\ \emph {et~al.}(2019)\citenamefont
  {Sirunyan} \emph {et~al.}}]{Sirunyan:2018mlv}%
  \BibitemOpen
  \bibfield  {author} {\bibinfo {author} {\bibfnamefont {A.~M.}\ \bibnamefont
  {Sirunyan}} \emph {et~al.} (\bibinfo {collaboration} {CMS}),\ }\href
  {\doibase 10.1140/epjc/s10052-019-6788-2} {\bibfield  {journal} {\bibinfo
  {journal} {Eur. Phys. J. C}\ }\textbf {\bibinfo {volume} {79}},\ \bibinfo
  {pages} {313} (\bibinfo {year} {2019})},\ \Eprint
  {http://arxiv.org/abs/1812.10534} {arXiv:1812.10534 [hep-ex]} \BibitemShut
  {NoStop}%
\bibitem [{\citenamefont {Lazzeroni}\ \emph {et~al.}(2013)\citenamefont
  {Lazzeroni} \emph {et~al.}}]{Lazzeroni:2012cx}%
  \BibitemOpen
  \bibfield  {author} {\bibinfo {author} {\bibfnamefont {C.}~\bibnamefont
  {Lazzeroni}} \emph {et~al.} (\bibinfo {collaboration} {NA62}),\ }\href
  {\doibase 10.1016/j.physletb.2013.01.037} {\bibfield  {journal} {\bibinfo
  {journal} {Phys. Lett.}\ }\textbf {\bibinfo {volume} {B719}},\ \bibinfo
  {pages} {326} (\bibinfo {year} {2013})},\ \Eprint
  {http://arxiv.org/abs/1212.4012} {arXiv:1212.4012 [hep-ex]} \BibitemShut
  {NoStop}%
\bibitem [{\citenamefont {Ambrosino}\ \emph {et~al.}(2009)\citenamefont
  {Ambrosino} \emph {et~al.}}]{Ambrosino:2009aa}%
  \BibitemOpen
  \bibfield  {author} {\bibinfo {author} {\bibfnamefont {F.}~\bibnamefont
  {Ambrosino}} \emph {et~al.} (\bibinfo {collaboration} {KLOE}),\ }\href
  {\doibase 10.1140/epjc/s10052-009-1217-6, 10.1140/epjc/s10052-009-1177-x}
  {\bibfield  {journal} {\bibinfo  {journal} {Eur. Phys. J.}\ }\textbf
  {\bibinfo {volume} {C64}},\ \bibinfo {pages} {627} (\bibinfo {year}
  {2009})},\ \bibinfo {note} {[Erratum: Eur. Phys. J.65,703(2010)]},\ \Eprint
  {http://arxiv.org/abs/0907.3594} {arXiv:0907.3594 [hep-ex]} \BibitemShut
  {NoStop}%
\bibitem [{\citenamefont {Cirigliano}\ and\ \citenamefont
  {Rosell}(2007)}]{Cirigliano:2007xi}%
  \BibitemOpen
  \bibfield  {author} {\bibinfo {author} {\bibfnamefont {V.}~\bibnamefont
  {Cirigliano}}\ and\ \bibinfo {author} {\bibfnamefont {I.}~\bibnamefont
  {Rosell}},\ }\href {\doibase 10.1103/PhysRevLett.99.231801} {\bibfield
  {journal} {\bibinfo  {journal} {Phys. Rev. Lett.}\ }\textbf {\bibinfo
  {volume} {99}},\ \bibinfo {pages} {231801} (\bibinfo {year} {2007})},\
  \Eprint {http://arxiv.org/abs/0707.3439} {arXiv:0707.3439 [hep-ph]}
  \BibitemShut {NoStop}%
\bibitem [{\citenamefont {Pich}(2014)}]{Pich:2013lsa}%
  \BibitemOpen
  \bibfield  {author} {\bibinfo {author} {\bibfnamefont {A.}~\bibnamefont
  {Pich}},\ }\href {\doibase 10.1016/j.ppnp.2013.11.002} {\bibfield  {journal}
  {\bibinfo  {journal} {Prog. Part. Nucl. Phys.}\ }\textbf {\bibinfo {volume}
  {75}},\ \bibinfo {pages} {41} (\bibinfo {year} {2014})},\ \Eprint
  {http://arxiv.org/abs/1310.7922} {arXiv:1310.7922 [hep-ph]} \BibitemShut
  {NoStop}%
\bibitem [{\citenamefont {Czapek}\ \emph {et~al.}(1993)\citenamefont {Czapek}
  \emph {et~al.}}]{Czapek:1993kc}%
  \BibitemOpen
  \bibfield  {author} {\bibinfo {author} {\bibfnamefont {G.}~\bibnamefont
  {Czapek}} \emph {et~al.},\ }\href {\doibase 10.1103/PhysRevLett.70.17}
  {\bibfield  {journal} {\bibinfo  {journal} {Phys. Rev. Lett.}\ }\textbf
  {\bibinfo {volume} {70}},\ \bibinfo {pages} {17} (\bibinfo {year}
  {1993})}\BibitemShut {NoStop}%
\bibitem [{\citenamefont {Britton}\ \emph {et~al.}(1992)\citenamefont {Britton}
  \emph {et~al.}}]{Britton:1992pg}%
  \BibitemOpen
  \bibfield  {author} {\bibinfo {author} {\bibfnamefont {D.~I.}\ \bibnamefont
  {Britton}} \emph {et~al.},\ }\href {\doibase 10.1103/PhysRevLett.68.3000}
  {\bibfield  {journal} {\bibinfo  {journal} {Phys. Rev. Lett.}\ }\textbf
  {\bibinfo {volume} {68}},\ \bibinfo {pages} {3000} (\bibinfo {year}
  {1992})}\BibitemShut {NoStop}%
\bibitem [{\citenamefont {Bryman}\ \emph {et~al.}(1983)\citenamefont {Bryman},
  \citenamefont {Dubois}, \citenamefont {Numao}, \citenamefont {Olaniyi},
  \citenamefont {Olin}, \citenamefont {Dixit}, \citenamefont {Berghofer},
  \citenamefont {Poutissou}, \citenamefont {Macdonald},\ and\ \citenamefont
  {Robertson}}]{Bryman:1982em}%
  \BibitemOpen
  \bibfield  {author} {\bibinfo {author} {\bibfnamefont {D.~A.}\ \bibnamefont
  {Bryman}}, \bibinfo {author} {\bibfnamefont {R.}~\bibnamefont {Dubois}},
  \bibinfo {author} {\bibfnamefont {T.}~\bibnamefont {Numao}}, \bibinfo
  {author} {\bibfnamefont {B.}~\bibnamefont {Olaniyi}}, \bibinfo {author}
  {\bibfnamefont {A.}~\bibnamefont {Olin}}, \bibinfo {author} {\bibfnamefont
  {M.~S.}\ \bibnamefont {Dixit}}, \bibinfo {author} {\bibfnamefont
  {D.}~\bibnamefont {Berghofer}}, \bibinfo {author} {\bibfnamefont {J.~M.}\
  \bibnamefont {Poutissou}}, \bibinfo {author} {\bibfnamefont {J.~A.}\
  \bibnamefont {Macdonald}}, \ and\ \bibinfo {author} {\bibfnamefont {B.~C.}\
  \bibnamefont {Robertson}},\ }\href {\doibase 10.1103/PhysRevLett.50.7}
  {\bibfield  {journal} {\bibinfo  {journal} {Phys. Rev. Lett.}\ }\textbf
  {\bibinfo {volume} {50}},\ \bibinfo {pages} {7} (\bibinfo {year}
  {1983})}\BibitemShut {NoStop}%
\bibitem [{\citenamefont {Aguilar-Arevalo}\ \emph {et~al.}(2015)\citenamefont
  {Aguilar-Arevalo} \emph {et~al.}}]{Aguilar-Arevalo:2015cdf}%
  \BibitemOpen
  \bibfield  {author} {\bibinfo {author} {\bibfnamefont {A.}~\bibnamefont
  {Aguilar-Arevalo}} \emph {et~al.} (\bibinfo {collaboration} {PiENu}),\ }\href
  {\doibase 10.1103/PhysRevLett.115.071801} {\bibfield  {journal} {\bibinfo
  {journal} {Phys. Rev. Lett.}\ }\textbf {\bibinfo {volume} {115}},\ \bibinfo
  {pages} {071801} (\bibinfo {year} {2015})},\ \Eprint
  {http://arxiv.org/abs/1506.05845} {arXiv:1506.05845 [hep-ex]} \BibitemShut
  {NoStop}%
\bibitem [{\citenamefont {Antonelli}\ \emph {et~al.}(2010)\citenamefont
  {Antonelli} \emph {et~al.}}]{Antonelli:2010yf}%
  \BibitemOpen
  \bibfield  {author} {\bibinfo {author} {\bibfnamefont {M.}~\bibnamefont
  {Antonelli}} \emph {et~al.} (\bibinfo {collaboration} {FlaviaNet Working
  Group on Kaon Decays}),\ }\href {\doibase 10.1140/epjc/s10052-010-1406-3}
  {\bibfield  {journal} {\bibinfo  {journal} {Eur. Phys. J.}\ }\textbf
  {\bibinfo {volume} {C69}},\ \bibinfo {pages} {399} (\bibinfo {year}
  {2010})},\ \Eprint {http://arxiv.org/abs/1005.2323} {arXiv:1005.2323
  [hep-ph]} \BibitemShut {NoStop}%
\bibitem [{\citenamefont {Cirigliano}\ \emph {et~al.}(2012)\citenamefont
  {Cirigliano}, \citenamefont {Ecker}, \citenamefont {Neufeld}, \citenamefont
  {Pich},\ and\ \citenamefont {Portoles}}]{Cirigliano:2011ny}%
  \BibitemOpen
  \bibfield  {author} {\bibinfo {author} {\bibfnamefont {V.}~\bibnamefont
  {Cirigliano}}, \bibinfo {author} {\bibfnamefont {G.}~\bibnamefont {Ecker}},
  \bibinfo {author} {\bibfnamefont {H.}~\bibnamefont {Neufeld}}, \bibinfo
  {author} {\bibfnamefont {A.}~\bibnamefont {Pich}}, \ and\ \bibinfo {author}
  {\bibfnamefont {J.}~\bibnamefont {Portoles}},\ }\href {\doibase
  10.1103/RevModPhys.84.399} {\bibfield  {journal} {\bibinfo  {journal} {Rev.
  Mod. Phys.}\ }\textbf {\bibinfo {volume} {84}},\ \bibinfo {pages} {399}
  (\bibinfo {year} {2012})},\ \Eprint {http://arxiv.org/abs/1107.6001}
  {arXiv:1107.6001 [hep-ph]} \BibitemShut {NoStop}%
\bibitem [{\citenamefont {Czarnecki}\ \emph {et~al.}(2019)\citenamefont
  {Czarnecki}, \citenamefont {Marciano},\ and\ \citenamefont
  {Sirlin}}]{Czarnecki:2019mwq}%
  \BibitemOpen
  \bibfield  {author} {\bibinfo {author} {\bibfnamefont {A.}~\bibnamefont
  {Czarnecki}}, \bibinfo {author} {\bibfnamefont {W.~J.}\ \bibnamefont
  {Marciano}}, \ and\ \bibinfo {author} {\bibfnamefont {A.}~\bibnamefont
  {Sirlin}},\ }\href {\doibase 10.1103/PhysRevD.100.073008} {\bibfield
  {journal} {\bibinfo  {journal} {Phys. Rev.}\ }\textbf {\bibinfo {volume}
  {D100}},\ \bibinfo {pages} {073008} (\bibinfo {year} {2019})},\ \Eprint
  {http://arxiv.org/abs/1907.06737} {arXiv:1907.06737 [hep-ph]} \BibitemShut
  {NoStop}%
\bibitem [{\citenamefont {Seng}\ \emph {et~al.}(2018)\citenamefont {Seng},
  \citenamefont {Gorchtein}, \citenamefont {Patel},\ and\ \citenamefont
  {Ramsey-Musolf}}]{Seng:2018yzq}%
  \BibitemOpen
  \bibfield  {author} {\bibinfo {author} {\bibfnamefont {C.-Y.}\ \bibnamefont
  {Seng}}, \bibinfo {author} {\bibfnamefont {M.}~\bibnamefont {Gorchtein}},
  \bibinfo {author} {\bibfnamefont {H.~H.}\ \bibnamefont {Patel}}, \ and\
  \bibinfo {author} {\bibfnamefont {M.~J.}\ \bibnamefont {Ramsey-Musolf}},\
  }\href {\doibase 10.1103/PhysRevLett.121.241804} {\bibfield  {journal}
  {\bibinfo  {journal} {Phys. Rev. Lett.}\ }\textbf {\bibinfo {volume} {121}},\
  \bibinfo {pages} {241804} (\bibinfo {year} {2018})},\ \Eprint
  {http://arxiv.org/abs/1807.10197} {arXiv:1807.10197 [hep-ph]} \BibitemShut
  {NoStop}%
\bibitem [{\citenamefont {Crivellin}\ \emph {et~al.}(2019)\citenamefont
  {Crivellin}, \citenamefont {Müller},\ and\ \citenamefont
  {Wiegand}}]{Crivellin:2019dun}%
  \BibitemOpen
  \bibfield  {author} {\bibinfo {author} {\bibfnamefont {A.}~\bibnamefont
  {Crivellin}}, \bibinfo {author} {\bibfnamefont {D.}~\bibnamefont {Müller}},
  \ and\ \bibinfo {author} {\bibfnamefont {C.}~\bibnamefont {Wiegand}},\ }\href
  {\doibase 10.1007/JHEP06(2019)119} {\bibfield  {journal} {\bibinfo  {journal}
  {JHEP}\ }\textbf {\bibinfo {volume} {06}},\ \bibinfo {pages} {119} (\bibinfo
  {year} {2019})},\ \Eprint {http://arxiv.org/abs/1903.10440} {arXiv:1903.10440
  [hep-ph]} \BibitemShut {NoStop}%
\bibitem [{\citenamefont {Descotes-Genon}\ \emph {et~al.}(2016)\citenamefont
  {Descotes-Genon}, \citenamefont {Hofer}, \citenamefont {Matias},\ and\
  \citenamefont {Virto}}]{Descotes-Genon:2015uva}%
  \BibitemOpen
  \bibfield  {author} {\bibinfo {author} {\bibfnamefont {S.}~\bibnamefont
  {Descotes-Genon}}, \bibinfo {author} {\bibfnamefont {L.}~\bibnamefont
  {Hofer}}, \bibinfo {author} {\bibfnamefont {J.}~\bibnamefont {Matias}}, \
  and\ \bibinfo {author} {\bibfnamefont {J.}~\bibnamefont {Virto}},\ }\href
  {\doibase 10.1007/JHEP06(2016)092} {\bibfield  {journal} {\bibinfo  {journal}
  {JHEP}\ }\textbf {\bibinfo {volume} {06}},\ \bibinfo {pages} {092} (\bibinfo
  {year} {2016})},\ \Eprint {http://arxiv.org/abs/1510.04239} {arXiv:1510.04239
  [hep-ph]} \BibitemShut {NoStop}%
\bibitem [{\citenamefont {Bona}\ \emph {et~al.}(2008)\citenamefont {Bona} \emph
  {et~al.}}]{Bona:2007vi}%
  \BibitemOpen
  \bibfield  {author} {\bibinfo {author} {\bibfnamefont {M.}~\bibnamefont
  {Bona}} \emph {et~al.} (\bibinfo {collaboration} {UTfit}),\ }\href {\doibase
  10.1088/1126-6708/2008/03/049} {\bibfield  {journal} {\bibinfo  {journal}
  {JHEP}\ }\textbf {\bibinfo {volume} {03}},\ \bibinfo {pages} {049} (\bibinfo
  {year} {2008})},\ \Eprint {http://arxiv.org/abs/0707.0636} {arXiv:0707.0636
  [hep-ph]} \BibitemShut {NoStop}%
\bibitem [{\citenamefont {Sirunyan}\ \emph {et~al.}(2017)\citenamefont
  {Sirunyan} \emph {et~al.}}]{Sirunyan:2017ygf}%
  \BibitemOpen
  \bibfield  {author} {\bibinfo {author} {\bibfnamefont {A.~M.}\ \bibnamefont
  {Sirunyan}} \emph {et~al.} (\bibinfo {collaboration} {CMS}),\ }\href
  {\doibase 10.1007/JHEP07(2017)013} {\bibfield  {journal} {\bibinfo  {journal}
  {JHEP}\ }\textbf {\bibinfo {volume} {07}},\ \bibinfo {pages} {013} (\bibinfo
  {year} {2017})},\ \Eprint {http://arxiv.org/abs/1703.09986} {arXiv:1703.09986
  [hep-ex]} \BibitemShut {NoStop}%
\bibitem [{\citenamefont {Aaboud}\ \emph {et~al.}(2017)\citenamefont {Aaboud}
  \emph {et~al.}}]{Aaboud:2017buh}%
  \BibitemOpen
  \bibfield  {author} {\bibinfo {author} {\bibfnamefont {M.}~\bibnamefont
  {Aaboud}} \emph {et~al.} (\bibinfo {collaboration} {ATLAS}),\ }\href
  {\doibase 10.1007/JHEP10(2017)182} {\bibfield  {journal} {\bibinfo  {journal}
  {JHEP}\ }\textbf {\bibinfo {volume} {10}},\ \bibinfo {pages} {182} (\bibinfo
  {year} {2017})},\ \Eprint {http://arxiv.org/abs/1707.02424} {arXiv:1707.02424
  [hep-ex]} \BibitemShut {NoStop}%
\bibitem [{\citenamefont {Wood}\ \emph {et~al.}(1997)\citenamefont {Wood},
  \citenamefont {Bennett}, \citenamefont {Cho}, \citenamefont {Masterson},
  \citenamefont {Roberts}, \citenamefont {Tanner},\ and\ \citenamefont
  {Wieman}}]{Wood:1997zq}%
  \BibitemOpen
  \bibfield  {author} {\bibinfo {author} {\bibfnamefont {C.~S.}\ \bibnamefont
  {Wood}}, \bibinfo {author} {\bibfnamefont {S.~C.}\ \bibnamefont {Bennett}},
  \bibinfo {author} {\bibfnamefont {D.}~\bibnamefont {Cho}}, \bibinfo {author}
  {\bibfnamefont {B.~P.}\ \bibnamefont {Masterson}}, \bibinfo {author}
  {\bibfnamefont {J.~L.}\ \bibnamefont {Roberts}}, \bibinfo {author}
  {\bibfnamefont {C.~E.}\ \bibnamefont {Tanner}}, \ and\ \bibinfo {author}
  {\bibfnamefont {C.~E.}\ \bibnamefont {Wieman}},\ }\href {\doibase
  10.1126/science.275.5307.1759} {\bibfield  {journal} {\bibinfo  {journal}
  {Science}\ }\textbf {\bibinfo {volume} {275}},\ \bibinfo {pages} {1759}
  (\bibinfo {year} {1997})}\BibitemShut {NoStop}%
\bibitem [{\citenamefont {Bennett}\ and\ \citenamefont
  {Wieman}(1999)}]{Bennett:1999pd}%
  \BibitemOpen
  \bibfield  {author} {\bibinfo {author} {\bibfnamefont {S.~C.}\ \bibnamefont
  {Bennett}}\ and\ \bibinfo {author} {\bibfnamefont {C.~E.}\ \bibnamefont
  {Wieman}},\ }\href {\doibase 10.1103/PhysRevLett.82.4153,
  10.1103/PhysRevLett.83.889, 10.1103/PhysRevLett.82.2484} {\bibfield
  {journal} {\bibinfo  {journal} {Phys. Rev. Lett.}\ }\textbf {\bibinfo
  {volume} {82}},\ \bibinfo {pages} {2484} (\bibinfo {year} {1999})},\ \bibinfo
  {note} {[Erratum: Phys. Rev. Lett.82,4153(1999); Erratum: Phys. Rev.
  Lett.83,889(1999)]},\ \Eprint {http://arxiv.org/abs/hep-ex/9903022}
  {arXiv:hep-ex/9903022 [hep-ex]} \BibitemShut {NoStop}%
\bibitem [{\citenamefont {Androić}\ \emph {et~al.}(2018)\citenamefont
  {Androić} \emph {et~al.}}]{Androic:2018kni}%
  \BibitemOpen
  \bibfield  {author} {\bibinfo {author} {\bibfnamefont {D.}~\bibnamefont
  {Androić}} \emph {et~al.} (\bibinfo {collaboration} {Qweak}),\ }\href
  {\doibase 10.1038/s41586-018-0096-0} {\bibfield  {journal} {\bibinfo
  {journal} {Nature}\ }\textbf {\bibinfo {volume} {557}},\ \bibinfo {pages}
  {207} (\bibinfo {year} {2018})},\ \Eprint {http://arxiv.org/abs/1905.08283}
  {arXiv:1905.08283 [nucl-ex]} \BibitemShut {NoStop}%
\bibitem [{\citenamefont {Caldwell}\ \emph {et~al.}(2009)\citenamefont
  {Caldwell}, \citenamefont {Kollar},\ and\ \citenamefont
  {Kroninger}}]{Caldwell:2008fw}%
  \BibitemOpen
  \bibfield  {author} {\bibinfo {author} {\bibfnamefont {A.}~\bibnamefont
  {Caldwell}}, \bibinfo {author} {\bibfnamefont {D.}~\bibnamefont {Kollar}}, \
  and\ \bibinfo {author} {\bibfnamefont {K.}~\bibnamefont {Kroninger}},\ }\href
  {\doibase 10.1016/j.cpc.2009.06.026} {\bibfield  {journal} {\bibinfo
  {journal} {Comput. Phys. Commun.}\ }\textbf {\bibinfo {volume} {180}},\
  \bibinfo {pages} {2197} (\bibinfo {year} {2009})},\ \Eprint
  {http://arxiv.org/abs/0808.2552} {arXiv:0808.2552 [physics.data-an]}
  \BibitemShut {NoStop}%
\bibitem [{\citenamefont {Kass}\ and\ \citenamefont
  {Raftery}(1995)}]{Kass:1995}%
  \BibitemOpen
  \bibfield  {author} {\bibinfo {author} {\bibfnamefont {R.~E.}\ \bibnamefont
  {Kass}}\ and\ \bibinfo {author} {\bibfnamefont {A.~E.}\ \bibnamefont
  {Raftery}},\ }\href {\doibase 10.1080/01621459.1995.10476572} {\bibfield
  {journal} {\bibinfo  {journal} {J. Am. Stat. Assoc.}\ }\textbf {\bibinfo
  {volume} {90}},\ \bibinfo {pages} {773} (\bibinfo {year} {1995})}\BibitemShut
  {NoStop}%
\bibitem [{\citenamefont {Glaser}\ \emph {et~al.}(2018)\citenamefont {Glaser}
  \emph {et~al.}}]{Glaser:2018aat}%
  \BibitemOpen
  \bibfield  {author} {\bibinfo {author} {\bibfnamefont {C.~J.}\ \bibnamefont
  {Glaser}} \emph {et~al.} (\bibinfo {collaboration} {PEN}),\ }in\ \href@noop
  {} {\emph {\bibinfo {booktitle} {{13th Conference on the Intersections of
  Particle and Nuclear Physics (CIPANP 2018) Palm Springs, California, USA, May
  29-June 3, 2018}}}}\ (\bibinfo {year} {2018})\ \Eprint
  {http://arxiv.org/abs/1812.00782} {arXiv:1812.00782 [hep-ex]} \BibitemShut
  {NoStop}%
\bibitem [{\citenamefont {Altmannshofer}\ \emph {et~al.}(2019)\citenamefont
  {Altmannshofer} \emph {et~al.}}]{Kou:2018nap}%
  \BibitemOpen
  \bibfield  {author} {\bibinfo {author} {\bibfnamefont {W.}~\bibnamefont
  {Altmannshofer}} \emph {et~al.} (\bibinfo {collaboration} {Belle-II}),\
  }\href {\doibase 10.1093/ptep/ptz106, 10.1093/ptep/ptaa008} {\bibfield
  {journal} {\bibinfo  {journal} {PTEP}\ }\textbf {\bibinfo {volume} {2019}},\
  \bibinfo {pages} {123C01} (\bibinfo {year} {2019})},\ \bibinfo {note}
  {[Erratum: PTEP2020,no.2,029201(2020)]},\ \Eprint
  {http://arxiv.org/abs/1808.10567} {arXiv:1808.10567 [hep-ex]} \BibitemShut
  {NoStop}%
\bibitem [{\citenamefont {Cerri}\ \emph {et~al.}(2019)\citenamefont {Cerri}
  \emph {et~al.}}]{Cerri:2018ypt}%
  \BibitemOpen
  \bibfield  {author} {\bibinfo {author} {\bibfnamefont {A.}~\bibnamefont
  {Cerri}} \emph {et~al.},\ }\href {\doibase 10.23731/CYRM-2019-007.867}
  {\bibfield  {journal} {\bibinfo  {journal} {CERN Yellow Rep. Monogr.}\
  }\textbf {\bibinfo {volume} {7}},\ \bibinfo {pages} {867} (\bibinfo {year}
  {2019})},\ \Eprint {http://arxiv.org/abs/1812.07638} {arXiv:1812.07638
  [hep-ph]} \BibitemShut {NoStop}%
\bibitem [{\citenamefont {Apollinari}\ \emph {et~al.}(2017)\citenamefont
  {Apollinari}, \citenamefont {Béjar~Alonso}, \citenamefont {Brüning},
  \citenamefont {Fessia}, \citenamefont {Lamont}, \citenamefont {Rossi},\ and\
  \citenamefont {Tavian}}]{ApollinariG.:2017ojx}%
  \BibitemOpen
  \bibfield  {author} {\bibinfo {author} {\bibfnamefont {G.}~\bibnamefont
  {Apollinari}}, \bibinfo {author} {\bibfnamefont {I.}~\bibnamefont
  {Béjar~Alonso}}, \bibinfo {author} {\bibfnamefont {O.}~\bibnamefont
  {Brüning}}, \bibinfo {author} {\bibfnamefont {P.}~\bibnamefont {Fessia}},
  \bibinfo {author} {\bibfnamefont {M.}~\bibnamefont {Lamont}}, \bibinfo
  {author} {\bibfnamefont {L.}~\bibnamefont {Rossi}}, \ and\ \bibinfo {author}
  {\bibfnamefont {L.}~\bibnamefont {Tavian}},\ }\href {\doibase
  10.23731/CYRM-2017-004} {\  (\bibinfo {year} {2017}),\
  10.23731/CYRM-2017-004}\BibitemShut {NoStop}%
\bibitem [{\citenamefont {Aicheler}\ \emph {et~al.}(2012)\citenamefont
  {Aicheler}, \citenamefont {Burrows}, \citenamefont {Draper}, \citenamefont
  {Garvey}, \citenamefont {Lebrun}, \citenamefont {Peach}, \citenamefont
  {Phinney}, \citenamefont {Schmickler}, \citenamefont {Schulte},\ and\
  \citenamefont {Toge}}]{Aicheler:2012bya}%
  \BibitemOpen
  \bibfield  {author} {\bibinfo {author} {\bibfnamefont {M.}~\bibnamefont
  {Aicheler}}, \bibinfo {author} {\bibfnamefont {P.}~\bibnamefont {Burrows}},
  \bibinfo {author} {\bibfnamefont {M.}~\bibnamefont {Draper}}, \bibinfo
  {author} {\bibfnamefont {T.}~\bibnamefont {Garvey}}, \bibinfo {author}
  {\bibfnamefont {P.}~\bibnamefont {Lebrun}}, \bibinfo {author} {\bibfnamefont
  {K.}~\bibnamefont {Peach}}, \bibinfo {author} {\bibfnamefont
  {N.}~\bibnamefont {Phinney}}, \bibinfo {author} {\bibfnamefont
  {H.}~\bibnamefont {Schmickler}}, \bibinfo {author} {\bibfnamefont
  {D.}~\bibnamefont {Schulte}}, \ and\ \bibinfo {author} {\bibfnamefont
  {N.}~\bibnamefont {Toge}},\ }\href {\doibase 10.5170/CERN-2012-007} {\
  (\bibinfo {year} {2012}),\ 10.5170/CERN-2012-007}\BibitemShut {NoStop}%
\bibitem [{\citenamefont {Abramowicz}\ \emph {et~al.}(2013)\citenamefont
  {Abramowicz} \emph {et~al.}}]{Behnke:2013lya}%
  \BibitemOpen
  \bibfield  {author} {\bibinfo {author} {\bibfnamefont {H.}~\bibnamefont
  {Abramowicz}} \emph {et~al.},\ }\href@noop {} {\  (\bibinfo {year} {2013})},\
  \Eprint {http://arxiv.org/abs/1306.6329} {arXiv:1306.6329 [physics.ins-det]}
  \BibitemShut {NoStop}%
\bibitem [{\citenamefont {Abada}\ \emph
  {et~al.}(2019{\natexlab{a}})\citenamefont {Abada} \emph
  {et~al.}}]{Abada:2019lih}%
  \BibitemOpen
  \bibfield  {author} {\bibinfo {author} {\bibfnamefont {A.}~\bibnamefont
  {Abada}} \emph {et~al.} (\bibinfo {collaboration} {FCC}),\ }\href {\doibase
  10.1140/epjc/s10052-019-6904-3} {\bibfield  {journal} {\bibinfo  {journal}
  {Eur. Phys. J.}\ }\textbf {\bibinfo {volume} {C79}},\ \bibinfo {pages} {474}
  (\bibinfo {year} {2019}{\natexlab{a}})}\BibitemShut {NoStop}%
\bibitem [{\citenamefont {Abada}\ \emph
  {et~al.}(2019{\natexlab{b}})\citenamefont {Abada} \emph
  {et~al.}}]{Abada:2019zxq}%
  \BibitemOpen
  \bibfield  {author} {\bibinfo {author} {\bibfnamefont {A.}~\bibnamefont
  {Abada}} \emph {et~al.} (\bibinfo {collaboration} {FCC}),\ }\href {\doibase
  10.1140/epjst/e2019-900045-4} {\bibfield  {journal} {\bibinfo  {journal}
  {Eur. Phys. J. ST}\ }\textbf {\bibinfo {volume} {228}},\ \bibinfo {pages}
  {261} (\bibinfo {year} {2019}{\natexlab{b}})}\BibitemShut {NoStop}%
\bibitem [{\citenamefont {Crivellin}\ \emph {et~al.}(2018)\citenamefont
  {Crivellin}, \citenamefont {Hoferichter},\ and\ \citenamefont
  {Schmidt-Wellenburg}}]{Crivellin:2018qmi}%
  \BibitemOpen
  \bibfield  {author} {\bibinfo {author} {\bibfnamefont {A.}~\bibnamefont
  {Crivellin}}, \bibinfo {author} {\bibfnamefont {M.}~\bibnamefont
  {Hoferichter}}, \ and\ \bibinfo {author} {\bibfnamefont {P.}~\bibnamefont
  {Schmidt-Wellenburg}},\ }\href {\doibase 10.1103/PhysRevD.98.113002}
  {\bibfield  {journal} {\bibinfo  {journal} {Phys. Rev.}\ }\textbf {\bibinfo
  {volume} {D98}},\ \bibinfo {pages} {113002} (\bibinfo {year} {2018})},\
  \Eprint {http://arxiv.org/abs/1807.11484} {arXiv:1807.11484 [hep-ph]}
  \BibitemShut {NoStop}%
\bibitem [{\citenamefont {Crivellin}\ \emph {et~al.}(2014)\citenamefont
  {Crivellin}, \citenamefont {Hoferichter},\ and\ \citenamefont
  {Procura}}]{Crivellin:2014cta}%
  \BibitemOpen
  \bibfield  {author} {\bibinfo {author} {\bibfnamefont {A.}~\bibnamefont
  {Crivellin}}, \bibinfo {author} {\bibfnamefont {M.}~\bibnamefont
  {Hoferichter}}, \ and\ \bibinfo {author} {\bibfnamefont {M.}~\bibnamefont
  {Procura}},\ }\href {\doibase 10.1103/PhysRevD.89.093024} {\bibfield
  {journal} {\bibinfo  {journal} {Phys. Rev.}\ }\textbf {\bibinfo {volume}
  {D89}},\ \bibinfo {pages} {093024} (\bibinfo {year} {2014})},\ \Eprint
  {http://arxiv.org/abs/1404.7134} {arXiv:1404.7134 [hep-ph]} \BibitemShut
  {NoStop}%
\bibitem [{\citenamefont {Crivellin}\ \emph
  {et~al.}(2017{\natexlab{b}})\citenamefont {Crivellin}, \citenamefont
  {Davidson}, \citenamefont {Pruna},\ and\ \citenamefont
  {Signer}}]{Crivellin:2017rmk}%
  \BibitemOpen
  \bibfield  {author} {\bibinfo {author} {\bibfnamefont {A.}~\bibnamefont
  {Crivellin}}, \bibinfo {author} {\bibfnamefont {S.}~\bibnamefont {Davidson}},
  \bibinfo {author} {\bibfnamefont {G.~M.}\ \bibnamefont {Pruna}}, \ and\
  \bibinfo {author} {\bibfnamefont {A.}~\bibnamefont {Signer}},\ }\href
  {\doibase 10.1007/JHEP05(2017)117} {\bibfield  {journal} {\bibinfo  {journal}
  {JHEP}\ }\textbf {\bibinfo {volume} {05}},\ \bibinfo {pages} {117} (\bibinfo
  {year} {2017}{\natexlab{b}})},\ \Eprint {http://arxiv.org/abs/1702.03020}
  {arXiv:1702.03020 [hep-ph]} \BibitemShut {NoStop}%
\bibitem [{\citenamefont {Egorov}\ \emph {et~al.}(2006)\citenamefont {Egorov}
  \emph {et~al.}}]{Egorov:2006bb}%
  \BibitemOpen
  \bibfield  {author} {\bibinfo {author} {\bibfnamefont {V.}~\bibnamefont
  {Egorov}} \emph {et~al.},\ }\bibfield  {booktitle} {\emph {\bibinfo
  {booktitle} {{Calculation of double-beta-decay matrix elements. Proceedings,
  Workshop, MEDEX'05, Corfu, Greece, September 26-29, 2005}}},\ }\href
  {\doibase 10.1007/s10582-006-0108-4} {\bibfield  {journal} {\bibinfo
  {journal} {Czech. J. Phys.}\ }\textbf {\bibinfo {volume} {56}},\ \bibinfo
  {pages} {453} (\bibinfo {year} {2006})}\BibitemShut {NoStop}%
\bibitem [{\citenamefont {Abazov}\ \emph {et~al.}(2015)\citenamefont {Abazov}
  \emph {et~al.}}]{Abazov:2014jti}%
  \BibitemOpen
  \bibfield  {author} {\bibinfo {author} {\bibfnamefont {V.~M.}\ \bibnamefont
  {Abazov}} \emph {et~al.} (\bibinfo {collaboration} {D0}),\ }\href {\doibase
  10.1103/PhysRevLett.115.041801} {\bibfield  {journal} {\bibinfo  {journal}
  {Phys. Rev. Lett.}\ }\textbf {\bibinfo {volume} {115}},\ \bibinfo {pages}
  {041801} (\bibinfo {year} {2015})},\ \Eprint {http://arxiv.org/abs/1408.5016}
  {arXiv:1408.5016 [hep-ex]} \BibitemShut {NoStop}%
\bibitem [{\citenamefont {Aaltonen}\ \emph {et~al.}(2014)\citenamefont
  {Aaltonen} \emph {et~al.}}]{Aaltonen:2014loa}%
  \BibitemOpen
  \bibfield  {author} {\bibinfo {author} {\bibfnamefont {T.~A.}\ \bibnamefont
  {Aaltonen}} \emph {et~al.} (\bibinfo {collaboration} {CDF}),\ }\href
  {\doibase 10.1103/PhysRevD.89.072005} {\bibfield  {journal} {\bibinfo
  {journal} {Phys. Rev. D}\ }\textbf {\bibinfo {volume} {89}},\ \bibinfo
  {pages} {072005} (\bibinfo {year} {2014})},\ \Eprint
  {http://arxiv.org/abs/1402.2239} {arXiv:1402.2239 [hep-ex]} \BibitemShut
  {NoStop}%
\bibitem [{\citenamefont {CKMfitter~Group}(html)}]{CKMfitter:2019}%
  \BibitemOpen
  \bibfield  {author} {\bibinfo {author} {\bibfnamefont {.}~\bibnamefont
  {CKMfitter~Group}},\ }\href@noop {} {\  (\bibinfo {year}
  {\path{http://ckmfitter.in2p3.fr/www/results/plots_summer19/num/ckmEval_results_summer19.html}})}\BibitemShut
  {NoStop}%
\bibitem [{\citenamefont {Charles}\ \emph {et~al.}(2005)\citenamefont
  {Charles}, \citenamefont {Hocker}, \citenamefont {Lacker}, \citenamefont
  {Laplace}, \citenamefont {Le~Diberder}, \citenamefont {Malcles},
  \citenamefont {Ocariz}, \citenamefont {Pivk},\ and\ \citenamefont
  {Roos}}]{Charles:2004jd}%
  \BibitemOpen
  \bibfield  {author} {\bibinfo {author} {\bibfnamefont {J.}~\bibnamefont
  {Charles}}, \bibinfo {author} {\bibfnamefont {A.}~\bibnamefont {Hocker}},
  \bibinfo {author} {\bibfnamefont {H.}~\bibnamefont {Lacker}}, \bibinfo
  {author} {\bibfnamefont {S.}~\bibnamefont {Laplace}}, \bibinfo {author}
  {\bibfnamefont {F.~R.}\ \bibnamefont {Le~Diberder}}, \bibinfo {author}
  {\bibfnamefont {J.}~\bibnamefont {Malcles}}, \bibinfo {author} {\bibfnamefont
  {J.}~\bibnamefont {Ocariz}}, \bibinfo {author} {\bibfnamefont
  {M.}~\bibnamefont {Pivk}}, \ and\ \bibinfo {author} {\bibfnamefont
  {L.}~\bibnamefont {Roos}} (\bibinfo {collaboration} {CKMfitter Group}),\
  }\href {\doibase 10.1140/epjc/s2005-02169-1} {\bibfield  {journal} {\bibinfo
  {journal} {Eur. Phys. J.}\ }\textbf {\bibinfo {volume} {C41}},\ \bibinfo
  {pages} {1} (\bibinfo {year} {2005})},\ \Eprint
  {http://arxiv.org/abs/hep-ph/0406184} {arXiv:hep-ph/0406184 [hep-ph]}
  \BibitemShut {NoStop}%
\bibitem [{\citenamefont {Buras}\ and\ \citenamefont
  {Girrbach}(2012)}]{Buras:2012fs}%
  \BibitemOpen
  \bibfield  {author} {\bibinfo {author} {\bibfnamefont {A.~J.}\ \bibnamefont
  {Buras}}\ and\ \bibinfo {author} {\bibfnamefont {J.}~\bibnamefont
  {Girrbach}},\ }\href {\doibase 10.1007/JHEP03(2012)052} {\bibfield  {journal}
  {\bibinfo  {journal} {JHEP}\ }\textbf {\bibinfo {volume} {03}},\ \bibinfo
  {pages} {052} (\bibinfo {year} {2012})},\ \Eprint
  {http://arxiv.org/abs/1201.1302} {arXiv:1201.1302 [hep-ph]} \BibitemShut
  {NoStop}%
\bibitem [{\citenamefont {Ciuchini}\ \emph {et~al.}(1998)\citenamefont
  {Ciuchini}, \citenamefont {Franco}, \citenamefont {Lubicz}, \citenamefont
  {Martinelli}, \citenamefont {Scimemi},\ and\ \citenamefont
  {Silvestrini}}]{Ciuchini:1997bw}%
  \BibitemOpen
  \bibfield  {author} {\bibinfo {author} {\bibfnamefont {M.}~\bibnamefont
  {Ciuchini}}, \bibinfo {author} {\bibfnamefont {E.}~\bibnamefont {Franco}},
  \bibinfo {author} {\bibfnamefont {V.}~\bibnamefont {Lubicz}}, \bibinfo
  {author} {\bibfnamefont {G.}~\bibnamefont {Martinelli}}, \bibinfo {author}
  {\bibfnamefont {I.}~\bibnamefont {Scimemi}}, \ and\ \bibinfo {author}
  {\bibfnamefont {L.}~\bibnamefont {Silvestrini}},\ }\href {\doibase
  10.1016/S0550-3213(98)00161-8} {\bibfield  {journal} {\bibinfo  {journal}
  {Nucl. Phys.}\ }\textbf {\bibinfo {volume} {B523}},\ \bibinfo {pages} {501}
  (\bibinfo {year} {1998})},\ \Eprint {http://arxiv.org/abs/hep-ph/9711402}
  {arXiv:hep-ph/9711402 [hep-ph]} \BibitemShut {NoStop}%
\bibitem [{\citenamefont {Buras}\ \emph {et~al.}(2000)\citenamefont {Buras},
  \citenamefont {Misiak},\ and\ \citenamefont {Urban}}]{Buras:2000if}%
  \BibitemOpen
  \bibfield  {author} {\bibinfo {author} {\bibfnamefont {A.~J.}\ \bibnamefont
  {Buras}}, \bibinfo {author} {\bibfnamefont {M.}~\bibnamefont {Misiak}}, \
  and\ \bibinfo {author} {\bibfnamefont {J.}~\bibnamefont {Urban}},\ }\href
  {\doibase 10.1016/S0550-3213(00)00437-5} {\bibfield  {journal} {\bibinfo
  {journal} {Nucl. Phys.}\ }\textbf {\bibinfo {volume} {B586}},\ \bibinfo
  {pages} {397} (\bibinfo {year} {2000})},\ \Eprint
  {http://arxiv.org/abs/hep-ph/0005183} {arXiv:hep-ph/0005183 [hep-ph]}
  \BibitemShut {NoStop}%
\bibitem [{\citenamefont {Aoki}\ \emph {et~al.}(2020)\citenamefont {Aoki} \emph
  {et~al.}}]{Aoki:2019cca}%
  \BibitemOpen
  \bibfield  {author} {\bibinfo {author} {\bibfnamefont {S.}~\bibnamefont
  {Aoki}} \emph {et~al.} (\bibinfo {collaboration} {Flavour Lattice Averaging
  Group}),\ }\href {\doibase 10.1140/epjc/s10052-019-7354-7} {\bibfield
  {journal} {\bibinfo  {journal} {Eur. Phys. J.}\ }\textbf {\bibinfo {volume}
  {C80}},\ \bibinfo {pages} {113} (\bibinfo {year} {2020})},\ \Eprint
  {http://arxiv.org/abs/1902.08191} {arXiv:1902.08191 [hep-lat]} \BibitemShut
  {NoStop}%
\bibitem [{\citenamefont {Algueró}\ \emph
  {et~al.}(2019{\natexlab{b}})\citenamefont {Algueró}, \citenamefont
  {Capdevila}, \citenamefont {Descotes-Genon}, \citenamefont {Masjuan},\ and\
  \citenamefont {Matias}}]{Alguero:2018nvb}%
  \BibitemOpen
  \bibfield  {author} {\bibinfo {author} {\bibfnamefont {M.}~\bibnamefont
  {Algueró}}, \bibinfo {author} {\bibfnamefont {B.}~\bibnamefont {Capdevila}},
  \bibinfo {author} {\bibfnamefont {S.}~\bibnamefont {Descotes-Genon}},
  \bibinfo {author} {\bibfnamefont {P.}~\bibnamefont {Masjuan}}, \ and\
  \bibinfo {author} {\bibfnamefont {J.}~\bibnamefont {Matias}},\ }\href
  {\doibase 10.1103/PhysRevD.99.075017} {\bibfield  {journal} {\bibinfo
  {journal} {Phys. Rev.}\ }\textbf {\bibinfo {volume} {D99}},\ \bibinfo {pages}
  {075017} (\bibinfo {year} {2019}{\natexlab{b}})},\ \Eprint
  {http://arxiv.org/abs/1809.08447} {arXiv:1809.08447 [hep-ph]} \BibitemShut
  {NoStop}%
\bibitem [{\citenamefont {Aad}\ \emph {et~al.}(2020)\citenamefont {Aad} \emph
  {et~al.}}]{Aad:2020otl}%
  \BibitemOpen
  \bibfield  {author} {\bibinfo {author} {\bibfnamefont {G.}~\bibnamefont
  {Aad}} \emph {et~al.} (\bibinfo {collaboration} {ATLAS}),\ }\href {\doibase
  10.1007/JHEP11(2020)005} {\bibfield  {journal} {\bibinfo  {journal} {JHEP}\
  }\textbf {\bibinfo {volume} {11}},\ \bibinfo {pages} {005} (\bibinfo {year}
  {2020})},\ \Eprint {http://arxiv.org/abs/2006.12946} {arXiv:2006.12946
  [hep-ex]} \BibitemShut {NoStop}%
\bibitem [{\citenamefont {Cirigliano}\ \emph {et~al.}(2019)\citenamefont
  {Cirigliano}, \citenamefont {Falkowski}, \citenamefont {González-Alonso},\
  and\ \citenamefont {Rodríguez-Sánchez}}]{Cirigliano:2018dyk}%
  \BibitemOpen
  \bibfield  {author} {\bibinfo {author} {\bibfnamefont {V.}~\bibnamefont
  {Cirigliano}}, \bibinfo {author} {\bibfnamefont {A.}~\bibnamefont
  {Falkowski}}, \bibinfo {author} {\bibfnamefont {M.}~\bibnamefont
  {González-Alonso}}, \ and\ \bibinfo {author} {\bibfnamefont
  {A.}~\bibnamefont {Rodríguez-Sánchez}},\ }\href {\doibase
  10.1103/PhysRevLett.122.221801} {\bibfield  {journal} {\bibinfo  {journal}
  {Phys. Rev. Lett.}\ }\textbf {\bibinfo {volume} {122}},\ \bibinfo {pages}
  {221801} (\bibinfo {year} {2019})},\ \Eprint
  {http://arxiv.org/abs/1809.01161} {arXiv:1809.01161 [hep-ph]} \BibitemShut
  {NoStop}%
\bibitem [{\citenamefont {Schael}\ \emph {et~al.}(2013)\citenamefont {Schael}
  \emph {et~al.}}]{Schael:2013ita}%
  \BibitemOpen
  \bibfield  {author} {\bibinfo {author} {\bibfnamefont {S.}~\bibnamefont
  {Schael}} \emph {et~al.} (\bibinfo {collaboration} {ALEPH, DELPHI, L3, OPAL,
  LEP Electroweak}),\ }\href {\doibase 10.1016/j.physrep.2013.07.004}
  {\bibfield  {journal} {\bibinfo  {journal} {Phys. Rept.}\ }\textbf {\bibinfo
  {volume} {532}},\ \bibinfo {pages} {119} (\bibinfo {year} {2013})},\ \Eprint
  {http://arxiv.org/abs/1302.3415} {arXiv:1302.3415 [hep-ex]} \BibitemShut
  {NoStop}%
\end{thebibliography}%

\end{document}